\documentclass[nofootinbib,eqsecnum,tightenlines,superscriptaddress,11pt]{revtex4}

\usepackage{hyperref}
\usepackage{graphicx}
\usepackage{amsmath,amssymb,amsfonts,amsthm,stmaryrd,mathtools,bm,physics}
\usepackage{color}
\allowdisplaybreaks[1]

\usepackage{tikz}

\def\be#1\ee{\begin{align}#1\end{align}}

\def\ba{\begin{eqnarray}}
\def\ea{\end{eqnarray}}
\def\nn{\nonumber}
\def\q{\quad}

\begin{document}

\title{Lorentzian quantum gravity via Pachner moves: one-loop evaluation}

\author{Johanna N.~Borissova}
\email{jborissova@perimeterinstitute.ca} 
\affiliation{Perimeter Institute, 31 Caroline Street North, Waterloo, ON, N2L 2Y5, Canada}
\affiliation{Department of Physics and  Astronomy, University of Waterloo, 200 University Avenue West, Waterloo, ON, N2L 3G1, Canada}
\author{Bianca Dittrich}
\email{bdittrich@perimeterinstitute.ca} 
\affiliation{Perimeter Institute, 31 Caroline Street North, Waterloo, ON, N2L 2Y5, Canada}

\begin{abstract}

Lorentzian quantum gravity is believed to cure the pathologies encountered in Euclidean quantum gravity, such as the conformal factor problem. We show that this is the case for the Lorentzian Regge path integral expanded around a flat background. We illustrate how a subset of local changes of the triangulation, so-called Pachner moves, allow to isolate the indefinite nature of the gravitational action at the discrete level. The latter can be accounted for by oppositely chosen deformed contours of integration. Moreover, we construct a discretization-invariant local path integral measure for 3D Lorentzian Regge calculus and point out obstructions in defining such a measure in 4D.  We see the work presented here as a first step to establish the existence of the non-perturbative Lorentzian path integral for Regge calculus and related frameworks such as spin foams.  

An extensive appendix  provides an overview of Lorentzian Regge calculus, using the recently established concept of the complexified Regge action, and derives useful geometric formulae and identities needed in the main text.

\end{abstract}

\maketitle
\tableofcontents

\section{Introduction}\label{Sec:Introduction}

 Path integrals for quantum gravity aim to define quantum transition amplitudes by formally summing over all spacetime histories satisfying suitable boundary conditions and weighted by the exponential of $\imath$ times the action in units of Planck's constant $\hbar$. Prominent path integral approaches for quantum gravity are spin foams~\cite{PerezLR}, quantum Regge gravity \cite{WilliamsReview}, group field theories~\cite{Oriti:2006se}, or (causal) dynamical triangulations~\cite{Loll:2019rdj}, and as such they depend on a variety of choices. These choices include the type of histories to sum over, for instance defining the set of spacetime geometries, specifying configuration variables to parametrize this set, and imposing additional regularity conditions on these. The other essential ingredients for gravitational path integrals are the choice of action dictating the dynamics and the measure over the space of geometries, i.e.~equivalence classes of metrics modulo diffeomorphisms.

Path integrals with oscillatory weights $\exp(\imath S)$ are challenging as the integrations are
typically over unbounded domains, leading to integrals which are not absolutely converging, and for which there are only few effective tools for their evaluation available \cite{Visser,Dittrich:2014mxa, Delcamp:2016dqo,Turok,Ding, ADP21,toappear}. This is one of the main motivations for turning to Euclidean quantum gravity
approaches \cite{LollLR}. These are based on the argument that a formal Wick rotation changes Lorentzian metrics into Euclidean ones and changes the Lorentzian weight $\exp(\imath S_L)$ into the Euclidean non-oscillatory weight $\exp(-S_E)$. This allows to apply e.g. Monte-Carlo simulations.  
 
But Euclidean quantum gravity approaches suffer from two crucial drawbacks: firstly the fact that the gravitational action is not bounded from below, and can be made arbitrarily negative by maximizing the kinetic term for the conformal factor~\cite{ConformalFactor}. To obtain a convergent Euclidean path integral it is argued that the conformal mode needs to be Wick rotated in the opposite direction from the other modes~\cite{ConformalFactor}. In praxis it is necessary to flip the sign in front of the conformal mode, but the latter is hard to identify in non-perturbative approaches. Secondly, the space of Lorentzian  geometries is very different from the space of Euclidean  geometries,  a one-to-one mapping does not exist \cite{CDT1,CDT2}.

This motivates facing the Lorentzian path integral \cite{deBoer:2022zka}. We here consider the Lorentzian path integral for Regge calculus \cite{Regge}. Regge calculus provides a discretization for the Einstein-Hilbert action and thus a regularization for the gravitational path integral.  
The  main questions we will analyze here are: \\
 $(a)$ Can we expect a convergent Lorentzian path integral, so that the conformal factor problem is cured?  Additionally, can we establish a relation between the deformation of the integration contour in the Lorentzian case and Euclidean quantum gravity? In other words, can we justify the procedure of Euclidean quantum gravity? \\ $(b)$ Another set of questions we will consider are  related to fixing the measure in the path integral by demanding invariance under a change of (bulk) discretization of the path integral and thus a regularization-independent path integral. This has been successfully implemented for three-dimensional Euclidean gravity in \cite{Dittrich:2011vz}, and has allowed interesting applications, e.g. the computation of one-loop partition functions and confirmation of a holographic duality in three-dimensional quantum gravity (without a cosmological constant) \cite{Bonzom:2015ans,Dittrich:2017hnl}. On the other hand, it has been shown in \cite{Dittrich:2014rha}, that four-dimensional Regge gravity does not admit a local discretization-independent measure. More specifically, there does not even exist a local measure, which is invariant under a specific class of changes of triangulation, which leave the action invariant.  These results reflect the topological nature of three-dimensional gravity on the one hand, and the fact that four-dimensional gravity has local propagating degrees of freedom on the other hand. For the latter we have to expect that discretizations break diffeomorphism symmetry \cite{Dittrich:2008pw, Bahr:2009ku}, which implies also a dependence of the path integral on the discretization and thus regulator \cite{Dittrich:2011ien,DittrichBook14,Asante:2022dnj}. We expect to find similar results in the Lorentzian case. Indeed, we will succeed in constructing an invariant measure for the three-dimensional Lorentzian path integral, and the Lorentzian case will add an interesting twist to the Euclidean discussion, related to how the conformal factor is treated in Euclidean quantum gravity.

The Regge action constitutes a complicated non-polynomial function in the edge lengths, which are taken as fundamental variables in Regge calculus. Additionally one has to implement generalized triangle inequalities. One can therefore not hope for an exact analytical evaluation of the path integral. We will therefore consider a perturbative approach and evaluate the path integral to one-loop order, on a flat background. We will see, that this allows us to be, on the one hand, most general, and on the other hand still derive explicit (and quite simple) expressions for the path integral. 

To achieve these simple expressions, we will break down the general discrete path integral into a sequence of so-called Pachner moves. Similar to applying a coarse gaining framework, this allows to perform the path integral in terms of smaller steps. We will see that these steps always involve only a one-dimensional integration (and for a certain class of Pachner moves, a gauge fixing).

The main technical task is then to derive an expression for the Hessian of the Regge action for the various Pachner move configurations. We will do so starting from some identities for a measure over flat-space geometries  \cite{Korepanov:2000jp,Korepanov:2000aj}, which previously have been applied to the construction of a triangulation-invariant partition function \cite{Baratin:2006yu,Baratin:2006gy,Baratin:2014era} for topological BFCG theory \cite{Girelli,Asante:2019lki}. These techniques were adjusted and applied to the computation of the Pachner move Hessians in Euclidean Regge gravity in \cite{Dittrich:2011vz}. The paper \cite{Dittrich:2014rha} streamlined this derivation by introducing a Caley-Menger determinant associated to the Pachner move configurations. We will pick up this technique and apply a further simplification, which will allow us to derive a simple expression for the Lorentzian Regge Hessian for all Pachner moves in general dimensions (larger than two). Using this new simplification we will in particular be able to determine the various signs appearing in the Pachner moves, which will be important to determine the deformations of the integration contour in the Lorentzian path integral, and thus the relation between Lorentzian and Euclidean path integrals.

~\\
Our paper is structured as follows. In Section~\ref{Sec:LorentzianReggeCalculus} we review the definition of complex dihedral angles and deficit angles needed to define the Lorentzian Regge action and Lorentzian Regge path integral. Section~\ref{Sec:PachnerMoves} discusses Pachner moves in general dimensions with primary focus on 3D and 4D, as well as on the properties of a Caley-Menger determinant associated to Pachner moves. In Section~\ref{Sec:LinearizedReggeCalculus} we expand the Lorentzian Regge action around a flat background and derive a simple form for the Hessians for all Pachner moves, which will show that these Hessians factorize. Section~\ref{Sec:PathIntegralMeasure} defines the path integral for Lorentzian Regge calculus in a flat background expansion to second order, for a given Pachner move, and collects identities on one-dimensional and multi-dimensional Gaussian integrals with imaginary exponent. In Sections~\ref{Sec:3DPachnerMoves} and \ref{Sec:4DPachnerMoves} we analyze the path integral for the $3-2$ and $4-1$ moves in 3D, as well as for the $5-1$ and $4-2$ moves in 4D, and discuss the relation between Lorentzian and Euclidean path integral. As expected, we will manage to construct a discretization-invariant local measure in 3D, and illustrate that such a measure in 4D does not exist. We finish with a discussion and outlook in Section~\ref{Sec:DiscussionOutlook}.

Our paper is supplemented by an extended appendix providing an in-depth  overview on the geometry of Lorentzian and Euclidean simplices. The reason for doing so is that such an cohesive overview, in particular with regard to Lorentzian simplices, seems to be missing from the literature, and a large number of geometrical identities will be needed to follow the discussion the main text. Our overview includes Caley-Menger determinants, generalized triangle inequalitites and length Gram matrices, as well as dual or angle Gram matrices. The appendix  reviews the steps towards defining and computing complex dihedral angles. This includes a discussion on how to project out a hinge and the definition of angles in Euclidean and Minkowskian planes. This allows us to construct  the complex Regge action which unifies the Euclidean and Lorentzian action into one compact form. Finally, we derive the Schl\"afli identity for Euclidean and Lorentzian triangulations and an extremely useful formula for the derivatives of dihedral angles, needed in the main text.

\section{Lorentzian (quantum) Regge calculus}\label{Sec:LorentzianReggeCalculus}

Regge calculus~\cite{Regge} provides a coordinate-free formulation of general relativity on a piecewise flat discretization of the spacetime manifold. The basic building blocks in a piecewise flat\footnote{Formulations with homogeneously curved building blocks also exist \cite{Bahr:2009qc,NewRegge}.} discretization of a $d$-dimensional spacetime  are $d$-simplices glued along shared $(d-1)$-subsimplices.  The framework has been originally defined for Euclidean spacetimes, a formulation for Lorentzian spacetimes has been provided by Sorkin \cite{Sorkin1974,Sorkin2019}. Recent work \cite{Ding,ADP21} extended Regge calculus to a complexified configuration space, which allows a unified framework for Euclidean and Lorentzian geometries. In particular, \cite{ADP21} finds a relationship between the violation of causality conditions and the branch cut structure of the action.

In (length) Regge calculus the metric tensor which serves as the fundamental variable in the continuum metric formulation of gravity, is replaced  by geometric degrees of freedom associated to the edges $e$ of the simplicial building blocks.\footnote{There exist also other variants of Regge caculus, which are important for spin foam gravity \cite{PerezLR,EffSF1}. On the one hand, there is area Regge calculus in four dimensions \cite{Barrettetal, ADHAreaR}, whose discrete equations of motion differ from length Regge calculs, but whose perturbative construction of the continuum limit agrees with the one of length Regge calculus \cite{HC1,HC2}.  On the other hand, there is area angle Regge calculus \cite{DittrichSpeziale}, which, through adjusting constraints, can interpolate between a discretization of a topological field theory and the equation of motion for length Regge calculus. Area and length Regge calculus are also available in a first-order formulation \cite{BarrettFO,NewRegge,ADHAreaR}.} On a Lorentzian triangulation these can be taken as the set of signed squared edge lengths $s_e \equiv \vec{e}\cdot \vec{e}$, defined by the norms of the edge vectors with respect to the Minkowski metric $\eta = \text{diag}(-1,+1,+1,+1)$. Thus $s_e >0$ for a space-like edge, $s_e<0$ for a time-like edge and $s_e=0$ for a null edge. We will call these quantities $s_e$ signed length squares. The geometry of a flat $d$-simplex $\sigma$ is completely characterized by the geometric data $\{s_e|e \subset \sigma\}$.

Curvature in a simplicial complex is concentrated at codimension two subsimplices $h \subset \sigma$, so-called ``hinges", and parametrized by deficit angles $\epsilon_{h}$. The latter provide a measure for the angular gap between the $d$-simplices $\sigma$ meeting at a given hinge $h$ which would arise after flattening the $d$-simplices to $d$-dimensional Minkowski space. The deficit angles can be deduced from successive projections of adjacent $d$-simplices onto planes orthogonal to the hinges, whereby the resulting solid angles correspond to the $d$-dimensional dihedral angles $\theta_{\sigma,h}$ at $h\subset \sigma$. 

We provide in Appendix \ref{SecSub:DefinitionsAngles} an overview how to define and compute the dihedral and deficit angles for Lorentzian (and also Euclidean) triangulations, using the formalism of complex angles in \cite{ADP21}. There are two versions $\theta^\pm_{\sigma,h}$ for the complex angles, in the following we will use $\theta^+_{\sigma,h}$, but drop the superscript $+$. This complex dihedral angle can be defined in terms of the signed length squares as

\be\label{eq:DihedralAngles}
\theta^+_{\sigma,h} = -\imath \log_- \qty(\frac{
	\frac{d^2}{  \mathbb{V}_{h}}  \frac{\partial \mathbb{V}_{\sigma}}{\partial s_{\bar{h}}}  -  \imath \,
	\sqrt{\!\!{}_{{}_+} \,\,    \frac{\mathbb{V}_{\rho_a} }{   \mathbb{V}_{h}  }\frac{\mathbb{V}_{\rho_b}}{\mathbb{V}_{h} }     -\qty( \frac{d^2}{  \mathbb{V}_{h}}  \frac{\partial \mathbb{V}_{\sigma}}{\partial s_{\bar{h}}} )^2
	} 
} 
{ \sqrt{\!\!{}_{{}_+} \,\,      \frac{\mathbb{V}_{\rho_a} }{   \mathbb{V}_{h}  } }   \sqrt{\!\!{}_{{}_+} \,\,  \frac{\mathbb{V}_{\rho_b}}{    \mathbb{V}_{h} }} } ) \, .
\ee
Here we indicate with $\rho_a$ and $\rho_b$  the two $(d-1)$-subsimplices  which share the hinge $h$ and $s_{\bar{h}}$ denotes the signed length square of the edge opposite to $h$ in the $d$-simplex $\sigma$. The signed volume square $\mathbb{V}_X$ of a simplex $X$ can be computed from its associated Caley-Menger determinant and gives the positive (negative) square of the volume if $X$ is spacelike (timelike), see Appendix \ref{SecSub:CaleyMenger}. 

For $z \in \mathbb{C}$ with $\arg(z) \in (-\pi,\pi)$ we adopt the principal branch for the logarithm $\log_-(z)$ and the square root $\sqrt{\!\!{}_{{}_+} \,\, z}$. But, along the branch cut when $z$ is a negative real number $-r<0$, we use $\log_-(-r)=\log(r)-\imath \pi$ and $\sqrt{\!\!{}_{{}_+} \,\, -r}=\imath \sqrt{r}$. Later, we will only deal with the derivatives of the dihedral angle. Here we have ${\bf d} \log_-(z)={\bf d}z/z$ and ${\bf d} \sqrt{\!\!{}_{{}_+} \,\, z}={\bf d}z/(2 \sqrt{\!\!{}_{{}_+} \,\, z})$. 
Thus we are only left with $\sqrt{\!\!{}_{{}_+}\,\,z }$, and as the square root adopts the principal value for the branch cut, we will henceforth omit the subscript $+$.

The (complex) deficit and boundary curvature angles are then given as 
\be\label{eq:DeficitAngles}
\epsilon^{\text{(bulk)}}_{h} =2\pi + \sum_{\sigma \supset h} \theta_{\sigma,h}\,, \quad\quad 
\epsilon^{\text{(bdry)}}_{h} = \pi k + \sum_{\sigma \supset h} \theta_{\sigma,h}\,.
\ee 
The sum in~\eqref{eq:DeficitAngles} runs over all $d$-simplices $\sigma$ in the $d$-dimensional  triangulation $\triangle^{(d)}$, which contain the hinge $h$. The parameter $k$ for the boundary curvature angle depends on the number of pieces glued together at this boundary, e.g.~$k=1$ if two pieces are glued together.

As detailed in  Appendix \ref{SecSub:ReggeAction}, the Lorentzian Regge action for a triangulation $\triangle^{(d)}$ of $d$-dimensional spacetime with vanishing cosmological constant can be written as~\cite{ADP21} 
\be\label{eq:LorentzianReggeAction}
\imath S_{\text{Regge}} = \sum_{h\subset \triangle^{(d)}}\sqrt{\mathbb{V}_{h}} \, \epsilon_{h}\,,
\ee
where we again use $\sqrt{z}=\sqrt{\!\!{}_{{}_+}\,\,z }$. Note that $S_{\text{Regge}}$ is real, if $(a)$ the generalized Lorentzian triangle inequalities are satisfied, see Appendix \ref{SecSub:CaleyMenger} and if, $(b)$, the triangulation has no light cone irregularities, see \cite{ADP21}.  Small perturbations around a triangulation of flat space will be light cone regular.

~\\

The Lorentzian path integral for Regge gravity can then be defined as 
\be\label{eq:LorentzianPathIntegral}
Z = \int_{s_{e{| e \subset \text{bdry}}}} \prod_{e\subset \text{bulk}} \dd{s_e} \mu(s_e)\exp{\imath S_{\text{Regge}}}\,.
\ee
Here $\mu(s_e)$ denotes a suitable measure factor depending on the configuration variables which are taken to be the signed squared edge lengths $s_e$. As boundary conditions we assume fixed signed length squares  for the edges in the boundary of the triangulation. In all expressions natural units and therefore $\hbar=1$ is implied.

 The configurations included in the integral~\eqref{eq:LorentzianPathIntegral} are required to satisfy the Lorentzian generalized triangle inequalities which guarantee the ``realizability" of a given set of edge data as edge data of a triangulation, where each $d$-simplex can be embedded into $d$-dimensional Minkowskian spacetime, see Appendix \ref{SecSub:CaleyMenger}. Note that the Lorentzian generalized triangle inequalities and the Euclidean generalized triangle inequalities specify disjoint data sets, if we demand also non-degenerate simplices. 
 
In this way (\ref{eq:LorentzianReggeAction}) gives $\imath$ times the Lorentzian Regge action for Lorentzian data. But it gives minus the Euclidean action for Euclidean data \cite{ADP21}, see also Appendix \ref{SecSub:ReggeAction}. One can thus easily adjust the results for the Hessian of the Regge action developed in the main text to Euclidean space times.

Apart from the generalized triangle inequalities one also needs to specify how to deal with Lorentzian data that feature light cone irregular structures. E.g.~one can have vertices, where  more or less than two light cones meet. A certain class of light cone irregular structures lead to branch cuts for the Regge action. Thus, if one decides to integrate over these structures one has also to specify the side of the branch cut. Note that (\ref{eq:LorentzianReggeAction}) implements a certain choice of branch cut, which, however does not need to be adopted for the path integral. See \cite{ADP21} for explicit examples and a more in-depth discussion. We will here eventually adopt a perturbation around a flat background and thus assume that such light cone irregular structures do not appear.

The dihedral angles~\eqref{eq:DihedralAngles} are complicated functions of the signed squared edge lengths.  Additionally one has to satisfy the generalized triangle inequalities and possibly implement (rather involved) conditions for a regular light cone structure.

Thus one can in general not compute the path integral~\eqref{eq:LorentzianPathIntegral} analytically.\footnote{The two-dimensional case might be an exception, as the Regge action is a topological invariant. Therefore there is only the measure term. In the following we will consider a perturbative expansion of the Regge action and therefore restrict to dimensions $d\geq 3$.} Even a symmetry reduction of the degrees of freedom to a Regge model describing cosmology and featuring only one integration variable still requires numerical integration techniques \cite{DGS,ADP21,toappear}.

In the following, we therefore consider Lorentzian Regge calculus linearized around a flat background solution, which is assumed to satisfy the generalized inequalities and to have a light cone regular structure.  Only perturbations around the classical background, viewed as quantum fluctuations, will be integrated over in the path integral.

\subsection{Strategy for the (approximate) evaluation of the path integral}

Instead of evaluating the entire path integral at once we will here pursue a strategy where the path integral evaluation is split into a number of basic steps. These steps will be associated to Pachner moves, which describe local changes of the (bulk) triangulation. Any two triangulations of a piecewise linear manifold can be related by a finite sequence of such Pachner moves~\cite{Pachner:1991}.  

Instead of analyzing the full path integral we therefore need to only analyze the various steps associated to the Pachner moves. This allows us to pursue the main aim of this paper, namely to answer the questions $(a)$ and $(b)$ raised in the introduction (Section \ref{Sec:Introduction}). To answer question $(a)$, whether the Lorentzian path integral is finite, we can now consider whether the steps associated to each of the Pachner moves lead to a finite expression or not. Similarly, we can consider question $(b)$, which asks whether we can construct a triangulation-invariant path integral, and in particular a local triangulation-invariant measure, for each of the Pachner moves.   

As outlined in the introduction, the answer to question $(b)$ will be positive for three-dimensional Regge calculus  and negative for four-dimensional Regge calculus. This has important implications for the evaluation of the Regge path integral: 

{\bf In the three-dimensional case} one finds that the Regge Hamilton-Jacobi function (i.e.~the Regge action evaluated on solutions to the equations of motion) is independent of the triangulation \cite{Dittrich:2011vz}. We will also succeed in constructing a triangulation-invariant (to one-loop) path integral measure. Thus, the partition function $Z$ will be invariant under changes of the triangulation, and in particular under Pachner moves. 

The three-dimensional Pachner moves can be divided into two sets. The first set can be interpreted as coarse-graining moves: here one removes edges from the triangulation, which for the path integral evaluation means to integrate out the associated edge length variables. Here we will show that these integrations lead to  finite results, if we do account for  remnant diffeomorphism symmetries associated to the bulk vertices of the triangulation. Moreover, choosing an appropriate measure, we will show that the partition function is invariant (to one-loop order) under such coarse graining moves. I.e., the result after integrating out the edge length variables associated to the coarse graining Pachner move is the same as defining the path integral for the same triangulation as before, but with the coarse-graining Pachner move applied to it.

In terms of formulae,  we can rewrite the path integral (\ref{eq:LorentzianPathIntegral}) as
\ba\label{2.5}
Z = \int_{s_{e{| e \subset \text{bdry}}}} \prod_{e\subset \text{bulk}'} \dd{s_e} \mu(\{s_e\}_{e\subset \text{bulk}' })\exp{\imath S^{\text {coarse}}_{\text{Regge}}} \, \int \prod_{e \subset \text{Pachner}} \dd{s_e} \mu^{\rm Pachner} (\{s_e\}) \exp{\imath S^{\text {Pachner}}_{\text{Regge}}} \,,\nn\\
\ea
where we have split the integrations for the edges $e \subset \text{Pachner}$ which are removed in the coarse graining Pachner move from the remaining bulk edges $e\subset \text{bulk}'$. Here $S^{\text {coarse}}_{\text{Regge}}$ is the action for the triangulation to which one has applied the coarse-graining Pachner move. $S^{\text {Pachner}}_{\text{Regge}}$ is the action associated to the Pachner move, given by the difference of the actions for the triangulations before and after applying the Pachner move. Thus the action for the initial (fine) triangulation is $S^{\text {fine}}_{\text{Regge}}=S^{\text {coarse}}_{\text{Regge}}+S^{\text {Pachner}}_{\text{Regge}}$ and (\ref{2.5}) is indeed just a rewriting of (\ref{eq:LorentzianPathIntegral}) applied to the initial (fine) triangulation. (Here we also assume that the measure term can be split into a term associated to the Pachner move and a remaining term.) If we choose the measure to be equal to the triangulation-invariant measure we will construct in the course of the article, one finds that, to one loop order,
\ba\label{2.6}
Z = \int_{s_{e{| e \subset \text{bdry}}}} \prod_{e\subset \text{bulk}'} \dd{s_e} \mu(\{s_e\}_{e\subset \text{bulk}' })\exp{\imath S^{\text {coarse}}_{\text{Regge}}}      \q .
\ea

The second set of Pachner moves arises from the inverse moves of the first set and can therefore be interpreted as refining moves. This corresponds to replacing (\ref{2.6}) by (\ref{2.5}).  This should not be a surprise, as the coarse graining and refining moves are inverses of each other.

Thus, to evaluate the path integral for a general bulk triangulation associated to a given boundary triangulation we can change the triangulation via Pachner moves and arrive in this way at a coarsest triangulation. (The notion of coarsest triangulation is not unique, but one can demand that the number of bulk edges should be minimal.)  Due to the triangulation independence of the three-dimensional Regge path integral (using the measure we will later construct), the resulting path integral for this coarsest triangulation is the same that one would have defined right away for this coarsest triangulation. 

Altogether we will show in this work, that we can replace any bulk triangulation with an arbitrary coarse triangulation and just evaluate the path integral for this coarse triangulation. To compute the  (one-loop order) path integral for such a coarse triangulation one needs to evaluate a multi-dimensional Gaussian path integral with purely imaginary argument of the exponent. This can be done by diagonalizing the Hessian of the action, see the discussion in Section~\ref{SecSub:IntegralGaussianImaginary}.
 The details of such a computation will depend very much on the boundary triangulation at hand and is beyond the scope of this article. See \cite{Bonzom:2015ans} for a similar computation of the partition function for the Euclidean solid torus (proving a holographic relationship to a theory defined on the boundary of the triangulation), which can now also be performed in Lorentzian signature.

{\bf In the four-dimensional case} the Regge Hamilton-Jacobi function is invariant only under a subset of the Pachner moves \cite{Dittrich:2011vz}, namely the so-called $5-1$ and $4-2$ Pachner moves. The Hamilton-Jacobi function is in general not invariant under $3-3$ Pachner moves. Regarding the path integral measure, \cite{Dittrich:2014rha} showed that for Euclidean Regge calculus, there does not exist a local invariant measure, even if one restricts only to the $5-1$ and $4-2$ Pachner moves. The arguments in \cite{Dittrich:2014rha} are algebraic and we can thus assume that they hold also for Lorentzian signature.

Thus in four dimensions the path integral does depend on the choice of triangulation. This difference between the four-dimensional and three-dimensional case is due to the fact that four-dimensional general relativity features local (propagating) degrees of freedom, whereas three-dimensional gravity does not \cite{Dittrich:2011ien}.

We will nevertheless discuss the  $5-1$ and $4-2$ Pachner moves and the corresponding procedures of integrating out length variables from the path integral. 

To apply the same strategy of reducing a triangulation to a coarsest triangulation as in the three-dimensional case one has  to start with a triangulation that can be constructed by applying only $1-5$ and $2-4$ Pachner moves from a coarsest triangulation.  As the $1-5$ is inverse to the $5-1$ move and the $2-4$ to the $4-2$ move one can thus arrive at the coarsest triangulation by applying coarse-graining Pachner moves and integrating out length variables accordingly.  


~\\
~\\
In principle we can evaluate the perturbative Regge path integral to one-loop order for an arbitrary triangulation by diagonalizing the Hessian of the Regge action and by applying the standard formulae for a multi-dimensional Gaussian path integral with purely imaginary exponent, see Section \ref{SecSub:IntegralGaussianImaginary}.  A main point of this work is however to investigate the fate of the conformal mode problem in Lorentzian signature.  Indeed, splitting the path integral into Pachner moves, we can interpret the integrations associated to the $4-1$ and $5-1$ moves as conformal modes. The reason for doing so, is that for Euclidean Regge calculus the Hessians associated to the (non-gauge) modes appearing in these moves come with the `wrong sign' \cite{Dittrich:2011vz}, and need to be rotated by hand. 

We will find that the sign for this Hessian in the Lorentzian case does again depend on whether we consider the three-dimensional or four-dimensional case. In the three-dimensional case we will find that there is a global sign (both for Euclidean and Lorentzian signature), which only depends on the type of Pachner move, but not on the length of the edges in the background triangulation.  In the four-dimensional case, whereas for Euclidean signature the Hessian for the $5-1$ move has a global sign, the signs for the Lorentzian signature case do depend on the background edge lengths. We conclude that there is no global notion of Wick rotation that connects the Lorentzian and Euclidean signature cases.

\section{Pachner moves and their properties}\label{Sec:PachnerMoves}

Pachner moves~\cite{Pachner:1991} describe local changes of the bulk triangulation which keep the boundary triangulation and thus the extrinsic geometry fixed.  Any two triangulations of a piecewise linear manifold can be related by a finite sequence of such Pachner moves~\cite{Pachner:1991}.

An $n_{\rm i}-n_{\rm f}$ Pachner move, with $n_{\rm i}+n_{\rm f}=d+2$, changes an initial complex of $n_{\rm i}$ $d$-simplices into a final complex of $n_{\rm f}$ $d$-simplices. In $d$ dimensions there are $d+1$ different types of moves denoted by $(d+1)-(1), (d)-(2),\cdots,(1)-(d+1)$.

One way to visualize an $n_{\rm i}-n_{\rm f}$ Pachner move in a $d$-dimensional triangulation is as follows, e.g.~\cite{DittrichHoehnCanSimp}: The initial sub-complex ${\cal C}_{\rm i}$ of $n_{\rm i}$ $d$-simplices involved in the Pachner move has to be identifiable with a corresponding sub-complex  ${\cal C}'$ of $n_{\rm i}$ $d$-simplices in a $(d+1)$-simplex $\sigma^{d+1}$.  The Pachner move can be visualized as a gluing of $\sigma^{d+1}$ onto the $d$-dimensional triangulation, such that ${\cal C}_{\rm i}$ is identified with the corresponding sub-complex of the $(d+1)$-simplex. In effect, one replaces ${\cal C}_{\rm i}$, the sub-complex of $n_{\rm i}$ $d$-simplices,  with ${\cal C}_{\rm f}=   \sigma^{(d+1)}/{\cal C}'$, i.e. the sub-complex obtained by removing the $n_{\rm i}$ $d$-simplices in ${\cal C}'$ from $\sigma^{(d+1)}$.

Here, we will consider the initial and final Pachner move configuration itself as $d$-dimensional triangulations with boundary. We start with a complex ${\cal C}_{\rm i}$ of $n_{\rm i}$ $d$-simplices and replace it with the complex ${\cal C}_{\rm f}=\sigma^{(d+1)}/{\cal C}_{\rm i}$, which contains $n_{\rm f}$ $d$-simplices. 

Now, for $1\leq n \leq (d+1)$, consider a sub-complex of $n$ $d$-simplices in a $(d+1)$-simplex as a $d$-dimensional triangulation with boundary. This $d$-dimensional triangulation has a unique lowest dimensional {\it bulk} sub-simplex. Its dimension is given by $(d+1-n)$.  E.g. for $n=(d+1)$ the lowest dimensional bulk sub-simplex is a vertex, and for $n=d$ it is an edge. For $n=1$ the bulk is given by one $d$-simplex.

Thus, we can also characterize a Pachner move by the lowest dimensional bulk simplex in either the initial or the final configuration. E.g. a $(d+1)-(1)$ Pachner move has a bulk vertex in the initial configuration and the lowest dimensional  bulk simplex in the final configuration is the final $d$-simplex itself. The bulk vertex in the initial configuration is connected to $(d+1)$ bulk edges.
A $(d)-(2)$ Pachner move has a bulk edge but no bulk vertex in the initial configuration, and the lowest dimensional sub-simplex in the final configuration is a $(d-1)$-simplex.  Pachner moves with $n_{\rm i}<d$ do not have bulk edges in their initial configuration. 

Later, we will be interested in particular in those Pachner moves which involve in their initial configuration bulk edges, but not in their final configuration. We can integrate out the length variables associated to the bulk edges from the initial configuration and in this way interpret the Pachner moves as coarse graining moves \cite{TimeEvol}.
We see, that for $d\geq 3$ such coarse graining Pachner moves (for length Regge calculus) only include the $(d+1)-(1)$ and $(d)-(2)$ moves. 

We will also need information about the boundary and bulk hinges, that is the boundary and bulk $(d-2)$- simplices, in the initial Pachner move configuration ${\cal C}_{\rm i}$ and in the final Pachner move configuration ${\cal C}_{\rm f}=\sigma^{(d+1)}/{\cal C}_{\rm i}$. Firstly, not that ${\cal C}_{\rm i}$ and $\sigma^{(d+1)}/{\cal C}_{\rm i}$, seen as $d$-dimensional triangulation with boundary, have by construction, the same boundary complex. Thus, the set of boundary hinges in the initial configuration agrees with the set of boundary hinges in the final configuration. 

Also by construction, we have that the set of bulk hinges in the initial configuration ${\cal C}_i$ is disjoint from the set of bulk hinges in the final configuration ${\cal C}_{\rm f}=\sigma^{(d+1)}/{\cal C}_i$. 

~\\
In this work we will deal with initial and final Pachner move configurations, which can be embedded into flat (Minkowski) space. The reason is that we will consider a perturbative evaluation of the path integral, and use the flatly embeddable configurations as backgrounds around which we allow quantum fluctuations (which can include curvature fluctuations).

Such flatly embeddable Pachner moves can be constructed as follows:
Start with a set of $(d+2)$ vertices ${0,1,\ldots,d+1}$ embedded into $d$-dimensional Minkowski space, such that they form a degenerate $(d+1)$-simplex. Assume that all the  $(d+1)$ $d$-simplices, which are sub-simplices of the degenerate $(d+1)$-simplex, are non-degenerate.  For an $n_{\rm i}-n_{\rm f}$ Pachner move one identifies $n_{\rm i}$ of the $(d+1)$ $d$-simplices as initial configuration and $n_{\rm f}$ of the $(d+1)$ $d$-simplices as final configuration.\footnote{To lighten notation we will however assume that all $d$-simplices in the initial and final configuration have positive orientation. (See \cite{Dittrich:2014rha} for how to allow positive and negative orientation.) Given a configuration of $(d+2)$ vertices this requirement constrains which $d$-simplices one can choose as initial and which as final.}

In the following we denote with $\sigma_{\overline{i}}$ the $d$-simplex obtained by removing the vertex $i$, and all adjacent simplices, from the complex. 
We will assume, without loss of generality, that the initial complex includes the $d$-simplices $\sigma_{\overline{n_{\rm f}}}, \ldots, \sigma_{\overline{d+1}}$ and the final complex includes the $d$-simplices $\sigma_{\overline{0}}, \ldots, \sigma_{\overline{n_{\rm f}-1}}$.



Below we will give a short description of Pachner moves in three and four dimensions.

\subsection{3D}\label{Pachner3D}

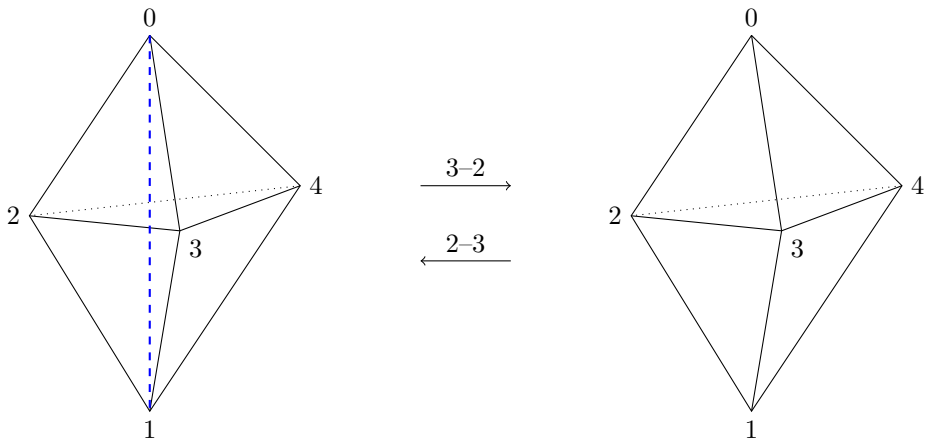
\begin{figure}[htb]
	\begin{tikzpicture}[scale = 2]
	\draw[] (0,1.5)--(1,0.5)--(0,-1)--(-0.8,0.3)--(0,1.5);
	\draw[] (1,0.5)--(0.2,0.2)--(-0.8,0.3);
	\draw[] (0,1.5)--(0.2,0.2)--(0,-1);
	\draw[dashed,blue,thick]  (0,1.5)--(0,-1);
	\draw[dotted]  (1,0.5)--(-0.8,0.3);
	\node [above] at (0,1.5) {$0$};
	\node [below] at (0,-1) {$1$};
	\node [left] at (-0.8,0.3) {$2$};
	\node [below right] at (0.2,0.2) {$3$};
	\node [right] at (1,0.5) {$4$};
	
	\draw[->] (1.8,0.5)--(2.4,0.5);
	\draw[->] (2.4,0.0)--(1.8,0.0);
	
	\node[above] at (2.1,0.5) {3--2};
	\node[above] at (2.1,0.0) {2--3};
	
	\begin{scope}[xshift=4cm]
	\draw[] (0,1.5)--(1,0.5)--(0,-1)--(-0.8,0.3)--(0,1.5);
	\draw[] (1,0.5)--(0.2,0.2)--(-0.8,0.3);
	\draw[] (0,1.5)--(0.2,0.2)--(0,-1);
	\draw[dotted]  (1,0.5)--(-0.8,0.3);
	\node [above] at (0,1.5) {$0$};
	\node [below] at (0,-1) {$1$};
	\node [left] at (-0.8,0.3) {$2$};
	\node [below right] at (0.2,0.2) {$3$};
	\node [right] at (1,0.5) {$4$};
	\end{scope}
	\end{tikzpicture}
	\caption{\label{Fig:32Move}3D Pachner move $3-2$ and its inverse $2-3$. In the initial configuration three tetrahedra share a bulk edge. Integrating out the bulk edge leads to a final configuration with two tetrahedra.}
\end{figure}

In three dimensions we can construct the flatly embedded Pachner moves from a set of vertices $\{0,1,2,3,4\}$ embedded into flat space.

The 3-simplices (or tetrahedra) in the initial complex  ${\cal C}_{\rm i}$ and final complex ${\cal C}_{\rm f}$ are given by
\ba
3-2: && {\cal C}_{\rm i}\supset\{\sigma_{\bar{2}},\sigma_{\bar{3}},\sigma_{\bar{4}}\}\,,\q\q\;
 {\cal C}_{\rm f}\supset\{\sigma_{\bar{0}},\sigma_{\bar{1}}\}\,, \nn\\\
4-1: && {\cal C}_{\rm i}\supset\{\sigma_{\bar{1}},\sigma_{\bar{2}},\sigma_{\bar{3}},\sigma_{\bar{4}}\}\,,\q 
{\cal C}_{\rm f}\supset\{\sigma_{\bar{0}}\}\, .
\ea

The $3-2$ move is depicted in Fig.~\ref{Fig:32Move}. The bulk edge $(01)$ shared by the three tetrahedra in the initial triangulation and is removed, and a bulk triangle in the final triangulation is 'inserted', leading to two tetrahedra.

\begin{figure}[htb]
	\begin{tikzpicture}[scale = 2]
	\draw[] (0,1.5)--(1,0.5)--(0.2,0.2)--(-0.8,0.3)--(0,1.5)--(0.2,0.2);
	\draw[dotted]  (1,0.5)--(-0.8,0.3);
	\draw[dashed,blue,thick]  (0,1.5)--(0.25,0.7);
	\draw[dashed,blue,thick]  (1,0.5)--(0.25,0.7);
	\draw[dashed,blue,thick]  (-0.8,0.3)--(0.25,0.7);
	\draw[dashed,blue,thick]  (0.2,0.2)--(0.25,0.7);
	\node [above right] at (0.25,0.7) {$0$};
	\node [above] at (0,1.5) {$1$};
	\node [left] at (-0.8,0.3) {$2$};
	\node [below] at (0.2,0.2) {$3$};
	\node [right] at (1,0.5) {$4$};
	
	\draw[->] (1.8,1.0)--(2.4,1.0);
	\draw[->] (2.4,0.5)--(1.8,0.5);
	
	\node[above] at (2.1,1.0) {4--1};
	\node[above] at (2.1,0.5) {1--4};
	
	\begin{scope}[xshift=4cm]
	\draw[] (0,1.5)--(1,0.5)--(0.2,0.2)--(-0.8,0.3)--(0,1.5)--(0.2,0.2);
	\draw[dotted]  (1,0.5)--(-0.8,0.3);
	\node [above] at (0,1.5) {$1$};
	\node [left] at (-0.8,0.3) {$2$};
	\node [below] at (0.2,0.2) {$3$};
	\node [right] at (1,0.5) {$4$};
	\end{scope}
	\end{tikzpicture}
	\caption{\label{Fig:41Move}3d Pachner move $4-1$ and its inverse $1-4$. In the initial configuration four tetrahedra share a bulk vertex. Integrating out the four bulk edges leads to a final configuration with one tetrahedron.}
\end{figure}
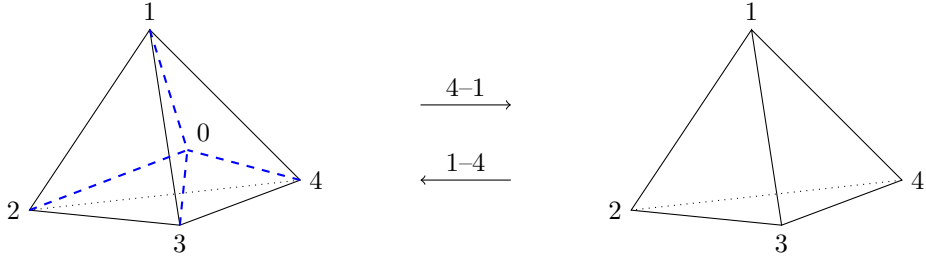

The $4-1$ move is depicted in Fig.~\ref{Fig:41Move}. Here the bulk vertex $0$ and all adjacent simplices are removed from the initial configuration, including the bulk edges $(0k)$, for $k\geq 1$. This leads to the final tetrahedron $\sigma_{\bar{0}}$.

\subsection{4D}\label{Pachner4D}

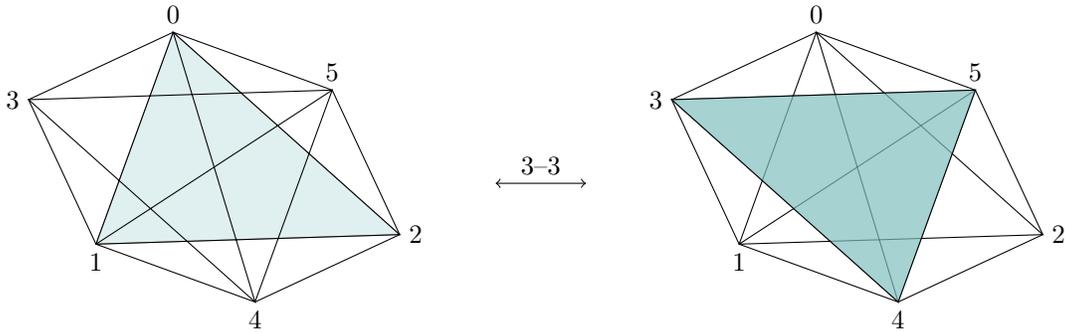
\begin{figure}[htb]
	\begin{tikzpicture}[scale = 1.5]
	\begin{scope}[rotate = -20]
	\draw[] (0,1)--(-1,0)--(0,-1)--(1.5,-1)--(2.5,0)--(1.5,1)--cycle;
	\draw[fill = teal!30,opacity=0.4] (0,1)--(0,-1)--(2.5,0)--cycle;
	\draw[]   (1.5,1)--(-1,0)--(1.5,-1)--(0,1)--(2.5,0)--(0,-1)--(0,1) (1.5,-1)--(1.5,1)--(0,-1);
	\node [left] at (-1,0) {$3$};
	\node [above] at (0,1) {$0$};
	\node [below] at (0,-1) {$1$};
	\node [below] at (1.5,-1) {$4$};
	\node [above] at (1.5,1) {$5$};
	\node [right] at (2.5,0) {$2$};
	\end{scope}
	\draw[<->,] (4,-0.4)--(3.2,-0.4);
	\node[above] at (3.6,-0.4) {3--3};
	\begin{scope}[xshift=5.7cm,rotate = -20]
	\draw[] (0,1)--(-1,0)--(0,-1)--(1.5,-1)--(2.5,0)--(1.5,1)--cycle;
	\draw[]   (1.5,1)--(-1,0)--(1.5,-1)--(0,1)--(2.5,0)--(0,-1)--(0,1) (1.5,-1)--(1.5,1)--(0,-1);
	\draw[fill = teal!50,opacity = 0.7]  (1.5,1)--(-1,0)--(1.5,-1)--cycle;
	\node [left] at (-1,0) {$3$};
	\node [above] at (0,1) {$0$};
	\node [below] at (0,-1) {$1$};
	\node [below] at (1.5,-1) {$4$};
	\node [above] at (1.5,1) {$5$};
	\node [right] at (2.5,0) {$2$};
	\end{scope}
	\end{tikzpicture}
	\caption{\label{Fig:33Move}4D Pachner move $3-3$. In the initial configuration three simplices share a triangle $(012)$. The Pachner move changes the initial configuration into a final configuration with three four-simplices sharing a triangle $(345)$. No bulk edges are involved.}
\end{figure}

In four dimensions we consider the $3-3$, the $4-2$ and the $5-1$ Pachner move. The Pachner moves involve the vertex set $\{0,1,\ldots,5\}$ and feature the following set of 4-simplices in their initial and final complexes
\ba
3-3: && {\cal C}_{\rm i}\supset\{\sigma_{\bar{3}},\sigma_{\bar{4}},\sigma_{\bar{5}}  \}\,,\q\q\q\;\;\;
 {\cal C}_{\rm f}\supset\{\sigma_{\bar{0}},\sigma_{\bar{1}},\sigma_{\bar{2}}\}\,, \nn\\\
 4-2: && {\cal C}_{\rm i}\supset\{\sigma_{\bar{2}},\sigma_{\bar{3}},\sigma_{\bar{4}},\sigma_{\bar{5}}  \}\,,\q\q\;
 {\cal C}_{\rm f}\supset\{\sigma_{\bar{0}},\sigma_{\bar{1}}\}\, ,\nn\\\
5-1: && {\cal C}_{\rm i}\supset\{\sigma_{\bar{1}},\sigma_{\bar{2}},\sigma_{\bar{3}},\sigma_{\bar{4}},\sigma_{\bar{5}}\}\,,\q 
{\cal C}_{\rm f}\supset\{\sigma_{\bar{0}}\}\, .
\ea

The lowest-dimensional bulk simplex in the initial configuration of the $3-3$ move, depicted in Fig.~\ref{Fig:33Move} is the triangle $(012)$, shared by all three initial 4-simplices. The lowest-dimensional bulk simplex in the final configuration is the triangle $(345)$, which is shared by all three final 4-simplices.

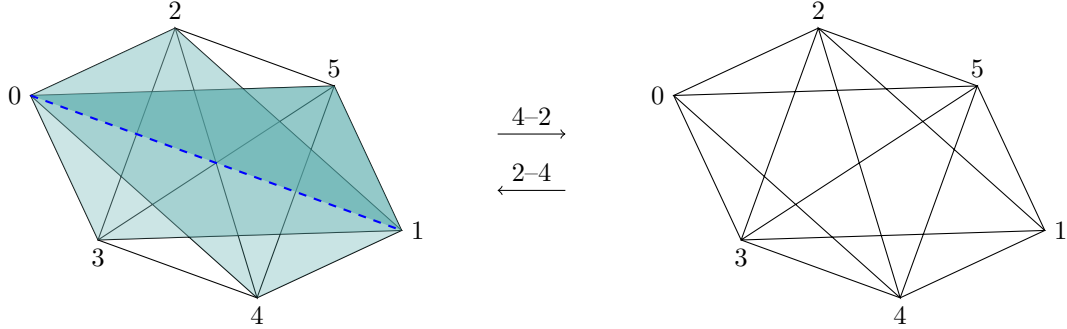
\begin{figure}[htb]
	\begin{tikzpicture}[scale = 1.5]
	\begin{scope}[rotate = -20]
	\draw[] (0,1)--(-1,0)--(0,-1)--(1.5,-1)--(2.5,0)--(1.5,1)--cycle;
	\draw[]   (1.5,1)--(-1,0)--(1.5,-1)--(0,1)--(2.5,0)--(0,-1)--(0,1) (1.5,-1)--(1.5,1)--(0,-1);
	\path[fill= teal!50,opacity= 0.5] (-1,0)--(2.5,0)--(0,1);
	\path[fill= teal!50,opacity= 0.4] (-1,0)--(2.5,0)--(0,-1);
	\path[fill= teal!70,opacity= 0.6] (-1,0)--(2.5,0)--(1.5,1);
	\path[fill=teal!70,opacity= 0.3] (-1,0)--(2.5,0)--(1.5,-1);
	\draw[thick,blue,dashed] (-1,0)--(2.5,0);    
	\node [left] at (-1,0) {$0$};
	\node [above] at (0,1) {$2$};
	\node [below] at (0,-1) {$3$};
	\node [below] at (1.5,-1) {$4$};
	\node [above] at (1.5,1) {$5$};
	\node [right] at (2.5,0) {$1$};
	\end{scope}
	
	\draw[->] (3.2,0)--(3.8,0);
	\draw[->] (3.8,-0.5)--(3.2,-0.5);
	
	\node[above] at (3.5,0) {4--2};
	\node[above] at (3.5,-0.5) {2--4};
	
	\begin{scope}[xshift=5.7cm,rotate = -20]
	\draw[] (0,1)--(-1,0)--(0,-1)--(1.5,-1)--(2.5,0)--(1.5,1)--cycle;
	\draw[]   (1.5,1)--(-1,0)--(1.5,-1)--(0,1)--(2.5,0)--(0,-1)--(0,1) (1.5,-1)--(1.5,1)--(0,-1);
	\node [left] at (-1,0) {$0$};
	\node [above] at (0,1) {$2$};
	\node [below] at (0,-1) {$3$};
	\node [below] at (1.5,-1) {$4$};
	\node [above] at (1.5,1) {$5$};
	\node [right] at (2.5,0) {$1$};
	\end{scope}
	\end{tikzpicture}
	\caption{\label{Fig:42Move} 4D Pachner move $4-2$ and its inverse $2-4$. In the initial configuration four four-simplices share one bulk edge. Integrating out the bulk edge leads to a final configuration with two four-simplices.}
\end{figure}

In the $4-2$ move, depicted in Fig.~\ref{Fig:42Move} the bulk edge $(01)$, shared by the initial 4-simplices, is removed from the configuration and a (bulk) tetrahedron $(2345)$, shared by the two final 4-simplices, is inserted.

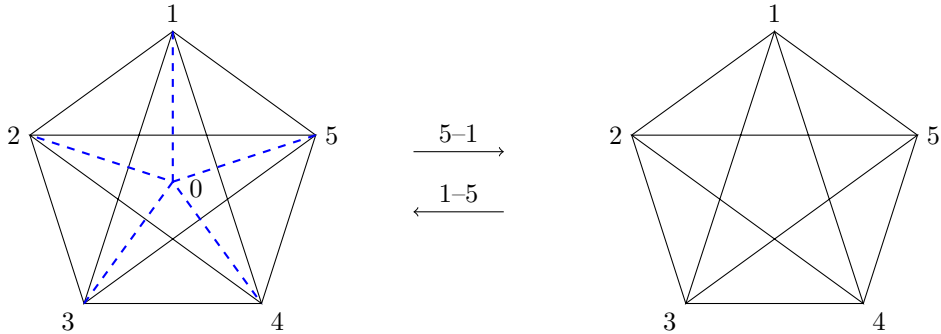
\begin{figure}[htb]
	\begin{tikzpicture}[scale = 2]
	\draw[] (0,1)--(-0.95,0.31)--(-0.59,-0.81)--(0.59,-0.81)--(0.95,0.31)--cycle;
	\draw[]   (-0.95,0.31)--(0.95,0.31)--(-0.59,-0.81)--(0,1)--(0.59,-0.81)--cycle;
	\draw[dashed, blue,thick]  (0,1)--(0,0)--(-0.95,0.31) (-0.59,-0.81)--(0,0)--(0.59,-0.81) (0.95,0.31)--(0,0);
	\node [right] at (0.05,-0.045) {$0$};
	\node [above] at (0,1) {$1$};
	\node [right] at (0.95,0.31) {$5$};
	\node [below right] at (0.59,-0.81) {$4$};
	\node [below left] at (-0.59,-0.81) {$3$};
	\node [left] at (-0.95,0.31) {$2$};
	
	\draw[->] (2.2,-0.2)--(1.6,-0.2);
	\draw[->] (1.6,0.2)--(2.2,0.2);
	
	\node[above] at (1.9,0.2) {5--1};
	\node[above] at (1.9,-0.2) {1--5};
	\begin{scope}[xshift=4cm]
	\draw[] (0,1)--(-0.95,0.31)--(-0.59,-0.81)--(0.59,-0.81)--(0.95,0.31)--cycle;
	\draw[]   (-0.95,0.31)--(0.95,0.31)--(-0.59,-0.81)--(0,1)--(0.59,-0.81)--cycle;
	\node [above] at (0,1) {$1$};
	\node [right] at (0.95,0.31) {$5$};
	\node [below right] at (0.59,-0.81) {$4$};
	\node [below left] at (-0.59,-0.81) {$3$};
	\node [left] at (-0.95,0.31) {$2$}; 
	
	\end{scope}
	\end{tikzpicture}
	\caption{\label{Fig:51Move}4D Pachner move $5-1$ and its inverse $1-5$. In the initial configuration five four-simplices share a bulk vertex. Integrating out the five bulk edges leads to a final configuration with one four-simplex.}
\end{figure}

For the $5-1$ move, depicted in Fig.~\ref{Fig:51Move}, the bulk vertex 0 and all adjacent simplices are removed from the initial configuration. This leaves the final 4-simplex $(12345)$.

\subsection{Degenerate Caley-Menger matrices}\label{SecSub:DegenerateCaleyMengerMatrix}

In Sec.~\ref{Sec:PachnerMoves} we discussed that flatly embeddable Pachner moves in $d$ dimensions can be constructed by embedding $(d+2)$ vertices into $d$-dimensional flat space.
These  $(d+2)$ vertices therefore form a degenerate $(d+1)$-simplex, i.e.~the signed $(d+1)$-volume square $\mathbb{V}^{d+1}$ of the vertex configuration has to vanish. The latter is given by
\ba\label{eq:SignedVolumeSquare}
\mathbb{V}^{d+1}&\,=\,& -\frac{ (-1)^{d+1} }{ 2^{(d+1)} ((d+1)!)^2} \det(C)\,
\ea
where $C$ is the so-called Caley-Menger matrix
\ba\label{eq:CaleyMengerMatrix}
C\,=\,& \left(
\begin{matrix}
	0 & 1 & 1 & 1 & \cdots & 1 \\
	1 & 0 & s_{01} & s_{02} & \cdots & s_{0(d+1)} \\
	1 & s_{01} & 0 & s_{12} & \cdots & s_{1(d+1)} \\
	\vdots & \vdots & \vdots & \vdots & \ddots & \vdots \\
	1 & s_{0(d+1)} & s_{1(d+1)} & s_{2(d+1)} & \cdots & 0
\end{matrix}
\right) \,.
\ea
Here $s_{ij}$ denotes the signed squared length of the vector connecting vertices $i$ and $j$ with $i,j=0,1,\ldots,d+1$. 
Thus the requirement that the $(d+2)$ vertices form a degenerate $(d+1)$-simplex is satisfied if the Caley-Menger determinant vanishes. This means that the matrix $ C$ has a null vector. As the $d$-simplices contained in the $(d+2)$ vertex configuration are required to be non-degenerate, $C$ has exactly one null vector whose form has been derived in~\cite{Dittrich:2014rha}. In the next subsection we provide a summary of this derivation.  Using this form of $C$ will allow us to derive an explicit and simple form for the Regge action Hessians associated to the Pachner moves, and in particular show that these Hessians factorize.

\subsubsection{Null vector}\label{SecSubSub:NullVector}

Denote by $c_{ab}$ the $(a,b)$-th minor of $C$ defined by the determinant of the submatrix of $C$ obtained by removing the $a$-th row and $b$-th column, with $a,b=1,\ldots,d+3$. Note that from~\eqref{eq:SignedVolumeSquare}, $c_{(i+2)(i+2)}$ for $i\in \{0,\dots,d+1\}$ is proportional to the signed $d$-volume square of the simplex obtained by removing vertex $i$ from the initial configuration,
\be\label{eq:MinorFunctionOfVolume}
c_{(i+2)(i+2)} = -(-1)^{d}2^d(d!)^2 \mathbb{V}_{\overline{i}}.
\ee
Any such simplex describes a non-degenerate $d$-simplex embedded in $d$-dimensional Minkowski space and thus has negative signed volume square, $ \mathbb{V}_{\overline{i}}<0$. Consequently the signs of the minors $c_{(i+2)(i+2)}$ are fixed by the global Lorentzian signature to $\text{sign}(c_{(i+2)(i+2)}) = (-1)^{d}$.

The minors of $C$ satisfy the relation~\cite{Kokkendorff:2007}
\ba
c_{aa}  c_{bb} - c_{ab} c_{ba} \propto \det(C) = 0\, .
\ea
Moreover, due to the symmetry of the Cayley-Menger matrix, $ c_{ab}=  c_{ba}$. Note also that  $c_{ab}\in \mathbb{R}$, as these are polynomials in the real entries of $C$. 
We can thus conclude that
\ba
c_{ab} = (-1)^{\text{sign}(a,b)}\sqrt{\abs{c_{aa}}}\sqrt{\abs{c_{bb}}} \q .
\ea
 Here  the signs $\text{sign}(a,b)=0,1$  depend a priori on the pair $(i,j)$. For the determinant of $C$ we use the Laplace formula and expand around an arbitrary row $i$,
\be\label{eq:DeterminantC}
\det(C) = \sum_{b=1}^{d+3}(-1)^{a+b}c_{ab} C_{ab} =  
\sum_{b=1}^{d+3}(-1)^{a+b}(-1)^{\text{sign}(a,b)}\sqrt{\abs{c_{aa}}}\sqrt{\abs{c_{bb}}} C_{ab} = 0\,.
\ee
The coefficients multiplying $C_{ab}$ in the sum~\eqref{eq:DeterminantC} can be understood as the components of a null vector. That a unique null vector exists follows from the previous discussions. We can thus conclude that $\text{sign}(a,b)$ factorizes in its dependence on $a$ and $b$. The single null vector of $C$ can be derived by expanding the determinant of $C$ with respect to different rows and takes the form~\cite{Dittrich:2014rha}
\be\label{eq:NullVector}
\vec{v}_C = \qty((-1)^s\sqrt{\abs{c_{11} }}, (-1)^{s_0}\sqrt{\abs{\mathbb{V}_{\overline{0}}}},\dots,(-1)^{s_{d+1}}\sqrt{\abs{\mathbb{V}_{\overline{d+1}}}})\,,
\ee
where we made use of \eqref{eq:MinorFunctionOfVolume}.
In the following subsection we will discuss the signs $s, s_0,\dots,s_{d+1}$ as these will be central for determining the components of the Hessian of the Regge action expanded around a flat background. As we can multiply the null vector with an arbitrary factor  we can only fix the relative signs, e.g. $(s_0+s)$ and $(s_0+s_i)$.

\subsubsection{Signs}\label{SecSubSub:Signs}

The relative signs $(s_0+s)$ and $(s_0+s_i)$ characterizing the null vector~\eqref{eq:NullVector} of the Caley-Menger matrix can be determined by acting with the matrix on $\vec{v}_C$. The condition from the first row and the conditions  from the remaining rows give respectively \cite{Dittrich:2014rha}
\ba\label{eq:SignsConditions}
\sum_{i=0}^{d+1}(-1)^{s_i}\sqrt{\abs{\mathbb{V}_{\overline{i}}}} &= 0\,, \nn\\
\forall j\geq 0: \quad (-1)^s \sqrt{\abs{c_{11}}} + \sum_{i=0}^{d+1}(-1)^{s_i} 2^{d/2}d! s_{ij} \sqrt{\abs{\mathbb{V}_{\overline{i}}}} &= 0 \,.
\ea
The first condition fixes the relative signs $(s_0+s_i)$, so that for fixed (positive) orientation  these are fully specified by the given type of Pachner move.\footnote{Here we assume that all simplices have positive orientation. It is possible to extend the discussion to allow for positive and negative orientation \cite{Dittrich:2014rha}. The signs will then depend on these orientations.}   For an $n_{\rm i}-n_{\rm f}$ Pachner move, we have
\ba\label{signsSol}
s_0 + s_i =
\begin{cases}
0 \, {\rm mod}\, 2, \,\, i=1, \ldots, n_{\rm f}-1, \\
1 \, {\rm mod}\, 2, \,\, i=n_{\rm f},\ldots, d+1,
\end{cases}
\ea
where we remind the reader that $\mathbb{V}_{\overline{0}}, \ldots, \mathbb{V}_{\overline{n_{\rm f}-1}}$ are the signed volume squares of the $d$-simplices after the move, and $\mathbb{V}_{\overline{n_{\rm f}}}, \ldots, \mathbb{V}_{\overline{d+1}}$ are the signed volume squares of the $d$-simplices before the move.

Note that the signs $(s_0+s_i)$ do not depend on the spacetime signature as only the (positive) volumes of the $d$-simplices enter the first condition. By contrast, the sign $(s_0+s)$ depends on the choice of Euclidean/Lorentzian signature, and, in general, on the geometric configuration under consideration. In fact, after fixing the signs $(s_0+s_i)$, the second condition in (\ref{eq:SignsConditions}) can be used for any $j$ to determine the remaining relative sign $(s_0+s)$ via
\be\label{Didentity}
\quad (-1)^{s_0+s} \sqrt{\abs{c_{11}}} = -  \sum_{i=0}^{d+1}(-1)^{s_0+s_i} 2^{d/2}d! s_{ij} \sqrt{\abs{\mathbb{V}_{\overline{i}}}} \,.
\ee
As the signed squared edge lengths can be positive or negative in Lorentzian signature, $s$ will generally depend on the spacelike/timelike nature of the edges, see Sec.~\ref{Sec:H4D}. Note also, that the sum on the right hand side of (\ref{Didentity}) actually does \emph{not} depend on the choice of $j$.

\subsubsection{Variation of the determinant}\label{SecSubSub:VariationDeterminant}

To compute the Hessian of the Regge action expanded around a flat background, we will need the variation of the determinant of the Caley-Menger matrix with respect to the signed squared edge lengths evaluated on the flat hypersurface $\{\det(C)=0\}$. Introducing the adjugate matrix $\text{Adj}(C)$ defined as the transpose of the matrix of cofactors $\text{Cof}(C)_{ab}\equiv (-1)^{a+b}c_{ab}$ of C, Jacobi's formula allows us to express the derivative of the determinant of $C$ as
\be
\pdv{\det(C)}{s_{ij}}=\tr(\text{Adj}(C)\pdv{C}{s_{ij}})\,.
\ee
Using the property $\text{Adj}(C)C = C\text{Adj}(C) = \det(C)\mathbb{I}$, it follows from $\det(C)=0$ that the components of the adjugate matrix are determined by the single null vector of $C$ in the form $\text{Adj}(C)_{ab} \propto v_C^a v_C^b$. Using in this form for the adjugate matrix the expression~\eqref{eq:NullVector} for $\vec{v}_C$ leads to
\be\label{eq:DeterminantVariation}
\eval{\pdv{\det(C)}{s_{ij}}}_{\det(C) = 0} =  
N\, (-1)^{s_i} (-1)^{s_j} \sqrt{\abs{\mathbb{V}_{\bar{i}}}}\sqrt{\abs{\mathbb{V}_{\bar{j}}}}\,,
\ee
where $N$ is a global factor, which for now will be left unspecified.

\section{Linearized Regge calculus}\label{Sec:LinearizedReggeCalculus}

\subsection{Flat background expansion}\label{SecSub:FlatBackgroundExpansion}

In the following we assume a classical background solution $\{s^{(0)}_e\}$ satisfying the generalized triangle inequalities and Regge equations of motion 
\ba\label{eq:ClassicalEOM}
\eval{\pdv{S_{\text{Regge}}}{s_e}}_{s^{(0)}_e} = 0\,.
\ea
Moreover, we assume that the background solution is locally flat and therefore the associated bulk deficit angles vanish,
\ba\label{eq:FlatnessBackground}
\eval{\epsilon^{(\text{bulk})}_{h}}_{s^{(0)}_e} = 0\,.
\ea
Now we expand the signed squared edge lengths as
\ba
s_e = s^{(0)}_e + \lambda_e\,,
\ea
where $\{\lambda_e\}$ represent small perturbations around the classical background. Then the Regge action~\eqref{eq:LorentzianReggeAction} to second order in the fluctuations takes the form
\ba\label{eq:ActionExpansion}
\imath S_{\text{Regge}} = \imath \eval{S_{\text{Regge}}}_{s^{(0)}_e} + \imath \eval{\pdv{S_{\text{Regge}}}{s_e}}_{s^{(0)}_e}\lambda_e + \frac{\imath}{2} \eval{\pdv[2]{S_{\text{Regge}}}{s_e}{s_e'}}_{s^{(0)}_e}\lambda_e \lambda_e' + \dots \,.
\ea
Here the zeroth order term represents the Regge action evaluated on the flat background, whereas the first-order term vanishes for bulk edges due to the equation of motion~\eqref{eq:ClassicalEOM}. The second-order term defines the Hessian of second derivatives of the Regge action with respect to the signed squared edge lengths,
\be\label{eq:HessianDefinition}
H_{ee'}\equiv \eval{\pdv[2]{S_{\text{Regge}}}{s_e'}{s_e}}_{s^{(0)}_e}\,.
\ee
We assume a background configuration around which the Regge action is real, $S_{\text{Regge}}\in \mathbb{R}$, and thereby exclude light cone irregularities~\cite{ADP21}. Using moreover the exchange symmetry of partial derivatives it then follows that the Hessian matrix is real and symmetric. Inserting the definition of the Regge action~\eqref{eq:LorentzianReggeAction} leads to
\be
H_{ee'} &= \frac{1}{\imath} \eval{\pdv[2]{}{s_{e'}}{s_e}}_{s^{(0)}_e} \sum_{h}\sqrt{\mathbb{V}_{h}}\epsilon_{h} \label{eq:HessianStart}\\
&=  \frac{1}{\imath} \eval{\pdv{}{s_{e'}}}_{s^{(0)}_e} \sum_{h}\qty(\pdv{\sqrt{\mathbb{V}_{h}}}{s_e}\epsilon_{h} + \sqrt{\mathbb{V}_{h}}\pdv{\epsilon_{h}}{s_e})\\
&=  \frac{1}{\imath} \eval{\sum_{h}\qty( \pdv[2]{\sqrt{\mathbb{V}_h}}{s_{e'}}{s_e}\epsilon_{h} + \pdv{\sqrt{\mathbb{V}_{h}}}{s_e} \pdv{\epsilon_{h}}{s_{e'}})}_{s^{(0)}_e}\,.\label{eq:HessianIntermediate}
\ee 
In the second line the last term was dropped using the Schl\"afli identity for Lorentzian triangulations, $\sum_h \sqrt{\mathbb{V}_h}\var{\theta_h}=0$. See Appendix~\ref{Sec:SchlaefliIdentity} for a proof. When evaluating \eqref{eq:HessianIntermediate} on a flat background, terms with $\epsilon_{h}$, where $h\subset \text{bulk}$, vanish due to~\eqref{eq:FlatnessBackground}. Moreover, terms with $\epsilon_{h}$, where $h\subset \text{boundary}$, will not be affected during Pachner moves as these keep the extrinsic geometry fixed.

\subsection{Hessian for Pachner moves}\label{SecSub:HessianForPachnerMoves}

In the following we associate an action to a given $n_{\rm i}-n_{\rm f}$ Pachner move\footnote{We assume that the Pachner move describes a change from an initial configuration with $n_{\rm i}$ $d$-simplices to a final configuration with $n_{\rm f}$ $d$-simplices, where $n_{\rm i}\geq n_{\rm f}$.} defined by the difference of the actions for the initial and final complex,
\be\label{eq:ActionPachnerMove}
S_{\text{Regge}}^{\text{Pachner}} =S_{\text{Regge}}^{({\rm i})}- S_{\text{Regge}}^{({\rm f})}\,.
\ee
The actions on the right hand side of~\eqref{eq:ActionPachnerMove} for the initial and final configurations of the Pachner move are defined in~\eqref{eq:LorentzianReggeAction}.  The action associated to the Pachner move is of the same form as in~\eqref{eq:LorentzianReggeAction},
\ba
\imath S_{\text{Regge}}^{\text{Pachner}} = \sum_{h} \sqrt{\mathbb{V}_h} \tilde{\epsilon}_{h}\, .
\ea
To define $\tilde \epsilon_h$ we remind the reader, that the initial and final simplicial complexes have the same boundary, and hence the same set of boundary hinges $h^{\rm bdry}$, and that the set of initial bulk hinges $h^{\rm iblk}$ and final bulk hinges $h^{\rm fblk}$ is distinct, see Section~\ref{Sec:PachnerMoves}.  We therefore have
\ba\label{eq:DeficitAnglesModified}
\tilde{\epsilon}_{h^{\rm iblk}} &=&2\pi + \sum_{\sigma^{\rm i} \supset h^{\rm iblk}} \theta_{\sigma^{\rm i},h^{\rm iblk}}, \q\q
\tilde{\epsilon}_{h^{\rm fblk}} \,=\,-2\pi - \sum_{\sigma^{\rm f} \supset h^{\rm fblk}} \theta_{\sigma^{\rm f},h^{\rm fblk}}, \nn\\
 \tilde{\epsilon}_{h^{\rm bdry}} &=& \qty(\pi k + \sum_{\sigma^{\rm i} \supset h^{\rm bdry}} \theta_{\sigma^{\rm i},\,h^{\rm bdry}}^{(i)}) - \qty(\pi k + \sum_{\sigma^{\rm f} \supset h^{\rm bdry}} \theta_{\sigma^{\rm f},\,h^{\rm bdry}}) \,.
\ea
With these definitions we obtain
\ba
\eval{\tilde{\epsilon}_{h}}_{s^{(0)}_e} = 0
\ea
for {\it all} (bulk and boundary) hinges on a flat background. Repeating the steps~\eqref{eq:HessianStart}-\eqref{eq:HessianIntermediate} leads to the Hessian associated to the given Pachner move,
\be\label{eq:HessianTildePachnerMoves}
H_{ee'}^{\text{Pachner}} = \frac{1}{\imath} \eval{\sum_{h}\pdv{\sqrt{\mathbb{V}_{h}}}{s_e} \pdv{\tilde{\epsilon}_{h}}{s_{e'}}}_{s^{(0)}_e}\,.
\ee
\\
Now recall that a Pachner move in $d$ dimensions is characterized by a set of $(d+2)$ vertices embedded into flat $d$-dimensional spacetime such that the involved $d$-simplices are non-degenerate. The condition of embedability in $d$-dimensional spacetime implies that the determinant of the Caley-Menger matrix vanishes, $\det(C)=0$, whereas the requirement that the $d$-simplices are non-degenerate implies that there exists exactly one null vector given by~\eqref{eq:NullVector}. Moreover, demanding flatness for the bulk geometry while keeping the extrinsic geometry at the boundary fixed leads to $\tilde{\epsilon}_{h}=0$ for all hinges in a flat background expansion. Thus, variations of the deficit angles and the Caley-Menger determinant are related and can be parametrized by associating an implicit function $f_h$ for each hinge,
\be\label{eq:DeficitAngleParametrization}
\tilde{\epsilon}_{h} = f_h \det(C)\,.
\ee
As a consequence, on the $\{\det(C)=0\}$ surface and in a flat background expansion, the variation of the deficit angle at a hinge can be expressed as 
\be\label{eq:DeficitAnglesModifiedVariation}
\eval{\delta \tilde{\epsilon}_{h}}_{s^{(0)}_e} = f_h \eval{\delta(\det(C))}_{s^{(0)}_e}\,\,\,\,
=\,\, \,\tilde f_h\, 
\sum_{i<j} (-1)^{s_i} (-1)^{s_j} \sqrt{\abs{\mathbb{V}_{\bar{i}}}}\sqrt{\abs{\mathbb{V}_{\bar{j}}}}\, \delta s_{ij}\,=:\, \tilde f_h \sum_e g(e) \delta s_{e}
\ee
where we used (\ref{eq:DeterminantVariation}) and reabsorbed the $N$ appearing there into  $\tilde f_h=N f_h$.

As the full dependence of the deficit angles on data from the hinge is now encoded in the function $\tilde f_h$, we can express the Hessian~\eqref{eq:HessianTildePachnerMoves} as
\be\label{eq:HessianTildePachnerMovesWithFh}
H_{ee'}^{\text{Pachner}} = \frac{1}{\imath}  g(e') \eval{\sum_{h}\pdv{\sqrt{\mathbb{V}_{h}}}{s_e}\tilde f_h }_{s^{(0)}_e} \,.
\ee
The variation of the Caley-Menger determinant with respect to the signed squared edge length $s_{mn}$ of an edge $e'=(mn)$ was derived in Sec.~\ref{SecSubSub:VariationDeterminant} and is given by the signed product of the volumes of the two $d$-simplices obtained by removing the vertices $m$ and $n$, cf.~\eqref{eq:DeterminantVariation}. Using the fact that the Hessian is symmetric under the exchange of edges $e=(ij)$ and $e'=(mn)$, we arrive at the first main result for the Hessian associated to a Pachner move in arbitrary spacetime dimensions,
\be\label{eq:HessianProportionality}
H_{(ij)(mn)}^{\text{Pachner}} = N^{\text{Pachner}} (-1)^{s_i+s_j+s_{m}+s_{n}}\sqrt{\abs{\mathbb{V}_{\bar{i}}}}\sqrt{\abs{\mathbb{V}_{\bar{j}}}}\sqrt{\abs{\mathbb{V}_{\bar{m}}}}\sqrt{\abs{\mathbb{V}_{\bar{n}}}}\,.
\ee
Here the signs $s_k$ have to satisfy (\ref{signsSol}). Thus it remains to determine the factor $N^{\text{Pachner}}$.\\

Finally, note that the Hessian matrix for an arbitrary Pachner move has a factorizing structure of the form
\be\label{eq:HessianFactorizing}
H_{(ij)(mn)}^{\text{Pachner}} \propto h_{(ij)}h_{(mn)}\,.
\ee

\subsection{Global factor}\label{SecSub:GlobalFactor}

The global factor in the Hessian~\eqref{eq:HessianProportionality} can be derived by computing $H_{ee'}^{\text{Pachner}}$ explicitly via equation~\eqref{eq:HessianTildePachnerMoves} for a selected pair of edges. To that end, we first note that the derivative of the dihedral angle at a hinge $h$ with respect to an edge $e=\bar{h}$ opposite of this hinge, has remarkably simple form
\be\label{eq:DihedralAngleDerivativeOpposite}
\pdv{\theta_{\sigma,h}}{s_{\bar{h}} }\,=\, -
\frac{1}{2} \frac{1}{d(d-1)} \frac{\sqrt{\mathbb{V}_{h} }}{\sqrt{\mathbb{V}_{\sigma} }}\,,
\ee
see Appendix~\ref{SecSub:DihedralAngleDerivative}.

To continue, we first discuss the global factor for Pachner moves in three dimensions and then for Pachner moves in four and higher dimensions. To that end, we remind the reader that we consider $n_{\rm i}-n_{\rm f}$ Pachner moves with $n_{\rm i}\geq n_{\rm f}$ and $n_{\rm i}+n_{\rm f}=d+2$. Thus, we have in three dimensions the $3-2$ and the $4-1$ Pachner moves and in four dimensions the $4-2$, $5-1$ and $3-3$ move.

 \subsubsection{3D}
We detailed the 3D Pachner moves in Sec.~\ref{Pachner3D}. With the notation from this section we 
will consider the derivative of $\tilde \epsilon_h$ with hinge edge $h=(01)$ with respect to the edge $e=(34)$
\ba
\pdv{\tilde \epsilon_{(01)}}{s_{34}} &=& -  (-1)^{s_0+s_2}  \pdv{\theta_{\bar{2},(01) } }{s_{34}} 
\, =\, 
 (-1)^{s_0+s_2} \frac{1}{12} 
\frac{\sqrt{\mathbb{V}_{(01)} }}{\sqrt{\mathbb{V}_{\bar{2}}}}\, ,
\ea
where, for the first equation, we used the solution (\ref{signsSol}) for the relative signs, to express the factor $(-1)$ as  $(-1)^{s_0+s_2}$ and equation (\ref{eq:DeficitAnglesModified}), which defines $\tilde \epsilon$. For the second equation we utilized (\ref{eq:DihedralAngleDerivativeOpposite}).

Therefore
\ba\label{eq:HessianTilde3D}
 H_{(01)(34)}^{\text{Pachner}} \,=\, \frac{1}{\imath} \eval{\pdv{\sqrt{\mathbb{V}_{(01)}}}{s_{01}} \pdv{\tilde{\epsilon}_{(01)}}{s_{34}}}_{s^{(0)}_e}\,
\,&=&
\frac{1}{24\imath}  (-1)^{s_0+s_2} 
 \frac{1}{\sqrt{\mathbb{V}_{\bar{2}}}}\nn\\
 \,&=&
- \frac{1}{24}  (-1)^{s_0+s_2}
\frac{1}{\sqrt{ \abs{\mathbb{V}_{\bar{2}}}}} \,.
\ea
In the second line, we used that  with our conventions $\sqrt{-1}=\imath$
 and that $\mathbb{V}_{\bar{2}}$ is always negative.

On the other hand, we have according to (\ref{eq:HessianProportionality}) 
\be\label{eq:HessianProportionality3D}
H_{(01)(34)}^{\text{Pachner}} = N^{\text{Pachner}} (-1)^{s_0+s_1+s_{3}+s_{4}}\sqrt{\abs{\mathbb{V}_{\overline{0}}}}\sqrt{\abs{\mathbb{V}_{\overline{1}}}}\sqrt{\abs{\mathbb{V}_{\overline{3}}}}\sqrt{\abs{\mathbb{V}_{\overline{4}}}}\,.
\ee
With the solution~\eqref{signsSol} for the relative signs, we have that for our three-dimensional Pachner moves
\ba\label{signs3D}
N^{\text{Pachner}} &=&
 \frac{1}{24}  (-1)^{s_{\text{Pachner}}} \frac{1}{\prod_{l}  \sqrt{ \abs{\mathbb{V}_{\bar{l}}}     }     } \,\,\, \text{with}
 \;\;
 (-1)^{s_{\text{Pachner}} }
=
\begin{cases}
+1 \, \,\,\text{for the}\,\,  3-2\,\, \text{move} \\
-1 \, \,\, \text{for the}\,\,  4-1\,\, \text{move} \, .
\end{cases}
\ea

We therefore obtain for the general Hessian matrix element
\ba\label{eq:Hessian3dResult}
H_{(ij)(mn)}^{\text{Pachner}} = \frac{1}{24} (-1)^{s_{\text{Pachner}} }  (-1)^{s_i+s_j+s_{m}+s_{n}}
\frac{
\sqrt{\abs{\mathbb{V}_{\bar{i}}}}\sqrt{\abs{\mathbb{V}_{\bar{j}}}}\sqrt{\abs{\mathbb{V}_{\bar{m}}}}\sqrt{ \abs{\mathbb{V}_{\bar{n}}}}
}
{\prod_{l} \sqrt{  \abs{\mathbb{V}_{\bar{l}}}      }    }
\,.
\ea


\subsubsection{4D}\label{Sec:H4D}

We described the 4D Pachner moves in Sec.~\ref{Pachner4D}. To determine the global pre-factor for the Hessian we will apply a slightly different strategy from the three-dimensional case.
To make use (only) of the simpler case (\ref{eq:DihedralAngleDerivativeOpposite}) for the derivative of the dihedral angle at an hinge $h$ with respect to the signed length squared of an opposite edge, we will  first determine the factors $\tilde f_h$ in (\ref{eq:DeficitAnglesModifiedVariation}).

To this end, we consider the derivative of $\tilde \epsilon_h$ with hinge triangle $h=(ijk)$ with respect to the signed length squared of edge $e=(mn)$, were $(ijkmnp)$ is an arbitrary permutation of $(012345)$. Note that  only the $d$-simplex $\sigma_{\bar{p}}$ includes the vertices $i,j,k$ and $m,n$. Thus
\ba
\pdv{\tilde \epsilon_{(ijk)}}{s_{mn}} &=& -  (-1)^{s_0+s_p}  \pdv{\theta_{\bar{p},(ijk) }}  {s_{mn}} 
\, =\, 
 (-1)^{s_0+s_p} \frac{1}{24} 
\frac{\sqrt{\mathbb{V}_{(ijk)}}}{\sqrt{\mathbb{V}_{\bar{p}}}}
\, =\, 
 (-1)^{s_0+s_p} \frac{1}{24\imath} 
\frac{\sqrt{\mathbb{V}_{(ijk)}}}{\sqrt{|  \mathbb{V}_{\bar{p}}|}} \q .
\ea
Here we again referred to (\ref{eq:DeficitAnglesModified}) and the solution~\eqref{signsSol} for the relative signs, which states that $(-1)^{s_0+s_p}=+1$ if $\sigma_{\bar{p}}$ is in the final configuration and $(-1)^{s_0+s_p}=-1$ if $\sigma_{\bar{p}}$ is in the initial configuration. In the last equation we used the fact, that $\sigma_{\bar{p}}$ is a Lorentzian $d$-simplex and has negative signed volume square.

We compare this case with equation~\eqref{eq:DeficitAnglesModifiedVariation}, which yields
\ba
\eval{\pdv{\tilde \epsilon_{(ijk)}}{s_{mn}} }_{s^{(0)}_e}&=&  \tilde f_{(ijk)}
 (-1)^{s_m+s_n}  \sqrt{\abs{\mathbb{V}_{\bar{m}}}}\sqrt{\abs{\mathbb{V}_{\bar{n}}}}  \, .
\ea
Thus
\ba
  \tilde f_{(ijk)} &=& \frac{(-1)^{s_0}}{24\imath}  \mathbb{V}_{(ijk)} 
  \frac{  (-1)^{s_i+s_j+s_k}
   \sqrt{\abs{\mathbb{V}_{\bar{i}}}}\sqrt{\abs{\mathbb{V}_{\bar{j}}}}    \sqrt{\abs{\mathbb{V}_{\bar{k}}}}        
  }
  {
  \prod_{l}  (-1)^{s_l}   \sqrt{ \abs{\mathbb{V}_{\bar{l}}}  }   
   }\, .
\ea
Now, according to~\eqref{eq:HessianTildePachnerMovesWithFh} we have

\be\label{eq:HessianfTilde}
 H_{ee'}^{\text{Pachner}} = \frac{1}{\imath}  g(e') \eval{\sum_{h}\pdv{\sqrt{\mathbb{V}_{h}}}{s_e}\tilde f_h }_{s^{(0)}_e} \,\, \q
 \text{with} \q\q g((mn))= (-1)^{s_m+s_n} 
  \sqrt{\abs{\mathbb{V}_{\bar{m}}}}\sqrt{\abs{\mathbb{V}_{\bar{n}}}} \q .
\ee

Hence we have for a general choice of $i\neq j$ and $m\neq n$
\ba\label{eq:Hessian4dResult}
H_{(ij)(mn)}^{\text{Pachner}}&=&   \frac{(-1)^{s_{\text{Pachner}} }  }{48} 
(-1)^{s_i+s_j+s_m+s_n} 
 \frac{
  \sqrt{\abs{\mathbb{V}_{\bar{i}}}}\sqrt{\abs{\mathbb{V}_{\bar{j}}}}
  \sqrt{\abs{\mathbb{V}_{\bar{m}}}}\sqrt{\abs{\mathbb{V}_{\bar{n}}}}
  }
  {
   \prod_{l}    \sqrt{ \abs{\mathbb{V}_{\bar{l}}} }
  }
\sum_{k \neq i,j} 
\pdv{ 
\mathbb{V}_{(ijk)}
}
{
s_{ij}
} \,
    (-1)^{s_k+s_0}  \sqrt{\abs{\mathbb{V}_{\bar{k}}}}  \,,\q \nn\\
\ea
with $(-1)^{s_{\text{Pachner}} }=(-1)\prod_{l} (-1)^{s_l}$. Our choice for  distributing the sign factors illustrates more obviously that the over-all result does not depend on the choice for $s_0$: all exponents $x$ in $(-1)^x$ involve an even number of $s_i$. 

We can determine $(-1)^{s_{\text{Pachner}} }$ for the various Pachner moves listed in~\eqref{Pachner4D} using the solutions for the relative signs in~\eqref{signsSol}:
\ba
(-1)^{s_{\text{Pachner}} }
=
\begin{cases}
+1 \, \,\,\text{for the}\,\,  3-3\,\, \text{move} \\
-1 \, \,\, \text{for the}\,\,  4-2\,\, \text{move} \\
+1 \, \,\, \text{for the}\,\,  5-1\,\, \text{move} \, .
\end{cases}
\ea

Comparing the Hessian in~\eqref{eq:Hessian4dResult} with the one in~\eqref{eq:HessianProportionality} we see that the factor 
\ba\label{eq:DFactor}
D_{ij}=  \sum_{k \neq i,j} 
\pdv{ 
\mathbb{V}_{(ijk)}
}
{
s_{ij}
} \,
    (-1)^{s_k+s_0}  \sqrt{\abs{\mathbb{V}_{\bar{k}}}}  \,\,=: \, D
\ea
cannot depend on $(ij)$.  This can be shown explicitly \cite{Dittrich:2014rha}:  Using~\eqref{eq:AreaTriangle} for the signed  area square of a triangle, $\mathbb{V}_{(ijk)}=\mathbb{A}_{(ijk)}$, we compute 
\ba\label{CompD}
D &=& \sum_{k \neq i,j} \pdv{\mathbb{V}_{(ijk)}}{s_{ij}} \,
(-1)^{s_k+s_0}  \sqrt{\abs{\mathbb{V}_{\bar{k}}}}\nn\\
& =& \frac{(-1)^{s_0}}{8}\sum_{k \neq i,j} (-1)^{s_k} \qty(s_{ik}+s_{jk}-s_{ij}) \sqrt{\abs{\mathbb{V}_{\bar{k}}}}\nn\\
&=& \frac{(-1)^{s_0}}{8}\sum_k  (-1)^{s_k} \qty(s_{ik}+s_{jk}) \sqrt{\abs{\mathbb{V}_{\bar{k}}}} - \frac{(-1)^{s_0}}{8}\qty((-1)^{s_j}s_{ij}\sqrt{\abs{\mathbb{V}_{\bar{j}}}} + (-1)^{s_i}s_{ij}\sqrt{\abs{\mathbb{V}_{\bar{i}}}}) \nn\\
&&\q\q\q\q\q\q\q\q\q\q\q\q\q\q\q - \frac{(-1)^{s_0}}{8}\sum_{k\neq i,j}(-1)^{s_k}s_{ij}\sqrt{\abs{\mathbb{V}_{\bar{k}}}}\nn\\
&=& \frac{(-1)^{s_0}}{8}\sum_k  (-1)^{s_k} \qty(s_{ik}+s_{jk}) \sqrt{\abs{\mathbb{V}_{\bar{k}}}} - \frac{(-1)^{s_0}}{8}s_{ij} \sum_k (-1)^{s_k} \sqrt{\abs{\mathbb{V}_{\bar{k}}}}\,.
\ea
\normalsize
 Using the first condition in~\eqref{eq:SignsConditions}, the last term vanishes. For the first expression we can use the second condition in~\eqref{eq:SignsConditions} to relate the sum to the minor $\sqrt{\abs{c_{11}}}$. This leads to
\be\label{Dfinal}
D \,\,=\,\,  \frac{1}{4}\sum_k  (-1)^{s_0+s_k} s_{ik} \sqrt{\abs{\mathbb{V}_{\bar{k}}}} \,\,=\,\,\frac{-(-1)^{s+s_0}}{4\times 2^{d/2}d!}\sqrt{\abs{c_{11}}}  \,,
\ee
where the first equation holds for an arbitrary choice of $i=0,\ldots,d+1$.

The sign of $D$ (and therefore $(-1)^{s+s_0}$) depends on the Pachner move under consideration and on the specifics of the configuration. 

Consider, for example, the $5-1$ Pachner move and two special cases, namely that the edges involved in this move are $(a)$ all space-like  and $(b)$ all time-like.\footnote{To construct such examples, take any configuration of six vertices embedded into four-dimensional Minkowski space, and assume that none of the vectors connecting the vertices has vanishing time coordinate or vanishing spatial coordinates. To produce a configuration where all edges are space-like, multiply the spatial coordinates of these vertices with a sufficiently large constant. For a configuration where all edges are time-like, multiply the time coordinates of these vertices with a sufficiently large constant.}  Using the solutions \eqref{signsSol}
for the relative sign, and (\ref{Dfinal}) for $i=0$, we see that $D$ is negative in case $(a)$, and positive in case $(b)$: 
\ba\label{D51}
D^{(5-1)} \,\,=\,\,  -\frac{1}{4}\sum_{k= 1}^5  s_{0k} \sqrt{\abs{\mathbb{V}_{\bar{k}}}} \q .
\ea

For the $4-2$ Pachner move, we can write $D$ as (see second line in \eqref{CompD})
\ba\label{D42}
D^{(4-2)}\,\,=\,\, -\frac{1}{8}\sum_{k= 2}^5 (s_{0k}+s_{1k}-s_{01}) \sqrt{\abs{\mathbb{V}_{\bar{k}}}} \q .
\ea
Thus the sign of $D^{(4-2)}$ depends on the geometric configuration. We can again consider special cases, e.g. $(a)$ all edges $(0k)$ and $(1k)$ with $k\geq 2$ are space-like, whereas $(01)$ is time-like.\footnote{To construct such a configuration, start with a space-like tetrahedron $(2345)$, which is in the plane orthogonal to the time axis and has its barycentre at the origin of the coordinate system. Choose the edge $(01)$ to lie along the time axis with its barycentre also at the origin of the coordinate system. Now scale the tetrahedron  $(2345)$ with a sufficiently large factor, so that all edges $(0k)$ and $(1k)$, with $k\geq 2$ are space-like.} In this case $D^{(4-2)}$ is negative. 

This extends to the case $(a')$, in which all triangles $(01k)$ with $k\geq 2$ are time-like and have only time-like edges. In this case the triangle inequalities demand that one edge has greater time-like length (or proper time) then the sum of the other two edges. Assume that this `longest' edge is $(01)$ for all triangles $(01k)$. This ensures that $(s_{0k}+s_{1k}-s_{01})$ is positive, and $D^{(4-2)}$ negative.  

On the other hand, consider the case $(b)$, in which all edges $(0k)$ and $(1k)$ with $k\geq 2$ are time-like, whereas $(01)$ is space-like.\footnote{ Choose the vertices $0$ and $1$ to lie in the $t=0$ plane and vertices $k$ with $k\geq 2$, such that the absolute value of their time coordinates is sufficiently large.}
 This yields a positive $D^{(4-2)}$.  This extends to the case $(b')$,  in which all triangles $(01k)$ with $k\geq 2$ are time-like but have only space-like edges, and where $(01)$ is the longest edge for all such triangles.

\subsubsection{Higher dimensions}

The calculation for the four-dimensional case can be straightforwardly generalized to higher dimensions: in $d\geq 4$ dimensions we start by considering the derivative of $\tilde \epsilon_h$ with $h=(ijk_1\cdots k_{d-3})$ with respect to the signed length squared of edge $e=(mn)$, where $(ijk_1 \cdots k_{d-3} mnp)$  is an arbitrary permutation of $(012\cdots (d+1))$. The general matrix element of the Hessian is thereby found as
\ba
H_{(ij)(mn)}^{\text{Pachner}}&=&   \frac{(-1)^{s_{\rm{Pachner}}}}{4 d(d-1)} 
(-1)^{s_i+s_j+s_m+s_n} 
 \frac{
  \sqrt{\abs{\mathbb{V}_{\bar{i}}}}\sqrt{\abs{\mathbb{V}_{\bar{j}}}}
  \sqrt{\abs{\mathbb{V}_{\bar{m}}}}\sqrt{\abs{\mathbb{V}_{\bar{n}}}}
  }
  {
   \prod_{l}   \sqrt{ \abs{\mathbb{V}_{\bar{l}}} }
  }
  \times \nn\\
  &&
\q \sum_{k_1<\cdots <k_{d-3}: k_\alpha \neq i,j} 
\pdv{ 
\mathbb{V}_{(ijk_1 \cdots k_{d-3})}
}
{
s_{ij}
} \,
    (-1)^{s_0+s_{k_1}+\cdots + s_{k_{d-3}}}  \sqrt{\abs{\mathbb{V}_{\overline{k_1}}}} \cdots   \sqrt{\abs{\mathbb{V}_{\overline{k_{d-3}}}}}  \,,\q 
\ea
with $(-1)^{s_{\text{Pachner}} }=(-1)\prod_{l} (-1)^{s_l}$. From the symmetry of the Hessian we can again conclude that the factor in the second line does not depend on $(ij)$. We have furthermore for even space-time dimensions $d$
\ba
(-1)^{s_{\text{Pachner}} }
=
\begin{cases}
+1 \, \,\,\text{if}\,\,  n_{\rm i},n_{\rm f} \,\, \text{are odd} \\
-1 \, \,\, \text{if}\,\, n_{\rm i},n_{\rm f}\,\, \text{are even} \, ,
\end{cases}
\ea
and for odd space-time dimensions $d$
\ba
(-1)^{s_0}(-1)^{s_{\text{Pachner}} }
=
\begin{cases}
-1 \, \,\,\text{if}\,\,  n_{\rm i} \,\, \text{is even} \\
+1 \, \,\, \text{if}\,\, n_{\rm i}\,\, \text{is odd} \, .
\end{cases}
\ea

\section{Path integral}\label{Sec:PathIntegralMeasure}

In the following we investigate the path integral~\eqref{eq:LorentzianPathIntegral} for a $d$-dimensional triangulation with fixed boundary edge lengths. Specifically, we will examine the path integral for Pachner moves, which can be interpreted as coarse graining moves. As we discussed in Sec.~\ref{Sec:PachnerMoves}, this includes only the $(d+1)-(1)$ and $(d)-(2)$ moves.

The path integral for  Regge calculus can in general not be computed analytically since the dihedral angles in the Regge action are complicated functions of the squared length variables and moreover the configuration to be summed over have to satisfy the generalized triangle inequalities. To circumvent these difficulties we expanded the Regge action around a flat classical background solution $\{s_e^{(0)}\}$ and  only quantize (i.e.~integrate over) the perturbations $\{\lambda_e\}$ around the classical background. 

Now, the zeroth and first order terms in the expansion of the Regge action (\ref{eq:ActionExpansion}) on a flat background are the same before and after the Pachner move and only involve boundary  variables. We remind the reader that we defined $S_{\text{Regge}}^{\text{Pachner}} =S_{\text{Regge}}^{({\rm i})}- S_{\text{Regge}}^{({\rm f})}$.  For the perturbative evaluation of the path integral we use the second-order expansion of the action ${}^{(2)}\!S_{\text{Regge}}$, which leads to
\be\label{eq:PathIntegralHessian}
Z^{\text{Pachner}}  =\exp(\imath \,\,  {}^{(2)}\!S_{\text{Regge}}^{({\rm f})} )  \int \prod_{\tilde{e}\subset \text{bulk}} \dd{\lambda}_{\tilde{e}} \mu^{\text{Pachner}}\exp(\imath \sum_{e,e'} \frac{1}{2}H_{ee'}^{\text{Pachner}} \lambda_e \lambda_{e'})\,.
\ee
Here $\mu^{\text{Pachner}}\qty(s_e^{(0)})$ denotes a measure factor assumed to depend only on the background variables. To compute integrals of the type~\eqref{eq:PathIntegralHessian} we first collect some general results on Gaussian integrals with imaginary argument in the exponential function.

\subsection{Gaussian integrals with imaginary exponent and factorizing Hessians}\label{SecSub:IntegralGaussianImaginary}

\begin{figure}[t]
	\centering
	\includegraphics[width=.49\textwidth]{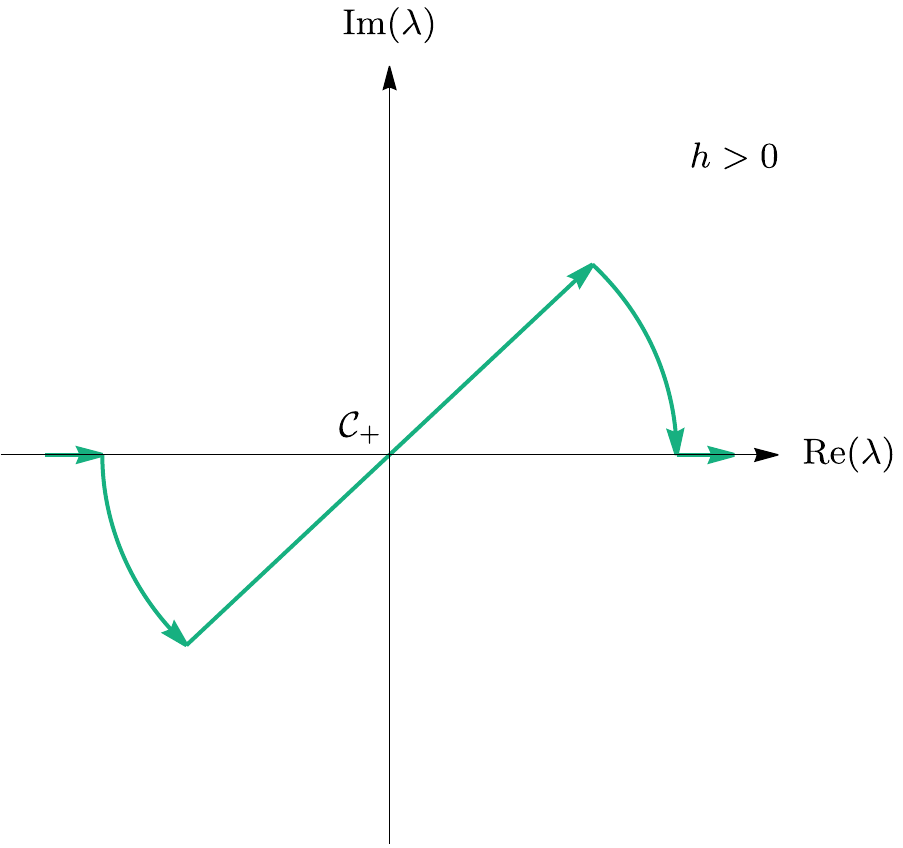}
	\hfill
	\includegraphics[width=.49\textwidth]{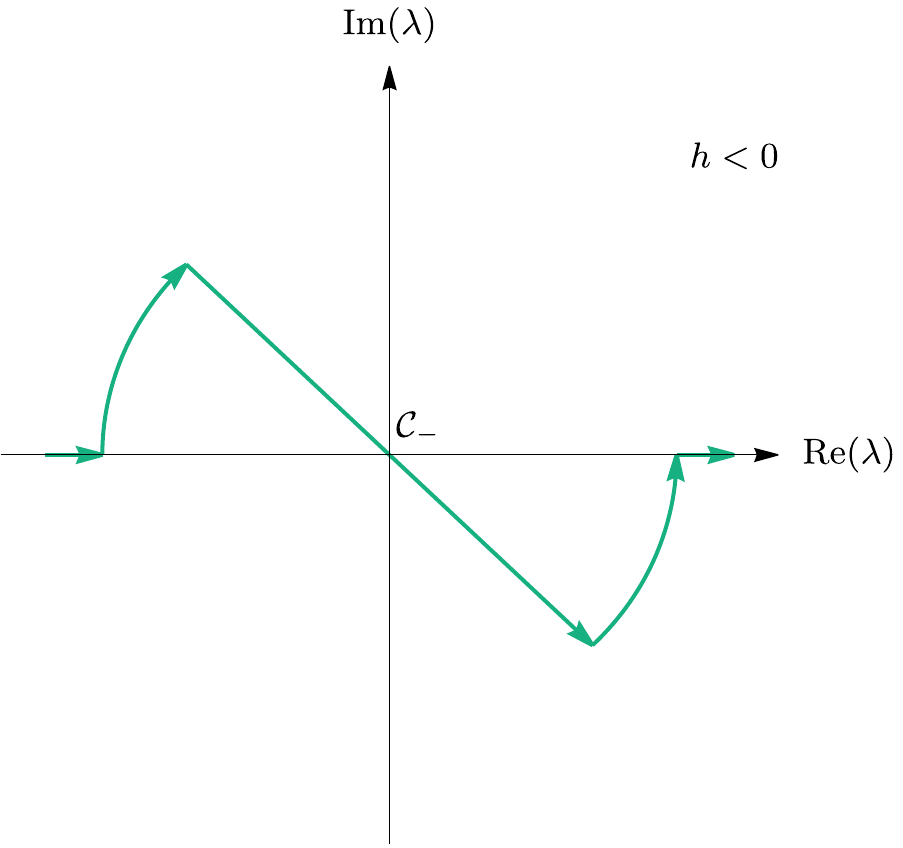}
	\caption{\label{Fig:IntegralContour} Choice of integration contours $\mathcal{C}_{\pm}$ for the integral $\int_{\mathbb{R}} d\lambda\, e^{\frac{\imath}{2} h \lambda^2}$ with $h>0$ (left) and $h<0$ (right), following the path of steepest descent.}
\end{figure}

Consider the one-dimensional Gaussian integral
\be\label{eq:IntegralGauss}
\int_{\mathbb{R}} d\lambda\, e^{\frac{\imath}{2} h \lambda^2} 
\ee
over the real axis, where $h\in \mathbb{R}\setminus\{0\}$ is a constant. \eqref{eq:IntegralGauss} is not well-defined as a Lebesgue integral. Nevertheless, the integral can be computed by deforming the integration contour in the complex plane and taking the path of steepest descent at the origin, as shown in Fig.~\ref{Fig:IntegralContour}, see e.g.~\cite{Baldazzi:2019kim}. Positive $h>0$ produce a factor of $e^{\imath \frac{\pi}{4}}$, whereas negative $h<0$ produce a factor of $e^{-\imath \frac{\pi}{4}}$. Altogether, applying the steepest descent method to compute~\eqref{eq:IntegralGauss} for $h\neq 0$ leads to
\be\label{eq:IntegralGaussCases}
\int_{\mathbb{R}} d\lambda\,  e^{\frac{\imath}{2} h \lambda^2}  = \begin{rcases}\begin{cases}
	e^{-\imath \frac{\pi}{4}} \sqrt{-\frac{2\pi}{h}} &,\quad h < 0\\
	e^{\imath \frac{\pi}{4}} \sqrt{\frac{2\pi}{h}} &,\quad h > 0
\end{cases}\end{rcases} = e^{\imath \frac{\pi}{4} \text{sign}(h)} \sqrt{\frac{2 \pi }{\abs{h}}} = \sqrt{\frac{2\pi \imath}{h}}\,.
\ee
The last equality holds if the complex square root function is defined by its principal value on the first Riemann sheet.\\

The result can be generalized to multidimensional Gaussian integrals with purely imaginary argument of the exponent, 
\be
\int_{\mathbb{R}^n} \dd{\vec{\lambda}} e^{\frac{\imath}{2} \vec{\lambda}^T H\vec{\lambda}} = \sqrt{\frac{2\pi \imath}{h_1}} \cdots    \sqrt{\frac{2\pi \imath}{h_n}}       \,.
\ee
where $H$ is assumed to be a regular real symmetric $n\times n$ matrix and $h_i,i=1,\ldots, n$ are its eigenvalues. 
If only $r<n$  of the variables $\vec{\lambda}=(\lambda_1,\dots,\lambda_r,\lambda_{r+1},\dots,\lambda_n)\equiv (\vec{\lambda}_{(r)}, \vec{\lambda}_{(n-r)})$ are integrated over, we can split the matrix $H$ into submatrices
\be\label{eq:HessianSplitNoGauge}
H = \left(
\begin{matrix}
	H_{r\times r} & \mathbb{H} \\
	\mathbb{H}^T & H_{(n-r)\times (n-r)}
\end{matrix}
\right)\,
\ee
and compute
\ba\label{eq:IntegralMasterFormulaNoGauge}
\int_{\mathbb{R}^r} \dd{\vec{\lambda}_{(r)}} e^{\frac{\imath}{2} \vec{\lambda}^T H \vec{\lambda}}& =&  
\sqrt{\frac{2\pi\imath}{h_1}}\cdots \sqrt{\frac{2\pi\imath}{h_r}}\,\,
e^{\frac{\imath}{2}\vec{\lambda}_{(n-r)}^T\tilde{H} \vec{\lambda}_{(n-r)}}\,,\nn\\
\tilde{H} & \equiv&  H_{(n-r)\times (n-r)} - \mathbb{H}^T {	H_{r\times r}}^{-1} \mathbb{H} \,,
\ea
where $h_i,\,i=1,\ldots,r$ are the eigenvalues of $H_{r\times r}$.
Note that here it is sufficient to demand only that $H_{r\times r}$ is regular. Integrals of this type will be needed to compute partition functions for the $(d+1)-(1)$ and the $(d)-(2)$ Pachner moves.

Let us start with the $(d)-(2)$ Pachner moves. Here we have only one bulk edge, thus we have in (\ref{eq:IntegralMasterFormulaNoGauge}) $r=1$.  Furthermore, we know from (\ref{eq:HessianProportionality}) that the Hessians for the Pachner moves are factorizing $H^{\rm Pachner} _{ee'}\propto h_e h_{e'}$. A straightforward computation shows that this implies for $\tilde H^{\rm Pachner} _{ee'}$ in (\ref{eq:IntegralMasterFormulaNoGauge})
\ba\label{Htildeis0}
\tilde H^{\rm Pachner} _{ee'}=0 \q .
\ea
Thus, integrating out the bulk edge in the $(d)-(2)$ Pachner moves changes the action only from the Regge action for the initial configuration $S^{({\rm i})}_{\rm Pachner}$ to the Regge action for the final configuration $S^{({\rm f})}_{\rm Pachner}$, see (\ref{eq:PathIntegralHessian}), without any additional terms. Therefore integrating out the bulk edges in the $(d)-(2)$ Pachner move configuration leaves the Regge action form-invariant.

For the $(d+1)-(1)$ moves we have $(d+1)$ bulk edges. But the factorizing form of the Pachner move Hessian implies that it has only one non-vanishing eigenvalue, and thus $(d+2)(d+1)/2-1$ null vectors. In particular, there are $d$ null vectors, which restrict to the $(d+1)$ bulk edges, and describe therefore gauge degrees of freedom. These correspond to a discrete remnant of diffeomorphism symmetry \cite{Rocek:1981ama,Dittrich:2008pw}, which exist on flat Regge backgrounds \cite{Bahr:2009ku, Bahr:2009qc}. Indeed, on a flat background, where all the (bulk) deficit angles are vanishing,  we can displace the bulk vertex $0$ in the embedding flat space, and in this way induce a change in the length squares of the bulk edges, without changing the bulk deficit angles and without changing the extrinsic curvature angles of the initial configuration, and thus without changing the value of the Regge action.

We will deal with these gauge symmetries by only integrating out one of the bulk edges, say $(01)$ (we therefore assume $H_{(01)(01)}\neq 0$), and by absorbing the remaining integrals into a measure over the gauge orbit.  We thus apply again (\ref{eq:IntegralMasterFormulaNoGauge}) with $r=1$. The factorizing form of the Hessian implies again, that $\tilde H^{\rm Pachner} _{ee'}=0$. This also shows, that the resulting action term does not depend on any of the variables we did not integrate over.

In summary, both types of coarse graining Pachner moves, the $(d+1)-(1)$ move and the $(d)-(2)$ move, leave the Regge action form-invariant. The intuitive reason for this form invariance is that, given a set of  boundary length squares satisfying the generalized triangle inequalities, a $d$-simplex and a neighboring pair of $d$-simplices can always be embedded into $d$-dimensional flat space. The initial configurations for the $(d+1)-(1)$ moves and $(d)-(2)$ moves result from inserting $(d+1)$ edges and one edge into the (flatly embeddable) $d$-simplex and the glued pair of $d$-simplices, respectively. There exist therefore flat solutions for the length squares of the bulk edges.\footnote{For special configurations there might exist also configuration with curvature, but these have to be considered as discretization artifacts, and are therefore expected to be situated outside a neighbourhood around the flat solutions. Here we are considering perturbations around flat solutions, such curved solutions are therefore not relevant. 
} In the process of removing these bulk edges we insert these flat solutions into the initial Regge action. This yields the Regge action for the final Pachner move configuration.

\normalsize

To furthermore study the form invariance of the measure, we will consider the various Pachner moves in more detail. Here we will restrict to three and four dimensions. Already in four dimensions we will see, that we cannot find a form-invariant measure. Additionally, there is the $3-3$ move, which does not leave the action form-invariant. In higher dimensions there are more and more Pachner moves, for which this applies.

\section{3D Pachner Moves}\label{Sec:3DPachnerMoves}

In three dimensions we have the $3-2$ and $4-1$ Pachner moves, which are of coarse graining type, and their respective inverses, the $2-3$ and $1-4$ moves. 

We discussed above that the (linearized) action is form-invariant for the coarse graining moves. That is solving for the bulk  lengths and inserting these solutions into the action, we find the Regge action for the final configurations of the move.  Since in three dimensions the Regge equation of motion impose flatness,  form invariance of the action applies also to the full action.

The statement of form invariance can be extended to the inverses of these coarse graining moves: here we just add to the action of the initial configuration the terms that arise from having new bulk hinges. 

Two triangulations with the same boundary can be connected via a finite sequence of Pachner moves. This form invariance of the action implies that the Hamilton-Jacobi function, defined as the action evaluated on the solution, will not depend on the choice of bulk triangulation, and also not on the choice of the (flat) background values for the bulk lengths.  The Hamilton-Jacobi function only depends on the choice of boundary and boundary data.  In the following we will study whether this statement extends to the path integral.

\subsection{$3-2$}\label{SecSub:InvarianceMeasure32Move}

For the path integral associated to the $3-2$ Pachner move we have
\be\label{eq:PathIntegral32Move}
Z^{(3-2)} = \exp(\imath \,\,  {}^{(2)}\!S_{\text{Regge}}^{({\rm f})} )  \int \dd{\lambda_{01}} \mu^{(3-2)} \exp(\imath \sum_{e,e'} \frac{1}{2}H_{ee'}^{(3-2)} \lambda_e \lambda_{e'})\,,
\ee
where we only integrate over the perturbation variable $\lambda_{(01)}$ associated to the one bulk edge $(01)$.  As we discussed in the previous section, due to the factorizing form of the Pachner move Hessian, we have $\tilde H_{ee'}^{(3-2)}=0$ for the exponential factor resulting from the Gaussian integral. The Gaussian integral leads furthermore to the factor 
\be
 \sqrt{\frac{2\pi \imath}{H^{(3-2)}_{(01)(01)}}} =e^{\imath \frac{\pi}{4}}\sqrt{48\pi } \abs{\frac{\mathbb{V}_{\overline{2}} \mathbb{V}_{\overline{3}}\mathbb{V}_{\overline{4}}}{\mathbb{V}_{\overline{0}}\mathbb{V}_{\overline{1}}}}^{\frac{1}{4}} =e^{\imath \frac{\pi}{4}} \sqrt{48\pi } \sqrt{\frac{V_{\overline{2}} V_{\overline{3}}V_{\overline{4}}}{V_{\overline{0}}V_{\overline{1}}}}\,,
\ee
where $V_{\overline{i}}$ denotes the positive volume of tetrahedron $\tau_{\bar{i}}$. Note that $H^{(3-2)}_{(01)(01)}>0$ (see \ref{signs3D}), and we therefore need to choose the integration contour $\mathcal{C}_+$ in Fig.~\ref{Fig:IntegralContour} for the Gaussian integral.

 Form invariance of the partition function can thus be achieved by choosing the measure factor as 
\be\label{eq:MeasureFactor32Move}
\mu^{(3-2)} =\frac{1}{\prod_{e\subset \text{bdry}}   \sqrt{ \sqrt{48\pi }}   }  \frac{1}{\prod_{e\subset \text{bulk}} \sqrt{48\pi }}\frac{\prod_\tau e^{-\imath \frac{\pi}{4}}}{\prod_{\tau}\sqrt{V_\tau}} \,.
\ee 
Here we wrote the measure factor in a form which allows generalization to arbitrary triangulations with boundary, and is consistent under gluing.

We also made a choice in associating the $e^{-\imath \frac{\pi}{4}}$ factors to the tetrahedra. We will later discuss that such a phase factor also appears for the Lorentzian Ponzano-Regge model. An alternative choice, which gives also form invariance for the $3-2$ move,  is to associated the phase factors to the edges of the triangulation.

\subsection{$4-1$}\label{SecSub:InvarianceMeasure41Move}

The path integral for the $4-1$ move takes the form
\be\label{eq:PathIntegral41Move}
Z^{(4-1)} = \exp(\imath \,\,  {}^{(2)}\!S_{\text{Regge}}^{({\rm f})} ) \int \prod_{i\in\{1,2,3,4\}} \dd{\lambda_{0i}} \mu^{(4-1)} \exp(\imath \sum_{e,e'} \frac{1}{2}H_{ee'}^{(4-1)} \lambda_e \lambda_{e'})\,.
\ee
As described in Sec.~\ref{SecSub:IntegralGaussianImaginary}, we have three gauge degrees of freedom, and we therefore only integrate explicitly over one variable, which without loss of generality, we choose as $\lambda_{01}$. We have again, that due to the factorizing form of the Pachner move Hessian, $\tilde H_{ee'}^{(4-1)}=0$. The Gaussian integral over $\lambda_{01}$ leads to a factor
\be
  \sqrt{\frac{2\pi \imath}{H^{(4-1)}_{(01)(01)}}} =e^{-\imath \frac{\pi}{4}}\sqrt{48\pi } \abs{\frac{\mathbb{V}_{\overline{2}} \mathbb{V}_{\overline{3}}\mathbb{V}_{\overline{4}}}{\mathbb{V}_{\overline{0}}\mathbb{V}_{\overline{1}}}}^{\frac{1}{4}} =e^{-\imath \frac{\pi}{4}}\sqrt{48\pi } \sqrt{\frac{V_{\overline{2}} V_{\overline{3}}V_{\overline{4}}}{V_{\overline{0}}V_{\overline{1}}}}\,.
\ee
Note that here $H^{(4-1)}_{(01)(01)}<0$ (see \ref{signs3D}), that is we have the opposite sign as compared to the $(3-2)$-move. We therefore need to choose the integration contour $\mathcal{C}_-$ in Fig.~\ref{Fig:IntegralContour} for the Gaussian integral.

Now, let us choose the measure $\mu^{(4-1)}$ to be of the same form\footnote{Again, one can also decide to associate the $e^{-\imath \frac{\pi}{4}}$ to the edges.} as the measure for the $3-2$ move \eqref{eq:MeasureFactor32Move}, i.e., 
\be\label{eq:MeasureFactor41Move}
\mu^{(4-1)} =\frac{1}{\prod_{e\subset \text{bdry}}   \sqrt{ \sqrt{48\pi }}   }  \frac{1}{\prod_{e\subset \text{bulk}} \sqrt{48\pi }}\frac{\prod_\tau e^{-\imath \frac{\pi}{4}}}{\prod_{\tau}\sqrt{V_\tau}} \,.
\ee 
Performing the integral over $\lambda_{01}$ we are left with an integral  \ba
Z^{(4-1)} = \exp(\imath \,\,  {}^{(2)}\!S_{\text{Regge}}^{({\rm f})} )
\frac{1}{\prod_{e\subset \text{bdry}}   \sqrt{ \sqrt{48\pi }}   }  \frac{e^{-\imath \frac{\pi}{4}}}{\sqrt{V_{\bar{0}}}} \int \prod_{i=2,3,4} \dd{\lambda_{0i}}  \frac{1} {\sqrt{48\pi }}
\frac{(-1)}{48\pi V_{\bar{1}}} \q .
\ea
We aim to identify the last factor as an integral over the gauge orbit. This gauge freedom results from the symmetry of the action under displacements of the vertex $0$ in the initial Pachner move configuration. It can be understood as a discrete remnant of the diffeomorphism symmetry in the continuum \cite{Rocek:1981ama,Dittrich:2008pw,Bahr:2009ku}.

Given a simplex $\sigma$ with vertices $(0,1,\ldots,d)$, embedded into $d$-dimensional Euclidean or Minkowskian spacetime, we have shown in Appendix \ref{SecSub:IntegrationMeasureGaugeOrbit}, that we have the following identity between integration measures 
\ba
\sum_{\text{orientation of $\sigma$}} \frac{\prod_{a=1}^d \dd s_{0a}}{2^d d! V_\sigma} \,=\, \prod_{\alpha=1}^d \dd v_0^\alpha  \, ,
\ea
where  $v_0^\alpha$ are the coordinates of the vertex $0$.  We can therefore interpret the integration over $\lambda_{0i},i=1,2,3$ as an integration over a gauge orbit. For our three-dimensional case, we obtain invariance of the path integral by setting 
\ba\label{6.9}
\int \prod_{i=2,3,4} \dd{\lambda_{0i}}  \frac{1} {\sqrt{48\pi }}
\frac{(-1)}{48\pi V_{\bar{1}}} \,\, \rightarrow \,\, 1    \, .
\ea
One can also include in the measure a factor of $(-1)$ for each vertex. We can then remove the $(-1)$ from the gauge contribution (\ref{6.9}). 

We mentioned that for the $3-2$ move, one can also associate the $e^{-\imath \frac{\pi}{4}}$ factor to the edges. This is also possible for the $4-1$ move, but one has  then to set a complex factor equal to 1:
\ba
e^{-\imath \frac{\pi}{4}}\int \prod_{i=2,3,4} \dd{\lambda_{0i}}  \frac{1} {\sqrt{48\pi }}
\frac{(-1)}{48\pi V_{\bar{1}}} \,\, \rightarrow \,\, 1    \q .
\ea
Alternatively, these phase factors can be again associated to the vertices of the triangulation.

\subsection{Summary for 3D Pachner moves}

We see that we can provide a measure for the three-dimensional path integral, which leads to form invariance at the one-loop level. Thus the (perturbative) path integral will, to one-loop order, not depend on the choice of bulk triangulation, but only on the boundary triangulation and data. This triangulation invariance is a reflection of diffeomorphism symmetry \cite{Bahr:2011uj,Dittrich:2011ien}.

The Ponzano-Regge model \cite{Ponzano:1969} constitutes a non-perturbative model of three-dimensional quantum gravity and is available in Euclidean signature (but implements $\exp(\imath S_{E})$ amplitudes and not the $\exp(-S_{E})$ amplitudes of Euclidean quantum gravity) and Lorentzian signature \cite{Davids,Freidel}.  The Ponzano-Regge model leads (with appropriate gauge fixing) also to triangulation-invariant amplitudes. 

This is the reason why the results we found here mirror closely the semi-classical limit of the amplitude associated to one tetrahedron, which is given by \cite{Ponzano:1969,Roberts:1998zka,Davids:1998bp,Davids}
\ba
\frac{1}{\sqrt{V_\tau}}\cos( \sum_{e\in \tau} |l_e| \tilde \psi_e  \pm \tfrac{\pi}{4})     \q .
\ea
Here $\tilde \psi_e$ is the Euclidean or Lorentzian {\it exterior} dihedral angle at $e\subset \tau$, and the $\pm$ sign is equal to $+1$ for the Euclidean case and equal to $-1$ for the Lorentzian case.  The cosine results from a sum over orientations of the tetrahedra, and the fact that the action associated to oppositely oriented tetrahedra differs by a global sign.  The $\tfrac{\pi}{4}$ contribution can again be understood\footnote{To determine the semi-classical limit one does apply a saddle point approximation \cite{Roberts:1998zka}.} to arise from a Gaussian integration with imaginary exponent, and the fact that the two saddle points come with Hessians of opposite sign.  In short, the $\tfrac{\pi}{4}$ contribution to the phase in the semi-classical limit of the Ponzano-Regge amplitude arises because of the sum over orientations.

It speaks to the power of our method that we also find this $\tfrac{\pi}{4}$ contribution, but via a very different mechanism. Here we do {\it not} sum over orientations. Rather, we  associated the $\tfrac{\pi}{4}$ phase to the tetrahedra because the signs for the Hessian matrix element  in the $3-2$ and $4-1$ move are $(a)$ independent from the specific configuration and $(b)$ opposite to each other.

Such a phase factor does not arise for Euclidean Regge calculus \cite{Dittrich:2011vz} with $\exp(-S_{E})$ amplitudes, because these signs are enforced to be the same for the $3-2$ and $4-1$ move: One identifies the non-gauge degree of freedom in the $4-1$ move as conformal mode, as the Hessian element has negative sign, and the path integral is a priori divergent. Therefore this sign is changed by hand, as suggested for the conformal mode in Euclidean (perturbative) quantum gravity \cite{ConformalFactor}. 

Working with $\exp(\imath S_{L})$ amplitudes we `solve' the conformal factor problem, as we $(a)$ work with an oscillatory amplitude and $(b)$ the integration contour is deformed according to the sign of the Hessian matrix element. The conformal factor problem is a main obstacle for non-perturbative lattice approaches to Euclidean quantum gravity, which often rely on Monte-Carlo simulations \cite{LollLR}.  These simulations tend to drive the systems to configurations which minimize the action, but due to the conformal mode problem the action is unbounded from below. For Regge calculus such configurations are so-called spikes \cite{AmbjornSpikes}. E.g.~in the $4-1$ move we obtain a spike configuration by choosing all bulk edge lengths to be very large.  

From this we can take away a lesson for (non-perturbative) Lorentzian Regge calculus: in evaluating the path integral we have to make sure to pick a deformation of the integration contour, which leads to a convergent result, e.g.~by using Picard-Lefschetz techniques \cite{PL,Turok,ADP21} or more implicit methods \cite{toappear}.

\section{4D Pachner Moves}\label{Sec:4DPachnerMoves}

 The set of coarse graining Pachner moves in four dimensions includes the $4-2$ and $5-1$ move. As discussed in Section~\ref{SecSub:IntegralGaussianImaginary} the linearized action is form invariant under these moves. In the same way as discussed for the 3D Pachner moves, this can be  extended to the $2-4$ and $1-5$ moves.  The full action is also form invariant, under the condition that one takes only the flat solution (which always exist) into account, and not curved solutions, which might arise as discretization artifacts.

In a $3-3$ move no edges are integrated out. This move only features one bulk hinge whose (linearized) deficit angle is completely determined by the boundary data, and is in general not vanishing.  The Hessian $H^{\text{Pachner}}_{ee'}$ in (\ref{eq:Hessian4dResult}) gives therefore directly the difference of the linearized Regge action before and after the move. The action is therefore in general\footnote{The initial action is equal the final action if the linearized deficit angle induced by the boundary data vanishes. This implies that the factor $D$ in (\ref{eq:Hessian4dResult}) vanishes.} not form-invariant under $3-3$ moves. This implies that the Hamilton-Jacobi function will in general depend on the choice of bulk triangulation. This is one feature of discretizations with broken diffeomorphism symmetry \cite{Bahr:2009ku, Dittrich:2011ien}. 

One can still ask, whether one can find a measure such that the path integral is form-invariant (to one-loop order) under $5-1$ or $4-2$ moves. In this context \cite{Dittrich:2011vz,Dittrich:2014rha} found already for Euclidean Regge calculus, that such a {\it local}\footnote{There are non-local constructions for such measures, see \cite{Asante:2021blx}.}  measure does not exist. 

Thus we do not expect to find such a measure in the Lorentzian case. Nevertheless, we will examine the path integral in order to understand the similarities and differences to the Euclidean case.

\subsection{$4-2$}\label{SecSub:42Move}

For the path integral associated to the $4-2$ Pachner move we have
\be\label{eq:PathIntegral42Move}
Z^{(4-2)} = \exp(\imath \,\,  {}^{(2)}\!S_{\text{Regge}}^{({\rm f})} )  \int \dd{\lambda_{01}} \mu^{(4-2)} \exp(\imath \sum_{e,e'} \frac{1}{2}H_{ee'}^{(4-2)} \lambda_e \lambda_{e'})\,,
\ee
where we  only integrate over the perturbation variable $\lambda_{(01)}$ associated to the one bulk edge $(01)$.  Due to the factorizing form of the Pachner move Hessian, we have $\tilde H_{ee'}^{(4-2)}=0$ for the exponential factor resulting from the Gaussian integral. The Gaussian integral results in the factor 
\be
 \sqrt{ \frac{2\pi \imath}{H^{(4-2)}_{(01)(01)}} } \,=\, 
 \sqrt{96\pi }e^{-\text{sign}(D)\imath \frac{\pi}{4}}
  \frac{1}{\sqrt{|D|}}
  \abs{\frac{\mathbb{V}_{\overline{2}} \mathbb{V}_{\overline{3}}\mathbb{V}_{\overline{4}}\mathbb{V}_{\overline{5}}}{\mathbb{V}_{\overline{0}}\mathbb{V}_{\overline{1}}}}^{\frac{1}{4}}
   = \sqrt{96\pi }e^{-\text{sign}(D)\imath \frac{\pi}{4}}  
  \frac{1}{\sqrt{|D|}} \sqrt{\frac{V_{\overline{2}} V_{\overline{3}}V_{\overline{4}}V_{\overline{5}}}{V_{\overline{0}}V_{\overline{1}}}}\,.
\ee
Choosing the measure
\be\label{eq:MeasureFactor42Move}
\mu^{(4-2)} =\frac{1}{\prod_{e\subset \text{bdry}}   \sqrt{ \sqrt{96\pi }}   }  \frac{1}{\prod_{e\subset \text{bulk}} \sqrt{96\pi }}
\frac{1
}
{\prod_{\sigma}\sqrt{V_\sigma}}\,\, \,,
\ee
we do not obtain full invariance, but rather remain with 
\ba\label{7.4}
Z^{(4-2)} = \frac{\exp(\imath \,\,  {}^{(2)}\!S_{\text{Regge}}^{({\rm f})} ) }{\prod_{e\subset \text{bdry}}   \sqrt{ \sqrt{96\pi }}  \sqrt{V_{\bar{0}}}\sqrt{V_{\bar{1}}}} \,\times\,  
  \frac{e^{-\text{sign}(D)\imath \frac{\pi}{4}}  }{\sqrt{|D|}} \q .
\ea
If there was only the first factor on the right hand side of (\ref{7.4}), we would have form invariance, but we also have a factor  $1/\sqrt{\imath D}$. It has been shown in \cite{Dittrich:2014rha}, that $D$ does not factorize over the simplices, and that therefore it is not possile not define a local measure (that is a measure which factorizes over simplices and sub-simplices) leading to form invariance.

We also discussed in Section \ref{Sec:H4D}, that the sign of $D$ is configuration-dependent. Different from the three-dimensional case we can therefore not absorb the phase $e^{-\text{sign}(D)\imath \frac{\pi}{4}}$ into terms which only depend on the number of simplices and possibly sub-simplices.

\subsection{$5-1$}\label{SecSub:51Move}

For the $5-1$ move the path integral is given by
\be\label{eq:PathIntegral51Move}
Z^{(5-1)} = \exp(\imath \,\,  {}^{(2)}\!S_{\text{Regge}}^{({\rm f})} ) \int \prod_{i\in\{1,2,3,4,5\}} \dd{\lambda_{0i}} \mu^{(5-1)} \exp(\imath \sum_{e,e'} \frac{1}{2}H_{ee'}^{(5-1)} \lambda_e \lambda_{e'})\,.
\ee
Due to the factorizing Hessian four of the five variables in the integral correspond to gauge degrees of freedom. Thus, we integrate only over $\lambda_{01}$, which leads to a factor 
\be
 \sqrt{ \frac{2\pi \imath}{H^{(5-1)}_{(01)(01)}} } \,=\, 
 \sqrt{96\pi }e^{\text{sign}(D)\imath \frac{\pi}{4}}
  \frac{1}{\sqrt{|D|}}
  \abs{\frac{\mathbb{V}_{\overline{2}} \mathbb{V}_{\overline{3}}\mathbb{V}_{\overline{4}}\mathbb{V}_{\overline{5}}}{\mathbb{V}_{\overline{0}}\mathbb{V}_{\overline{1}}}}^{\frac{1}{4}}
   = \sqrt{96\pi }e^{\text{sign}(D)\imath \frac{\pi}{4}}  
  \frac{1}{\sqrt{|D|}} \sqrt{\frac{V_{\overline{2}} V_{\overline{3}}V_{\overline{4}}V_{\overline{5}}}{V_{\overline{0}}V_{\overline{1}}}}\,.
\ee

Choosing the measure
\be\label{eq:MeasureFactor51Move}
\mu^{(5-1)} =\frac{1}{\prod_{e\subset \text{bdry}}   \sqrt{ \sqrt{96\pi }}   }  \frac{1}{\prod_{e\subset \text{bulk}} \sqrt{96\pi }}
\frac{1
}
{\prod_{\sigma}\sqrt{V_\sigma}}\,\, \,,
\ee
we remain with 
\ba
Z^{(5-1)} = \frac{\exp(\imath \,\,  {}^{(2)}\!S_{\text{Regge}}^{({\rm f})} ) }{\prod_{e\subset \text{bdry}}   \sqrt{ \sqrt{96\pi }}  \sqrt{V_{\bar{0}}}\sqrt{V_{\bar{1}}}} \,\times\,  
  \frac{e^{\text{sign}(D)\imath \frac{\pi}{4}}  }{\sqrt{|D|}} \int \prod_{i=2,3,4,5} \dd{\lambda_{0i}}  \frac{1} {(96\pi)^2}
\frac{1}{V_{\bar{1}}} \, .
\ea
With the same argument as for the $4-1$ moves we can associate 
\ba
\int \prod_{i=2,3,4,5} \dd{\lambda_{0i}} \frac{1} {(96\pi)^2}
\frac{1}{V_{\bar{1}}} \q \rightarrow \q 1
\ea
to an integral over a gauge orbit and set it to one.  We still do not obtain form invariance, as we obtained an additional factor of $\sqrt{\imath/D}$.

\subsection{Summary for 4D Pachner moves}

As expected, we find that the path integrals for the four-dimensional Pachner moves have much less invariance properties than for the three-dimensional Pachner moves. 

This starts at the classical level: whereas the action is form-invariant under $4-2$ and $5-1$ Pachner moves, it is not invariant under $3-3$ moves. Restricting the path integral to the $4-2$ and $5-1$ Pachner moves we cannot find a local measure that would lead to form invariance. 

This is  because of the factor $D$ appearing in the Hessian, which does not factorize over simplices. This issue already exist for Euclidean Regge calculus \cite{Dittrich:2014rha}. In the Lorentzian case we additionally encounter the phase $e^{-\text{sign}(D)\imath \frac{\pi}{4}} $ for the $(4-2)$ move and the phase $e^{\text{sign}(D)\imath \frac{\pi}{4}} $ for the $5-1$ move. 
We saw in Section \ref{Sec:H4D}, that the sign of $D$ can depend on the type of the Pachner move and also on the configuration, and we can therefore not absorb these phase factors into contributions that only depend on the number of simplices or sub-simplices.

As we discussed already for the Pachner moves in three dimensions, in the Euclidean case configurations for which the Hessian matrix element has negative sign are interpreted as conformal mode degrees of freedom, and the sign is changed by hand. E.g.~one can show that the sign for the $5-1$ move in Euclidean gravity is always negative \cite{Dittrich:2014rha}.

The Lorentzian path integral has the advantage that such sign rotations by hand are not necessary. It would be indeed quite problematic as the sign of $D$ appears to be much more configuration dependent for the Lorentzian case. We  find that the different signs lead to different deformations of the integration contour and that one can always make the integral convergent and thus avoid the conformal factor problem. 

A similar mechanism was found for the non-perturbative Regge path integral applied to cosmology \cite{ADP21}. In this case, the deformation of the contour was defined via Picard-Lefschetz theory \cite{PL}, and different boundary data led to different deformations of the contour. But if one isolates the behaviour of the action for large bulk edge lengths, the contours fell only into two classes, and emulated either a left-handed or right-handed Wick rotation, depending on whether the conformal factor dominated in the action or not. This mechanism of resolving the conformal factor problem is therefore the same as we encountered here for the Pachner moves.

We saw that in the three-dimensional Lorentzian case, the signs of the Hessian matrix elements  only dependent on the type of Pachner move, and not on the other specifics of the configuration. This holds also for three-dimensional Euclidean Regge gravity \cite{Dittrich:2011vz}. Thus in this case one can argue for a relation between Euclidean and Lorentzian gravity via Wick rotation: the different types of Pachner moves separate the cases with positive and negative sign for the Hessian element respectively. These come with separate deformations of the integration contour, and confirm the description of `rotating by hand' the sign for the conformal factor (in the $4-1$ move) for three-dimensional Euclidean quantum gravity \cite{ConformalFactor,Dittrich:2011vz}. As discussed in Section~\ref{SecSub:42Move}, there is a $\pi/4$ phase factor that is missed by this description, but it can be easily reconstructed.

In contrast, in the four-dimensional Lorentzian case, the signs of the Hessian matrix elements does not only depend on the type of Pachner move, but also on the further specifics of the configuration. Different configurations for e.g. the $5-1$ Pachner move might therefore require different deformations of the integration contour. But the sign for the (bulk) diagonal Hessian elements in the $5-1$ move is in the  Euclidean case always the same, namely negative. We can therefore {\it not} match the deformations of the Lorentzian contour with the sign rotations for the conformal factor in the Euclidean case.

\section{Discussion and outlook}\label{Sec:DiscussionOutlook}

In this work we considered the path integral for Lorentzian Regge calculus, to one-loop order.  We split this path integral into smaller steps defined by Pachner moves.  The advantage of doing so is that we only have to consider one-dimensional integrals for each Pachner move, and that we can derive an explicit and factorizing form for the Hessian of the action associated to such moves.  We then analyzed the path integral for the three-dimensional and four-dimensional case in more depth. We found considerable differences between these two cases.

For the three-dimensional case, we can define a  path integral measure that leads to form invariance (to one-loop order) of the Lorentzian path integral, and therefore to discretization independence. This will allow to evaluate path integrals for three-dimensional Lorentzian Regge gravity, by choosing the coarsest bulk triangulation. This has been utilized for Euclidean quantum gravity in \cite{Bonzom:2015ans}, in order to compute partition functions for the solid torus and to extend holographic dualitities to finite boundary \cite{Asante:2019ndj}. Such strategies can now also be applied to Lorentzian gravity.

We have also seen that in the three-dimensional case the different types of Pachner moves separate the conformal mode from the remaining modes. That is the $4-1$ and $3-2$ Pachner move Hessians have opposite signs in the Lorentzian and in the Euclidean case.  The non-gauge degree of freedom in the $4-1$ move can be identified as the conformal mode. The opposite signs for the Hessian require different deformations of the integration contour for the Lorentzian path integral, corresponding to Wick rotations in opposite directions. This justifies a posteriori the change of sign for the conformal mode in Euclidean, three-dimensional, quantum gravity \cite{ConformalFactor}.

On the other hand we do find an additional (complex) contribution to the measure in the Lorentzian case, as compared to the Euclidean case \cite{Dittrich:2011vz}. This additional contribution can be absorbed into a shift by $\pi/4$ for the phase of the amplitude associated to the tetrahedra. Interestingly, such a phase shift is also found for the asymptotics of the Lorentzian Ponzano-Regge model \cite{Davids}. This illustrates the power of the method applied here, that is to construct invariant path integral measures by demanding invariance under changes of the discretization, see also \cite{Bahr:2011uj,Asante:2021blx}.

In the four-dimensional case we cannot find a local path integral measure, that would lead to form invariance of the path integral. This holds even for Pachner moves for which the classical action is form invariant. This does {\it not} mean that one cannot achieve discretization or regularization invariance. One way out is to construct so-called perfect actions or discretizations \cite{Bahr:2009qc} that are typically non-local \cite{Bahr:2010cq}. E.g. \cite{Asante:2018kfo,Asante:2021blx} constructs an invariant measure (to one-loop measure) for a particular family of discretizations of Euclidean four-dimensional  quantum gravity, based on Regge gravity. This can now be extended to the Lorentzian case. Another way out is to abandon the notion of fundamental building blocks (and therefore locality) and use the consistent boundary formalism \cite{Dittrich:2012jq,DittrichBook14,Asante:2022dnj}. This framework suggests to start with a discrete approximation to the path integral, e.g. Regge gravity, and provides iterative methods to achieve results, which depend less and less on the choice of discretization.

In the four-dimensional case we also do not find such a straightforward relation between the deformation of the integration contour for the Lorentzian path integral and  the change of sign for the conformal mode in the Euclidean path integral, as for three-dimensional gravity. The sign for the (bulk) Hessian matrix elements depends in the Lorentzian case not only on the type of the Pachner move, but also on the specifics of the configuration. E.g. we find that the sign for the $5-1$ Pachner move can be positive or negative for the Lorentzian bulk Hessian matrix element. In contrast, one finds that the Euclidean bulk Hessian matrix element for the $5-1$ move has negative sign and therefore requires a change of sign. Thus, we have to apply in the Lorentzian case deformations of the integration contour, which depend on the configuration. But we have  to always change the sign for the Euclidean Hessian for the $5-1$ move.

Luckily, in the Lorentzian case we do not need to apply a `change of sign by hand' for the conformal mode. The deformation of the integration contour is rather determined by the convergence properties of the integrand, and leads for our quadratic actions always to a convergent path integral.  We can thus take as a lesson for the non-perturbative path integral, that one either needs to have a well-defined procedure to establish the correct deformation contour, e.g. Picard-Lefschetz theory \cite{PL}, which has already been applied in \cite{ADP21} to Lorentzian Regge cosmology. Or it is necessary to find evaluation methods for the Lorentzian path integral that are independent of such a contour deformations \cite{toappear}.

This brings us to possible extensions and further applications of the results presented in this work. We considered here the path integral to one-loop order. It would be interesting to see how much can be said about the non-perturbative path integral. We cannot hope for an exact analytical evaluation. But, we can hope to establish the asymptotic behaviour of the complex action for the various coarse graining Pachner moves. Indeed, we have seen that these Pachner moves (with gauge fixing) always lead to just a one-dimensional integration.  We can therefore hope that the asymptotic behaviour for large edge lengths will be sufficient to determine the required deformation of the integration contour, and to establish an integration measure such that the path integral (with gauge fixing) converges. This could help to establish the existence of the Lorentzian non-perturbative path integral for (Regge) gravity. Knowing the (approximate) behaviour of this path integral under coarse graining Pachner moves would help to understand crucial properties of Lorentzian renormalization. See \cite{Bonzom:2013ofa,Riello,Banburski:2014cwa,Dona:2023myv} for a related strategy in the context of  spin foams.

The strategy presented in this work can also be generalized to other forms of Regge gravity, e.g. for homogeneously curved simplices \cite{Bahr:2009qc,NewRegge},  as well as to effective spin foams \cite{EffSF1,EffSF3}. In Area Regge gravity (defined for four-dimensional triangulations) one uses areas instead of length as fundamental degrees of freedom \cite{Barrettetal,ADHAreaR}. One thus avoids the factor $D$ in the Hessian, whose appearance in length Regge calculus hinders the definition of a local path integral measure.  On the other hand, the action includse the inverse of the Jacobian associated to the change of variables from length to areas.  Effective spin foams \cite{EffSF1,EffSF3} are based on the Area Regge action, a perturbative path integral evaluation could therefore rely on similar techniques as presented here. Non-perturbative evaluations of the effective spin foam path integral have been already achieved for the $3-3$ move configuration \cite{EffSF1,EffSF3} and also for  triangulations with a bulk edge \cite{EffSF2}. These in particular show that effective spin foams implement discrete gravitational equations of motion.  For the EPRL spin foam model \cite{EPRL} one can evaluate  $3-3$ Pachner move configurations numerically \cite{Gozzini,Dona:2022yyn} and configurations with internal edges via a saddle point approximations \cite{Qu1,Qu2}.  The classical aspects of this class of models can be described by the Area-Angle Regge action \cite{DittrichSpeziale}. This action can also turned to describe a topological model, namely $\text{SU}(2)$ BF theory. A Pachner move analysis of this action, along the lines presented here, might therefore allow to construct a path integral measure with enhanced invariance properties.

\appendix

\section{Geometry of Lorentzian and Euclidean simplices}

The following is an overview on the geometry of Lorentzian and Euclidean simplices. We will largely follow the conventions of \cite{ADP21}, but will also provide extended material not covered in \cite{ADP21}. We will include the definition of Euclidean and Lorentzian Regge actions (and thus of Euclidean and Lorentzian dihedral and deficit angles), but utilize for this the notion of complex angles and complex Regge action from \cite{ADP21}. Additionally, we provide here a proof for the Euclidean and Lorentzian Schl\"afli identity (using the conventions employed in this paper),  and a formula for the derivatives of  dihedral angles, which are needed in the main text.   

To this end we will prove a number of geometrical identities for Lorentzian and Eucludian simplices. This work will pay off once we come to the proof of the Schl\"afli identity and to the formula detailing the derivative of the dihedral angle.

~\\

In the following we consider a $d$-simplex $\sigma$ embedded in either Minkowskian or Euclidean spacetime equipped with a Cartesian coordinate system. We denote the coordinates of the simplex vertices with $v_i^\alpha$ where $i=0,1,\ldots, d$ and $\alpha=1,\ldots,d$. Below we will choose $v_0$ as reference vertex and consider the set of edge vectors originating from $v_0$, that is 
\be
x_a^\alpha =v_a^\alpha -v_0^\alpha \q , \q\q a=1,\ldots, d \q .
\ee
These edge vectors form a basis of the Minkowskian or Euclidean spacetime.  We can encode these edge vectors into a quadratic matrix $X$ with elements
\ba
X_{a\alpha}\,=\, x_a^\alpha \q .
\ea
We assume that we have a non-degenerate simplex, and that the basis $\{x_a\}_{a=1}^d$ is positively oriented. The  (absolute) volume is then given by
\ba\label{A3}
V \,=\, \frac{1}{d!} \det X  \q .
\ea

Denoting by $\eta_{\alpha\beta}$ either the Minkowskian or Euclidean spacetime the signed length squared of the edges are given by
\ba
s_{ij} = \sum_{\alpha,\beta} (v_i^\alpha  -v_j^\alpha)\eta_{\alpha\beta}  (v_i^\beta  -v_j^\beta)
=:\langle  v_i -v_j \,|\, v_i  -v_j\rangle  \q .
\ea

\subsection{Caley-Menger determinants and generalized triangle inequalities}\label{SecSub:CaleyMenger}

The signed  volume square of a $d$-simplex can be also computed via a Caley-Menger determinant, see e.g.~\cite{Visser2}. This gives the signed volume square directly in terms of the signed length squares:
\ba\label{App:CaleyMenger}
\mathbb{V}^{d}&\,=\,& -\frac{ (-1)^{d} }{ 2^{d} (d!)^2}  \det \left(
\begin{matrix}
	0 & 1 & 1 & 1 & \cdots & 1 \\
	1 & 0 & s_{01} & s_{02} & \cdots & s_{0d} \\
	1 & s_{01} & 0 & s_{12} & \cdots & s_{1d} \\
	\vdots & \vdots & \vdots & \vdots & \ddots & \vdots \\
	1 & s_{0d} & s_{1d} & s_{2d} & \cdots & 0
\end{matrix}
\right) \,.
\ea
For example, we have for the signed area square of a triangle
\be\label{eq:AreaTriangle}
\mathbb{A}_{ijk} = -\frac{1}{16}\qty(s_{ij}^2+s_{ik}^2+s_{jk}^2 - 2(s_{ij}s_{ik}+s_{ij}s_{jk}+s_{ik}s_{jk}))\,.
\ee

Given the signed squared edge lengths of a $d$-simplex $\sigma$, we say it is either realizable as a Euclidean simplex or a Lorentzian simplex, if there exist a geometry-preserving embedding of this simplex in either Euclidean or Minkowskian spacetime, respectively. 

A Euclidean simplex is realizable (and non-degenerate) if its signed  volume square and the signed  volume squares of all its sub-simplices are strictly positive. 

A Lorentzian simplex $\sigma$ is realizable (and non-degenerate) if its signed  volume square is strictly negative and if the following requirement for all its sub-simplices is satisfied \cite{Visser, EffSF3},
\ba\label{eq:GeneralizedInequalitiesB}
\rho \subset \sigma, \mathbb{V}_{\rho} \leq 0 \;\; \Longrightarrow \;\; \forall  \rho' \supset \rho: \mathbb{V}_{\rho'} \leq 0 \,.
\ea
Note that we set the signed volume square associated to any vertex equal to 1.  The condition (\ref{eq:GeneralizedInequalitiesB}) ensures that if a given sub-simplex $\rho$ is time-like or null, then all the sub-simplices containing this sub-simplex $\rho$ are also time-like or null. 

\subsection{Gram matrix of edge vectors and Gram matrix of normals}

Above we defined the edge vectors $x_a$, where $a=1,\ldots, d$, which start from the vertex $v_0$, and form a basis for ${\mathbb R}^d$. 
The  metric tensor can be expressed in this basis as 
 \ba \label{gram1}
 G_{ab} &=&  \langle x_a\,|\, x_b\rangle \,=\,  \tfrac12 \left( s_{a0} + s_{b0} -s_{ab} \right) \q .
 \ea
 The latter equation follows from inverting the relations
 \ba
s_{ab} =& \langle x_a-x_b\,|\, x_a-x_b\rangle&= G_{aa}-2G_{ab}+G_{bb} \, ,\nn\\
s_{a0} =&\langle x_a\,|\, x_a\rangle   &= G_{aa} \,.
\ea

Note that $G$ coincides with the Gram matrix of the set of edge vectors $\{x_a\}_{a=1}^d$. Thus we can write $G$ also as
\ba\label{A9}
G\,=\, X \cdot \eta \cdot X^T\,,
 \ea
where we understand $(\eta)_{\alpha\beta}=\eta_{\alpha\beta}$ as matrix. 
~\\
The adjugate $\text{Adj}(X)$ to a matrix $X$ satisfies 
\ba
X\cdot \text{Adj}(X) \,=\, \text{Adj}(X)\cdot X \,=\, \det (X) \, \mathbb{I} \, = d! V  \, \mathbb{I} \, .
\ea
Thus, the columns of $\text{Adj}(X)$ are the normal one-forms\footnote{The corresponding normal vectors are inward pointing.}, which satisfy
\ba\label{A11}
\sum_\alpha  n_{a \alpha} x_b^\alpha \,=\, d! V \, \delta_{ab} \q \text{where} \q n_{a\alpha} =\text{Adj}(X)_{\alpha a}\, .
\ea
To determine the normalization of the normal one-forms we split $x_a$ into a part parallel to the normal vector and a part orthogonal to it:
\ba
x_a^\alpha  \,=\,  \sum_\beta h_a \frac{1}{|n_a|}  n_{a\beta} \eta^{\alpha\beta} + y^\alpha \q \text{with}\q \sum_\alpha n_{a\alpha}y^\alpha=0  \, ,
\ea
where $\eta^{\alpha\beta}$ denotes the inverse (Minkowskian or Euclidean) metric and $|n_a|=\sqrt{ |\sum_{\alpha,\beta} n_{a\alpha}\eta^{\alpha\beta}n_{a\beta}|}$.  The coefficient $h_a$ defines the height of the $d$-simplex $\sigma$ with respect to the sub-simplex $\sigma_{\bar{a}}$, which is obtained from $\sigma$ by removing the vertex $v_a$ and all its adjacent sub-simplices.

Using (\ref{A11}) we have
\ba
\sum_{\alpha\beta}  n_{a \alpha} \eta^{\alpha\beta} n_{a \beta}  \,=\,    h_a \, \text{sign}(n_a) |n_a| \,=\,   d! V \q .
\ea
The height $h_a$ in a $d$-simplex can also be defined by
\ba
h_a= \frac{d!}{(d-1)!}\frac{V}{V_{\bar{a}}} \, ,
\ea
where $V_{\bar{a}}$ is the (absolute) volume of the sub-simplex $\sigma_{\bar{a}}$.  Thus $|n_a|=(d-1)! V_{\bar{a}}$. With this normalization the closure relation for the normals associated to the faces of a simplex holds. That is, we can define the normal to the face $\sigma_{\bar{0}}$ as
\ba
n_{0} =- \sum_{a=1}^d n_{a }  \q \Rightarrow \q  \sum^d_{i=0} n_i \,=\, 0 \, .
\ea

~\\
Let us now consider the Gram matrix $\tilde G$ associated to the set of normals $\{n_a\}_{a=1}^d$, that is
\ba\label{A16}
\tilde G_{ab} &=& \sum_{\alpha\beta} n_{a\alpha}\eta^{\alpha\beta} n_{b\beta} \q\q \text{or}\q\q \tilde G\,=\, \text{Adj}(X)^T \cdot \eta^{-1} \cdot \text{Adj}(X) \q .
\ea
Using (\ref{A9}) we can conclude
\ba
G \cdot \tilde G \,=\, \tilde G \cdot G \,=\, \left( \det X\right)^2 \mathbb{I}
\ea
As $\det G=(\det X)^2 \det \eta$ we have
\ba\label{A18}
\boxed{\tilde G \,=\, \det(\eta^{-1}) \text{Adj}(G) }\, .
\ea

Later we will need the matrix elements of  $\text{Adj}(G) $. We remind the reader that the adjugate of $G$ can be also defined from the matrix of minors $\text{Mr}(G)$ of $G$. The matrix element  $(\text{Mr}(G))_{ab}$ is defined as determinant of the matrix $G^{\bar{a}\bar{b}}$, which is obtained from $G$ by removing the $a$-th row and $b$-th column. The adjugate of $G$ is then given by
\ba\label{A19}
({\rm Adj}(G))_{ba}=(-1)^{a+b} (\text{Mr}(G))_{ab} \, .
\ea 

In the following we will denote with  $\mathbb V$ the signed volume square of $\sigma$ and with $\mathbb  V_a$ the signed volume square of the sub-simplex $\sigma_{\bar{a}}$. The matrix elements of the adjugate of $G$ can then be expressed as
\ba\label{AdjA}
\boxed{({\rm Adj}(G))_{ab}= \begin{cases} ((d-1)!)^2 \, \mathbb  V_a &\; \text{for } a = b \\ 
-(d!)^2\, \frac{\partial \mathbb V}{\partial s_{ab}} & \;\text{for } a \neq b  \end{cases}} \, .
\ea
To see this, notice that in the case $a=b$ we have  $\text{Mr}(G)_{aa}=\text{Adj}(G)_{aa}$ and  $\text{Mr}(G)_{aa}$ is given by the determinant of the (edge vector) Gram matrix associated to the sub-simplex $\sigma_{\bar{a}}$.  $\text{Adj}(G)_{aa}$ gives therefore $((n-1)!)^2  \mathbb  V_{\bar{a}}$.

In the case $i\neq j$ we can use Jacobi's formula for the variation of the determinant of a square matrix:
\ba
\frac{\partial \text{det} G}{\partial s_{ab}} =\text{tr} \left( \text{Adj}(G) \frac{\partial G}{\partial s_{ab}} \right)
\ea
for $b>a$. The result follows by using that $\text{det} G=(d!)^2  \mathbb  V$ and that the matrix elements of $G$ are given in terms of edge lengths by (\ref{gram1}). 

~\\

\subsection{Projecting out a hinge}

The hinge $\sigma_{\overline{ab}}$ is a $(d-2)$-simplex obtained from the $d$-simplex $\sigma$ by removing the vertices $v_a$ and $v_b$, as well as all adjacent sub-simplices.

The dihedral angle $\theta_{ab}$ at a hinge $\sigma_{\overline{ab}}$ is the angle between the hyper-planes defined by the sub-simplices $\sigma_{\bar{a}}$ and $\sigma_{\bar{b}}$, which intersect in $\sigma_{\overline{ab}}$. To find this dihedral angle, we need to project the simplex metric onto the plane orthogonal to the hinge.

To do so, we assume without loss of generality, that the hinge is $\sigma_{\overline{(d-1)(d)}}$.
The Gram matrix $G$ for the $d$-simplex can then be written as 
\be \label{A22}
G = \begin{pmatrix} H & B \\  B^T& \bar H \end{pmatrix}   = \begin{pmatrix} H & B_{d-1} & B_d \\  B^T_{d-1} & \bar H_{(d-1)(d-1)} & \bar H_{(d-1)(d)} \\ B^T_n & \bar H_{(d-1)(d)} & \bar H_{(d)(d)} \end{pmatrix} \,,
\ee
where $H$ contains the metric components of  the hinge and $\bar H$ is the $2\times 2$ metric associated to the triangle defined by $\{v_0,v_{(n-1)},v_n\}$.

To find the metric orthogonal to the hinge we project out from  $x_c=v_c-v_0$ with $c=(d-1)$ or $c=d$  all parts parallel to the remaining vectors $x_a$ with $a=1,\ldots,(d-2)$.  These projections are given by
\ba
x^\text{proj}_c&=& x_c - \sum_{a,b=1}^{d-2}  \langle x_c\,| x_a\rangle (H^{-1})_{ab}\,\, x_b  \, .
\ea
It is then straightforward to compute the contractions between the $\{x^\text{proj}_{(d-1)}, x^\text{proj}_d\}$, which can be encoded in the Gram matrix
\be\label{projMetric}
P =  \begin{pmatrix}  P_{(d-1)(d-1)} & P_{(n-1)(n)} \\  P_{(n-1)(n)} & P_{(n)(n)} \end{pmatrix} \,\,=\,\,  \bar H - B^T ( H^{-1} ) B  \, .
\ee

We now consider the minors $(\text{Mr}(G))_{cc'}$ of $G$ with $c=(d-1),(d)$ and $c'=(d-1),(d)$. Using the form (\ref{A22}) for $G$ these minors are given by 
\be\label{B10}
({\rm Mr}(G))_{cc'} = \det \begin{pmatrix} H & B_c \\  B^T_{c'}& \bar H_{cc'} \end{pmatrix}  = \det H \left( \bar H_{cc'} - B^T_{c'} H^{-1} B_c \right) = \det H \, P_{cc'} \,  \, ,
\ee
where in the second equation we used an identity for the determinant of block matrices\footnote{\label{f11}$\det M = \det(A) \det(D-CA^{-1}B)$ for a matrix $M=\begin{pmatrix} A & B \\ C & D \end{pmatrix}$.}, and in the third equation we used the form of $P$ in (\ref{projMetric}). 
The matrix $H$ is the metric associated to the hinge and hence we have $\det H =  ((n-2)!)^2 \mathbb V_{\overline{(d-1)(d)}}$.  Thus, using the explicit form for the adjugate in (\ref{AdjA}) and its relation to the minors in (\ref{A19}), we obtain for the metric of the projected triangle 
\be
P = \frac{(d-1)^2}{\mathbb V_{\overline{(d-1)(d)}}} \begin{pmatrix} \mathbb V_{\overline{(d-1)}} & d^2 \frac{\partial \mathbb V}{\partial s_{(d-1)(d)}} \\  d^2 \frac{\partial \mathbb V}{\partial s_{(d-1)(d)}} &\mathbb V_{\overline{(d)}} \end{pmatrix}  \q.
\ee
The elements of $P$ give us the inner product between the vectors $p_{d-1}$ and $p_{d}$, which are obtained by projecting $\sigma_{\overline{(d-1)}}$ and $\sigma_{\overline{(d)}}$ onto the plane orthogonal to the hinge $\sigma_{\overline{(d-1)(d)}}$.

We can however generalize this conclusion for the vertex pair $(v_{(d-1)},v_{(d)})$ to any vertex pair $(v_i,v_j)$ in $\sigma$, and conclude that the inner product between the corresponding vectors $p_i,p_j$ is given by
\be\label{A27}
\boxed{\langle p_i\,|\,p_i\rangle   =(d-1)^2 \frac{ \mathbb V_{\bar{i}}}{\mathbb V_{\overline{ij}}} , \q 
\langle p_j\,|\,p_j\rangle   =(d-1)^2 \frac{ \mathbb V_{\bar{j}}}{\mathbb V_{\overline{ij}}} , \q 
\langle p_i\,|\,p_j\rangle    = \frac{d^2(d-1)^2}{\mathbb V_{\overline{ij}}} \frac{\partial \mathbb V}{\partial s_{ij}}\; }\, .
\ee

To close this subjection we derive an identity which will allow us to express the derivative of the volume in terms of products of volumina. To this end note that (\ref{A22}) and (\ref{projMetric}), together with the identity for the determinant of block matrices in Footnote \ref{f11}, implies
\ba
\det G\,=\, \det H \,\, \det P \q \Rightarrow \q
(d!)^2 \mathbb{V}\,=\, ((d-2)!)^2 \mathbb{V}_{\overline{ij}} 
\frac{(d-1)^4}{\mathbb{V}_{\overline{ij}}^2}
 \left( \mathbb V_{\bar{i}} \mathbb V_{\bar{j}}-d^4 \left(  \frac{\partial \mathbb V}{\partial s_{ij}} \right)^2\right)\,
\ea
and thus
\ba\label{A29}
\boxed{\left(  \frac{\partial \mathbb V}{\partial s_{ij}} \right)^2\,=\, \frac{1}{d^4} \mathbb V_{\bar{i}} \mathbb V_{\bar{j}} -\frac{1}{ d^2 (d-1)^2}     \mathbb{V}  \mathbb{V}_{\overline{ij}} } \, .
\ea

\subsection{Defining angles in Euclidean and Minkowskian planes}\label{SecSub:DefinitionsAngles}

Next we will discuss how to define angles of a (convex) wedge spanned by two vectors $a,b$ in the Euclidean or Minkowskian plane. 

In the Euclidean case the angle $\psi_E\in [0,\pi]$ is defined by
\ba
\psi_E \,=\, \cos^{-1}( \langle \hat a \,|\, \hat b \rangle) \, ,
\ea
where $\hat a=a/|a|$ and $|a|=\sqrt{\langle \hat a \,|\, \hat a \rangle}$. It can be understood to measure the distance between two points on a unit circle, which goes through the end points of $\hat a$ and $\hat b$ and is centred at the source point of these vectors.

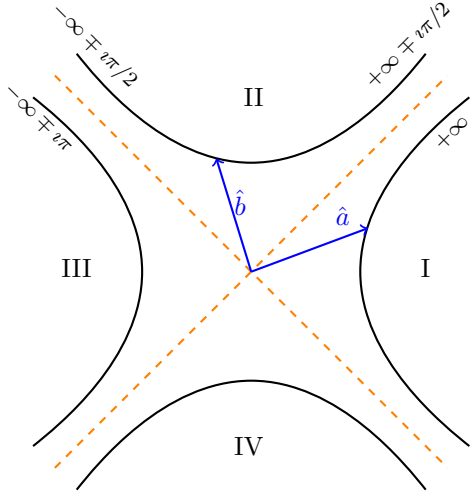
\begin{figure}[t]
\begin{tikzpicture}[scale=0.58]
\draw[dashed, orange,thick] (-4.5,-4.5)--(4.5,4.5) (-4.5,4.5)--(4.5,-4.5);

\draw[thick] (-4,-5) parabola bend (0,-2.5) (4,-5);
\draw[thick]  (4,5) parabola bend (0,2.5)(-4,5);
\draw[thick,rotate=90 ] (-4,-5) parabola bend (0,-2.5) (4,-5) ;
\draw[thick,rotate=90 ] (4,5) parabola bend (0,2.5)(-4,5);

\draw [blue,thick,->] (0,0)--(2.68,1.0) ;
\draw [blue,thick,->] (0,0) -- (-0.8,2.6) ;

\node[rotate = 45] at (4.6,3.2) {\scriptsize $+\infty $};
\node[rotate = 45] at (3.6,5.2) {\scriptsize $+\infty \mp \imath \pi/2$};
\node[rotate = -45] at (-3.6,5.2) {\scriptsize $-\infty \mp \imath \pi/2$};
\node[below, rotate = -45] at (-4.6,3.8) {\scriptsize $-\infty \mp \imath \pi$};

\node[below] at (4,0.5) {I};
\node[left] at (0.5,4) {II};
\node[below] at (-4,0.5) {III};
\node[left] at (0.5,-4) {IV};

\node[below] at (2.1,1.7) {$ \color{blue} \hat a$};
\node[right] at (-0.6,1.6) {$\color{blue} \hat b$};

\end{tikzpicture}
\caption{\label{Fig1} The different cases for angles in the Minkowski plane.  The value of the Lorentzian angle $\psi_{L\pm}$ between $\hat a$ and $\hat b$, for $\hat a$ fixed in quadrant I and the end point of $\hat b$ moving along the hyperbola in quadrant I, goes to $+\infty$. If $\hat b$ crosses the light ray between quadrant I and II  the angle $\psi_{L\pm}$ picks up a contribution $\mp \imath \pi/2$.  Moving the end point of $\hat b$ along the hyperbola of quadrant II to the left the real part of the angle changes from positive values to negative values, whereas the imaginary part remains constant. Crossing the second light ray into quadrant III, the angle acquires an imaginary part $\mp \imath \pi$.}
\end{figure}

In the case of the Minkowskian plane the unit circle is replaced with four disconnected hyperbolae branches, see Fig.~\ref{Fig1}.  To characterize the distance between points of different branches one allows the angles to be complex  \cite{Sorkin1974,Sorkin2019}. The imaginary part comes in multiples of $\mp \imath \pi/2$ and counts how many light rays are included in the wedge spanned by the vectors $a$ and $b$. Here the sign appears as a choice of convention and leads to two different angles denoted by $\psi_{L\pm}$. We will see, that these two angles arise from different choices of paths for the analytical continuation of the Euclidean angle to Lorentzian data. 

 The two different choices  lead to the same Lorentzian Regge action for data which have a light cone regular structure. A  two-dimensional triangulation with a light cone regular structure has always four light rays (or two light cones) meeting at a vertex \cite{Sorkin2019,ADP21}. Complexifying the configuration space of length (square) variables the Regge action has branch cuts along data describing light cone irregular triangulations \cite{ADP21}. In this case the two choices mentioned above give the values for the Regge action at the two opposite sides of the branch cut. This also means that the two choices reproduce equivalent analytical continuations of the (complexified) Regge action \cite{ADP21}.

To start with we specify the angles $\psi_{L\pm}$  case by case (as done in \cite{Sorkin2019}), depending on which quadrant the two vectors $a$ and $b$ are positioned in, see Figure \ref{Fig1}.\footnote{
 Here we will work with non-oriented angles. That is the angle for a wedge spanned by two vectors $a$ and $b$ is the same for clock-wise or anti-clockwise ordering of the vectors. We will furthermore assume that the wedge is convex, the angle can in this case  be uniquely specified by the vectors $a$ and $b$. The definition of angles can be extended to non-convex wedges, by demanding that the angles are additive \cite{Sorkin2019}.  We will exclude  the case that one or both of the vectors is null, see \cite{Sorkin2019} for an extensive discussion of this case.
 }

 Let us denote a normalized vector by $\hat a=a/|a|$ where $|a|=\sqrt{|a\cdot a|}$. The Lorentzian angles can then be defined as
\ba\label{ad1}
\psi_{L\pm}  &=& \cosh^{-1} ( \langle \hat a \,|\, \hat b\rangle )\q \q \q\q\q \text{if}  \q\q a \, \, \text{in quadrant I} \; \;\&\;\; b\,\,  \text{in quadrant I}  \nn\\
\psi_{L\pm}  &=& \sinh^{-1} (\langle \hat a \,|\, \hat b\rangle  ) \mp \tfrac{ \pi}{2} \imath   \q\q  \q \! \text{if} \q\q a \, \, \text{in quadrant I} \; \;\&\;\; b\,\,  \text{in quadrant II}  \nn\\
\psi_{L\pm}   &=& -\cosh^{-1} (-\langle \hat a \,|\, \hat b\rangle  ) \mp {\pi \imath }  \,\q \text{if} \q\q a \, \, \text{in quadrant I}  \;\;\&\;\; b\,\,  \text{in quadrant III}\nn\\
\psi_{L\pm}  &=& -\cosh^{-1} (-\langle \hat a \,|\, \hat b\rangle  ) \;  \q\q\q \text{if}  \q\q a \, \, \text{in quadrant II} \; \;\&\;\; b\,\,  \text{in quadrant II}\nn\\
\psi_{L\pm}  &=& \cosh^{-1} (\langle \hat a \,|\, \hat b\rangle  ) \mp \pi \imath \; \, \,\; \q\q \text{if}\q\q a \, \, \text{in quadrant II} \; \;\&\;\; b\,\,  \text{in quadrant IV} \,\,  . 
\ea
Here we define $\cosh^{-1}(x) \in \mathbb{R}_+$ with $x \geq 1$.  

The Euclidean and  Lorentzian angle arise as special cases from the following definition of  complex angles \cite{Sorkin2019,ADP21,Ding}
\ba\label{complexangle}
\theta^\pm  &=&   -\imath \log_\mp \frac{  \langle a\,|\, b\rangle \mp \imath \sqrt{\!\!{}_{{}_\pm} \,\,\langle a\,|\, a\rangle\langle b\,|\, b\rangle - \langle a\,|\, b\rangle^2 }} { \sqrt{\!\!{}_{{}_\pm}\,\,\langle a\,|\, a\rangle} \sqrt{\!\!{}_{{}_\pm}\,\,\langle b\,|\, b\rangle} }            \q .
\ea
Here  $\sqrt{\!\!{}_{{}_\pm}z}$ and $\log_\pm z$ are defined via their principal branch values for $z=re^{\imath\phi}$ and $r>0$ and $\phi \in (-\pi,\pi)$. For the branch cut along the real negative axis, we define for $r>0$
\ba
\log_\pm (-r) \,=\,  \log(r) \pm \imath \pi      \q ,\q\q \q \sqrt{\!\!{}_{{}_\pm}-r} \,=\, \pm \imath \sqrt{r} \, .
\ea

With these choices the complex angles reproduce the Euclidean and Minkowskian angles as follows \cite{ADP21}:
\be\label{A34}
\boxed{\q
\eval{\theta^\pm}_{\text{Eucl.}}\,=\, \mp \psi_E \q ,\q\q\q\q \eval{\theta^\pm}_{\text{Mink.}}\,=\, -\imath\psi_{L\pm}    
\q}\, .
\ee
For Euclidean data  $\langle a\,|\, a\rangle\langle b\,|\, b\rangle - \langle a\,|\, b\rangle^2$ is positive and $\langle a\,|\, a\rangle, \langle b\,|\, b\rangle$ are positive. Thus, for generic\footnote{The cases that $a$ and $b$ are parallel or anti-parallel can also be reproduced correctly, see \cite{ADP21}.} Euclidean data,  we do not encounter the branch cuts and  the result (\ref{A34}) follows. 

For Minkowskian data $\langle a\,|\, a\rangle\langle b\,|\, b\rangle - \langle a\,|\, b\rangle^2$ is negative and thus the numerator in (\ref{complexangle}) is the same for $\theta^+$ and $\theta^-$ and is real. If $a$ is time-like and $b$ is space-like, the denominator in (\ref{complexangle}) will be imaginary and have opposite signs for $\theta^\pm$. Thus one avoids the branch cut for the logarithm, but the argument in the logarithm is imaginary and has opposite signs for $\theta^\pm$, leading to $\theta^\pm=\imath\psi_{L\pm}$.

If $a$ and $b$ are both space-like or both time-like, one can show that the argument of the logarithm is positive real if $a$ and $b$ are from the same quadrant, and is negative real if $a$ and $b$ are from opposite quadrants. Thus, for the case that $a$ and $b$ are from opposite quadrants we encounter the branch cut of the logarithm, and the values for $\theta^\pm=\imath\psi_{L\pm}$ differ. For the case that $a$ and $b$ are from the same quadrant, the values for $\theta^\pm=\imath\psi_{L\pm}$ agree. See \cite{ADP21} for a more explicit analysis of all these cases.

Now, the dihedral angle $\theta_{\sigma,h}^\pm$ in a $d$-simplex $\sigma$, at a hinge $h$, is given by the angle between the vectors which result by projecting the $(d-1)$-dimensional sub-simplices $\rho_a$ and $\rho_b$, which share $h$, onto the plane orthogonal to $h$. We computed the inner product between these vectors (for $h$ given by $\sigma_{\overline{ij}}$) in (\ref{A27}).  We thus obtain
\be\label{A35}
\boxed{
\,\,\theta_{\sigma,h}^\pm = -\imath \log_\mp \qty(\frac{
	\frac{d^2}{  \mathbb{V}_{h}}  \frac{\partial \mathbb{V}_{\sigma}}{\partial s_{\bar{h}}}  \mp  \imath \,
	\sqrt{\!\!{}_{{}_\pm} \,\, \frac{\mathbb{V}_{\rho_a} }{   \mathbb{V}_{h}  }\frac{\mathbb{V}_{\rho_b}}{\mathbb{V}_{h} }-\qty( \frac{d^2}{  \mathbb{V}_{h}}  \frac{\partial \mathbb{V}_{\sigma}}{\partial s_{\bar{h}}} )^2
	} 
} 
{ \sqrt{\!\!{}_{{}_\pm}\,\, \frac{\mathbb{V}_{\rho_a} }{   \mathbb{V}_{h}  } }  
 \sqrt{\!\!{}_{{}_\pm} \,\,\frac{\mathbb{V}_{\rho_b}}{    \mathbb{V}_{h} }} } ) 
 \,\,} \q .
\ee
Here $s_{\overline{h}}$ is the signed length squared of the edge in $\sigma$, which is opposite form the hinge $h$.

Note that  (\ref{A35}) gives the dihedral angle in the form $\theta=-\imath \log( x/y \,\mp\, (\imath\sqrt{y^2-x^2})/y)$. We can thus conclude, that
\be\label{A36}
\,\,\cos(\theta_{\sigma,h}^\pm) \,=\, d^2    
 \frac{1}{   \mathbb{V}_{h}     
  \sqrt{\!\!{}_{{}_\pm}\,\, \frac{\mathbb{V}_{\rho_a} }{   \mathbb{V}_{h}  } }  
 \sqrt{\!\!{}_{{}_\pm} \,\,\frac{\mathbb{V}_{\rho_b}}{    \mathbb{V}_{h} }}         } 
  \frac{\partial \mathbb{V}_{\sigma}}{\partial s_{\bar{h}}}   
\,\, \, .
\ee
Using the fact, that if $ \mathbb{V}_{h}$ is negative, so are (according to the generalized triangle inequalities) $\mathbb{V}_{\rho_a}$ and $\mathbb{V}_{\rho_b}$ we can write this expression as
\be\label{A37}
\boxed{
\,\,\cos(\theta_{\sigma,h}^\pm) \,=\, d^2    
 \frac{1}{    
  \sqrt{\!\!{}_{{}_\pm}\,\, \mathbb{V}_{\rho_a} }  
 \sqrt{\!\!{}_{{}_\pm} \,\,\mathbb{V}_{\rho_b}   }     } 
  \frac{\partial \mathbb{V}_{\sigma}}{\partial s_{\bar{h}}}   
}\,\, \, .
\ee

Applying (\ref{A29}) to the argument of the square root in the numerator of (\ref{A35}), we also obtain
\be\label{A38}
\boxed{
\,\, \sin(\theta_{\sigma,h}^\pm) \,=\,
\mp  \frac{d}{(d-1)}
  \frac{ \sqrt{\!\!{}_{{}_\pm}\,\, \mathbb{V}_{h}} 
  \sqrt{\!\!{}_{{}_\pm}\,\,  \mathbb V_\sigma       }       
  }
  {  
  \sqrt{\!\!{}_{{}_\pm}\,\, \mathbb{V}_{\rho_a}       } 
  \sqrt{\!\!{}_{{}_\pm}\,\,  \mathbb{V}_{\rho_b} }                                 
  }
\,\,} \, ,
\ee
where we made again use of the fact, that $\mathbb{V}_{h}<0$ implies that all other signed volume squares appearing in the equation above are negative.

We regain equation~(\ref{A29}), by plugging (\ref{A37}) and (\ref{A38}) into the relation $\cos^2(\theta)+\sin^2(\theta)=1$.

\subsection{The Regge action}\label{SecSub:ReggeAction}

Given the notion of complex dihedral angle $\theta_{\sigma,h}^\pm$ in (\ref{A35}) we define the notion of complex deficit angle at a hinge $h$ as the difference between the sum of the dihedral angles attached to $h$ and the flat space value. The latter is given by $\mp 2\pi$. Thus, we define the bulk deficit angles
\ba
\epsilon^{\pm\text{(bulk)}}_{h} =2\pi \pm \sum_{\sigma \supset h} \theta^\pm_{\sigma,h}
\ea
and the boundary curvature angles
\ba
\epsilon^{\pm\text{(bdry)}}_{h} =\pi k \pm \sum_{\sigma \supset h} \theta^\pm_{\sigma,h} \, ,
\ea
where the choice of $k$ depends on how many boundary pieces one glues together at $h$. The complex Regge action is then given by
\be\label{eq:LorentzianReggeActionApp}
\boxed{\,\,
\imath S^\pm_{\text{Regge}} \,=\, \sum_{h\subset \triangle^{(d)}}\sqrt{\!\!{}_{{}_\pm}\,\,       \mathbb{V}_{h}}\,\epsilon^\pm_{h}
\,\,}\, ,
\ee
where $\triangle^{(d)}$ denotes the $d$-dimensional triangulation.  

Note that the Euclidean and Lorentzian Regge actions are defined as \cite{Regge,Hartle,Sorkin1974,ADP21}
\ba
\,\,S_{ E} = -\sum_h |V_h| (\epsilon_{ E})_h    \,\, ,\q\q 
 S_{L\pm} = \sum_{h:{\rm t-like}}  |V_h| (\epsilon_E)_h + \sum_{h:{\rm s-like}}  |V_h| (\epsilon_{L\pm})_h \, ,
\ea
where 
\ba
(\epsilon_E)_h = \pi k - \sum_{\sigma \supset h} \psi_{E}^{\sigma,h} \q,\q\q
(\epsilon_{L\pm})_h = \mp \pi k - \sum_{\sigma \supset h} \psi_{L\pm}^{\sigma,h} \q,\q\q
\ea
and $k=2$ for a bulk hinge, and appropriately chosen for a boundary hinge.

Using  (\ref{A34}) we can conclude that
\be\label{A44}
\boxed{\q
\eval{\imath S^\pm_{\text{Regge}}}_{\text{Eucl. data}} \,=\, - S_E \q ,\q\q
\eval{\imath S^\pm_{\text{Regge}}}_{\text{Lor. data} }\,=\, \pm \imath S_{L\pm} \q
}\, .
\ee
The reason for the  appearance of the global sign can be understood as follows: The complex Regge action (\ref{eq:LorentzianReggeActionApp}), with the dihedral angles defined in (\ref{A35}), is analytic around generic Euclidean data. In contrast to the Euclidean case, we might encounter the branch cuts of the square roots and logarithms, if we consider Lorentzian data.  The different signs appear because we approach these branch cuts from opposite sides. Consider for instance a triangulation for which we can introduce a global Wick rotation (see \cite{ADP21} for an example). Let the Wick rotation angle $\phi=0$ represent Euclidean data and $\phi=\pm \pi$ Lorentzian data. For Euclidean data, that is for $\phi=0$, we have $S^+_{\text{Regge}}=S^-_{\text{Regge}}$. Thus $S^{+}_{\text{Regge}}$ and $S^{-}_{\text{Regge}}$ determine the same analytical continuation. $S^+_{\text{Regge}}$ evaluated at Lorentzian data represents the analytical continuation by increasing $\phi$ from $\phi=0$ to $\phi=+\pi$. Similarly, $S^-_{\text{Regge}}$ evaluated at Lorentzian data represents the analytical continuation by decreasing $\phi$ from $\phi=0$ to $\phi=-\pi$. In fact, the full analytical continuation leads at least to a double cover for $\phi$, that is $\phi$ can be extended to $\left(-2\pi,2\pi\right]$, with $\phi=2\pi$ representing another copy of Euclidean data. Here, if the Lorentzian data have a light cone regular structure, the analytically continued complex Regge action evaluates to $+S_E$.  Light cone irregular Lorentzian data can lead to branch cuts and thus to an even more involved structure of the Riemann surface. We refer to \cite{ADP21} for a more detailed analysis and explicit examples.

\subsection{Schl\"afli identity for Euclidean and Lorentzian simplices}\label{Sec:SchlaefliIdentity}

The Schl\"afli identity for an Euclidean or Lorentzian $d$-simplex $\sigma$ states that 
\ba\label{Schlafli}
\sum_{i<j}   \sqrt{\!\!{}_{{}_\pm}\,\, \mathbb{V}_{\overline{ij}}}  \,\, {\dd}\theta^\pm_{ij}    =0 \, ,
\ea
for variations ${\dd}$ of the simplex geometry. A basis of such variations is given by ${\dd} s_{kl}$.

Here $i,j$ run from $0$ to $d$ and denote the vertices of $\sigma$.  $\mathbb{V}_{\overline{ij}}$ is the signed volume square of the hinge $\sigma_{\overline{ij}}$, which is obtained by removing from $\sigma$ the vertices $i$ and $j$ and all adjacent sub-simplices. $\theta^\pm_{ij}$ is the complex dihedral angle at $\sigma_{\overline{ij}}$.

Note that with (\ref{Schlafli}) holding for $d$-simplices, we will have for a more general triangulation
\be\label{eq:SchlaefliLorentzian}
\sum_{h\subset \triangle^{(d)}} \sqrt{\!\!{}_{{}_\pm}\,\,         \mathbb{V}_h}\,  \sum_{\sigma \supset h} {\dd}{\theta^\pm_{h,\sigma} }=0 \, .
\ee
This identity can be regarded as the discrete analog of the term containing the variation of the Ricci tensor in the variation of the continuum Einstein-Hilbert action. The latter produces only a vanishing total divergence and thereby guarantees that the resulting field equations are of second order.

To prove (\ref{Schlafli}) we will make use of many formulae derived in this appendix. We start with (\ref{A16}), which states that $\tilde G_{ab}$ gives the inner product between the normal one forms $n_a$ associated to the faces $\sigma_{\overline{a}}$. As we choose $v_0$ as reference vertex, we defined $a$ to run from $1$ to $d$. But we can also include the normal $n_0$ into the Gram matrix, and thus work with $\tilde G_{ij}$.  Using (\ref{A18}), (\ref{AdjA}) and (\ref{A37}), we can conclude that (as the form of the following equation is invariant under a relabeling of the vertices)
\ba
\boxed{
\tilde G_{ij} \,=\, \det(\eta^{-1})
\begin{Bmatrix} ((d-1)!)^2 \, \mathbb  V_{\bar{i}} &\; \text{for } i = j \\ 
-(d!)^2\, \frac{\partial \mathbb V}{\partial s_{ij}} & \;\text{for } i \neq j  \end{Bmatrix} 
\,=\, - ((d-1)!)^2 \det(\eta^{-1})  \cos(\theta_{ij}^\pm) 
 \sqrt{\!\!{}_{{}_\pm}\,\, \mathbb{V}_{\bar{i}} }  
 \sqrt{\!\!{}_{{}_\pm} \,\,\mathbb{V}_{\bar{j}}   }  
 }\, .\nn\\
\ea
Here we defined $\cos\theta^\pm_{ii}=-1$ (and accordingly $\sin\theta^\pm_{ii}=0$). Now, starting from the closure relation for the normals results in
\ba\label{A48}
\sum_{i=0}^d n_i\,=\,0 \q \Rightarrow \q  \sum_{i=0}^d \tilde G_{ij}=\,0\,  \q \Rightarrow \q  \sum_{i=0}^d  \cos(\theta_{ij}^\pm)  \sqrt{\!\!{}_{{}_\pm}\,\, \mathbb{V}_{\bar{i}} }  \,=\,0 \q \text{for}\,\, j=0, \ldots,d\,.
\ea
Applying ${\bf d}$ to the last equation in (\ref{A48}) we obtain
\ba\label{A49}
-\sum_{i=0}^d \sqrt{\!\!{}_{{}_\pm}\,\, \mathbb{V}_{\bar{i}} } \,\, \sin(\theta_{ij}^\pm)\, {\dd}\theta_{ij}^\pm\, \,\,+\,\,  \sum_{i=0}^d  \cos(\theta_{ij}^\pm)  {\dd}\sqrt{\!\!{}_{{}_\pm}\,\, \mathbb{V}_{\bar{i}} } \,=\,0 \, .
\ea
We multiply (\ref{A49}) with $\sqrt{\!\!{}_{{}_\pm}\,\, \mathbb{V}_{\bar{j}} }$ and sum over $j$. This summation annihilates the second term in (\ref{A49}). We are therefore left with
\ba
\sum_{i,j=0}^d \sqrt{\!\!{}_{{}_\pm}\,\, \mathbb{V}_{\bar{i}} } \sqrt{\!\!{}_{{}_\pm}\,\, \mathbb{V}_{\bar{j}} }  \,\,\sin(\theta_{ij}^\pm)\, {\dd}\theta_{ij}^\pm\,=\, 0 
 \q\stackrel{\rm (\ref{A38})}{\Longrightarrow}\q
 \sum_{i,j=0}^d \sqrt{\!\!{}_{{}_\pm}\,\, \mathbb{V}_{\overline{ij}} } \,\,{\dd}\theta_{ij}^\pm\,=\, 0 \, .
\ea
As ${\dd}\theta_{ii}^\pm =0$ the last equation implies the Schl\"afli identity (\ref{Schlafli}).

\subsection{Derivatives of the dihedral angle }\label{SecSub:DihedralAngleDerivative}

In (\ref{eq:DihedralAngleDerivativeOpposite}) used the derivative of the dihedral angle $\theta_{ij}$ at a hinge $\sigma_{\overline{ij}}$ in a simplex $\sigma$, with respect to the signed length squared $s_{ij}$ of the edge opposite to this hinge.



To prove this formula we will use again (\ref{A37}) and (\ref{A38}).  We take the derivative of equation (\ref{A38}), which defines $\sin \theta^\pm_{ij}$ and utilize that $\partial \mathbb{V}_{\bar{i}}/\partial s_{ij}=\partial \mathbb{V}_{\bar{j}}/\partial s_{ij}=0$ and $\partial \mathbb{V}_{\overline{ij}}/\partial s_{ij}=0$. We obtain
\ba
\,\,\cos(\theta_{ij}^\pm) \pdv{\theta^\pm_{ij}} {s_{ij}} &=& \mp  \frac{d}{2(d-1)}
  \frac{ \sqrt{\!\!{}_{{}_\pm}\,\, \mathbb{V}_{\overline{ij}}}    
  }
  {  
  \sqrt{\!\!{}_{{}_\pm}\,\, \mathbb{V}_{\bar{i}}       } 
  \sqrt{\!\!{}_{{}_\pm}\,\,  \mathbb{V}_{\bar{j}} }      
   \sqrt{\!\!{}_{{}_\pm}\,\,  \mathbb V       }                             
  }
  \pdv{\mathbb{V}}{s_{ij}} \nn\\
 &\stackrel{\rm (\ref{A37})}{=}&
  \mp  \frac{1}{2(d-1)d}   
 \frac{ \sqrt{\!\!{}_{{}_\pm}\,\, \mathbb{V}_{\overline{ij}}}    
   } {
   \sqrt{\!\!{}_{{}_\pm}\,\, \mathbb{V}    }
         }
 \,   \cos(\theta_{ij}^\pm)    \q .
\ea

Thus 
\ba
\boxed{\,\,
 \pdv{\theta^\pm_{ij}} {s_{ij}} \,\,=\,\,
   \mp  \frac{1}{2(d-1)d}   
 \frac{ \sqrt{\!\!{}_{{}_\pm}\,\, \mathbb{V}_{\overline{ij}}}    
   } {
   \sqrt{\!\!{}_{{}_\pm}\,\, \mathbb{V}    } 
         }
  %
  \,\,     } \, .
\ea

As $\theta^\pm$ reproduces $\mp\psi_E$ for Euclidean data, we also reproduce the formula for the derivative of the Euclidean dihedral angle in \cite{Dittrich:2007wm,Dittrich:2011vz}.

Using (\ref{A37}) and (\ref{A38}) we can also obtain a formula for the derivative of the dihedral angle with respect to the other signed length squared variables
\ba
\,\,\cos(\theta_{ij}^\pm) \pdv{\theta^\pm_{ij}} {s_{kl}} &=& \mp  \frac{d}{2(d-1)} 
 \frac{ \sqrt{\!\!{}_{{}_\pm}\,\, \mathbb{V}_{\overline{ij}}} 
  \sqrt{\!\!{}_{{}_\pm}\,\,  \mathbb V       }       
  }
  {  
  \sqrt{\!\!{}_{{}_\pm}\,\, \mathbb{V}_{\bar{i}}       } 
  \sqrt{\!\!{}_{{}_\pm}\,\,  \mathbb{V}_{\bar{j}} }                                 
  }
\left(
\frac{1}{ \mathbb{V}}  \pdv{ \mathbb{V}}{ s_{kl}}+
\frac{1}{ \mathbb{V}_{\overline{ij}}}  \pdv{ \mathbb{V}_{\overline{ij}}}{ s_{kl}}  
-
\frac{1}{ \mathbb{V}_{\overline{i}}}  \pdv{ \mathbb{V}_{\overline{i}}}{ s_{kl}} -
\frac{1}{ \mathbb{V}_{\overline{j}}}  \pdv{ \mathbb{V}_{\overline{j}}}{ s_{kl}} 
\right) \, ,
\ea
which allows us to conclude
\be
\boxed{\,\pdv{\theta^\pm_{ij}} {s_{kl}}\,\,=\,\,
\frac{1}{2}  \frac{\sin(\theta_{ij}^\pm)}{\cos(\theta_{ij}^\pm)} \left(
\frac{1}{ \mathbb{V}}  \pdv{ \mathbb{V}}{ s_{kl}}+
\frac{1}{ \mathbb{V}_{\overline{ij}}}  \pdv{ \mathbb{V}_{\overline{ij}}}{ s_{kl}} 
-
\frac{1}{ \mathbb{V}_{\overline{i}}}  \pdv{ \mathbb{V}_{\overline{i}}}{ s_{kl}} -
\frac{1}{ \mathbb{V}_{\overline{j}}}  \pdv{ \mathbb{V}_{\overline{j}}}{ s_{kl}} 
\right)  \,}     \, .
\ee

\section{Integration measure for gauge orbits}\label{SecSub:IntegrationMeasureGaugeOrbit}

Here we derive an identity between an integration measure in terms of the signed length squares of an (embedded) $d$-simplex and a measure in terms of the coordinates of one of its vertices \cite{Baratin:2006yu,Dittrich:2011vz}. 

We consider a $d$-simplex $\sigma$ embedded into flat Minkowskian or Euclidean space and vertex coordinates ${v_0^\alpha, v_1^\alpha, \ldots, v_d^\alpha}$. The signed length squares for the edges $(0a)$, with $a=1,\ldots,d$ are given by 
$
s_{0a}\,=\, \sum_{\alpha,\beta} (v_0^\alpha-v_a^\alpha)\eta_{\alpha\beta}(v_0^\beta-v_a^\beta)
$.
We can express the integration measure as
\ba
\prod_{a=1}^d \dd s_{0a}  \,=\, \prod_{\alpha=1}^d \dd v_0^\alpha \left|  \det\left(  \pdv{s_{0a}}{v^\beta_0}  \right)   \right| \q .
\ea
The determinant of the Jacobian gives
\ba
\left|  \det\left(  \pdv{s_{0a}}{v^\beta_0}  \right)   \right|  
&=&
2^d |\det(\eta)| \,\, |\det(v_0^\gamma-v_a^\gamma)| 
\,\stackrel{\rm (\ref{A3})}{=}\, 2^d d! \, V \,,
\ea
where $V$ is the absolute volume of the simplex. One can then write
\ba
\sum_{\text{orientation of $\sigma$}}\frac{\prod_{a=1}^d \dd s_{0a}}{2^d d! V} \,\,=\,\, \prod_{\alpha=1}^d \dd v_0^\alpha \q .
\ea
The sum over the two orientations of $\sigma$ appears because of the following:  Given the vertex $v_0$ we assume that it is not in the hyper-plane defined by the vertices $v_1,\ldots, v_d$ (which is the case if $\sigma$ is not degenerate). Constructing the mirror image of $v_0$ with respect to this hyper-plane we obtain a $d$-simplex, which has the same edge lengths as before but opposite orientation.

As we consider a perturbative evaluation, we do however not include a sum over orientations in our path integral. In the main text, we therefore ignore this sum over orientations, but nevertheless associate the integral over the $s_{0a}$ to an integral over the gauge orbit, resulting from displacements of the vertex $v_0$. In fact, if one does not include the sum  over orientations into the action, the action is only invariant under such displacements as long as the vertex is inside the coarser simplex.

\begin{acknowledgments}

We thank Dongxue Qu for collaboration during the initial phase of the project and  Seth Asante and Jos\'e Padua-Arg\"uelles for helpful discussions. JNB is supported by an NSERC grant awarded to BD.
Research at Perimeter Institute is supported in part by the Government of Canada through the Department of Innovation, Science and Economic Development Canada and by the Province of Ontario through the Ministry of Colleges and Universities.

\end{acknowledgments}


\begingroup

\begin{thebibliography}{98}\footnotesize
	
\bibitem{PerezLR}
A.~Perez,
``The Spin Foam Approach to Quantum Gravity,''
Living Rev.\ Rel.\  {\bf 16} (2013) 3
[arXiv: 1205.2019].

\bibitem{WilliamsReview} 
T.~Regge and R.~M.~Williams,
``Discrete structures in gravity,''
J. Math. Phys. \textbf{41} (2000), 3964-3984
[arXiv:gr-qc/0012035 [gr-qc]].

\bibitem{Oriti:2006se}
D.~Oriti,
``The Group field theory approach to quantum gravity,''
[arXiv:gr-qc/0607032 [gr-qc]].

\bibitem{Loll:2019rdj}
R.~Loll,
``Quantum Gravity from Causal Dynamical Triangulations: A Review,''
Class. Quant. Grav. \textbf{37} (2020) no.1, 013002
[arXiv:1905.08669 [hep-th]].

\bibitem{Visser} 
K.~Tate and M.~Visser,
``Fixed-Topology Lorentzian Triangulations: Quantum Regge Calculus in the Lorentzian Domain,''
JHEP \textbf{11} (2011), 072
[arXiv:1108.4965 [gr-qc]].

\bibitem{Dittrich:2014mxa}
B.~Dittrich, S.~Mizera and S.~Steinhaus,
``Decorated tensor network renormalization for lattice gauge theories and spin foam models,''
New J. Phys. \textbf{18} (2016) no.5, 053009
[arXiv:1409.2407 [gr-qc]].

\bibitem{Delcamp:2016dqo}
C.~Delcamp and B.~Dittrich,
``Towards a phase diagram for spin foams,''
Class. Quant. Grav. \textbf{34} (2017) no.22, 225006
[arXiv:1612.04506 [gr-qc]].

\bibitem{Turok} J.~Feldbrugge, J.~L.~Lehners and N.~Turok,
``Lorentzian Quantum Cosmology,''
Phys. Rev. D \textbf{95} (2017) no.10, 103508
[arXiv:1703.02076 [hep-th]].

\bibitem{Ding}
D.~Jia,
``Complex, Lorentzian, and Euclidean simplicial quantum gravity: numerical methods and physical prospects,''
Class. Quant. Grav. \textbf{39} (2022) no.6, 065002
[arXiv:2110.05953 [gr-qc]].

\bibitem{ADP21}
S.~K.~Asante, B.~Dittrich and J.~Padua-Arg\"uelles,
``Complex actions and causality violations: Applications to Lorentzian quantum cosmology,''
[arXiv:2112.15387 [gr-qc]].

\bibitem{toappear} B.~Dittrich and J.~Padua-Arg\"uelles,
``Lorentzian quantum cosmology from effective spin foams,''
[arXiv:2306.06012 [gr-qc]].

\bibitem{LollLR}
R.~Loll,
``Discrete approaches to quantum gravity in four-dimensions,''
Living Rev. Rel. \textbf{1} (1998), 13
[arXiv:gr-qc/9805049 [gr-qc]].

\bibitem{ConformalFactor}
G.~W.~Gibbons, S.~W.~Hawking and M.~J.~Perry,
``Path Integrals and the Indefiniteness of the Gravitational Action,''
Nucl. Phys. B \textbf{138} (1978), 141-150.

\bibitem{CDT1}
J.~Ambjorn and R.~Loll,
``Nonperturbative Lorentzian quantum gravity, causality and topology change,''
Nucl. Phys. B \textbf{536} (1998), 407-434
[arXiv:hep-th/9805108 [hep-th]].

\bibitem{CDT2}
J.~Ambjorn, J.~Jurkiewicz and R.~Loll,
``A Nonperturbative Lorentzian path integral for gravity,''
Phys. Rev. Lett. \textbf{85} (2000), 924-927
[arXiv:hep-th/0002050 [hep-th]].

\bibitem{deBoer:2022zka}
J.~de Boer, B.~Dittrich, A.~Eichhorn, S.~B.~Giddings, S.~Gielen, S.~Liberati, E.~R.~Livine, D.~Oriti, K.~Papadodimas and A.~D.~Pereira, \textit{et al.}
``Frontiers of Quantum Gravity: shared challenges, converging directions,''
[arXiv:2207.10618 [hep-th]].

\bibitem{Regge}
T.~Regge,
``General relativity without coordinates,''
Nuovo Cim. \textbf{19} (1961), 558-571

\bibitem{Dittrich:2011vz}
B.~Dittrich and S.~Steinhaus,
``Path integral measure and triangulation independence in discrete gravity,''
Phys. Rev. D \textbf{85} (2012), 044032
[arXiv:1110.6866 [gr-qc]].

\bibitem{Bonzom:2015ans}
V.~Bonzom and B.~Dittrich,
``3D holography: from discretum to continuum,''
JHEP \textbf{03} (2016), 208
[arXiv:1511.05441 [hep-th]].

\bibitem{Dittrich:2017hnl}
B.~Dittrich, C.~Goeller, E.~Livine and A.~Riello,
``Quasi-local holographic dualities in non-perturbative 3d quantum gravity I \textendash{} Convergence of multiple approaches and examples of Ponzano\textendash{}Regge statistical duals,''
Nucl. Phys. B \textbf{938} (2019), 807-877
[arXiv:1710.04202 [hep-th]].

\bibitem{Dittrich:2014rha}
B.~Dittrich, W.~Kami\'nski and S.~Steinhaus,
``Discretization independence implies non-locality in 4D discrete quantum gravity,''
Class. Quant. Grav. \textbf{31} (2014) no.24, 245009
[arXiv:1404.5288 [gr-qc]].

\bibitem{Dittrich:2008pw}
B.~Dittrich,
``Diffeomorphism symmetry in quantum gravity models,''
Adv. Sci. Lett. \textbf{2}, 151
[arXiv:0810.3594 [gr-qc]].

\bibitem{Bahr:2009ku}
B.~Bahr and B.~Dittrich,
``(Broken) Gauge Symmetries and Constraints in Regge Calculus,''
Class. Quant. Grav. \textbf{26} (2009), 225011
[arXiv:0905.1670 [gr-qc]].

\bibitem{Dittrich:2011ien}
B.~Dittrich,
``How to construct diffeomorphism symmetry on the lattice,''
PoS \textbf{QGQGS2011} (2011), 012
[arXiv:1201.3840 [gr-qc]].

\bibitem{DittrichBook14} 
B.~Dittrich,
``The continuum limit of loop quantum gravity - a framework for solving the theory,''
[arXiv:1409.1450 [gr-qc]].

\bibitem{Asante:2022dnj}
S.~K.~Asante, B.~Dittrich and S.~Steinhaus,
``Spin foams, Refinement limit and Renormalization,''
[arXiv:2211.09578 [gr-qc]].

\bibitem{Korepanov:2000jp}
I.~G.~Korepanov,
``Multidimensional analogues of the geometric s \ensuremath{<}--\ensuremath{>} t duality,''
Theor. Math. Phys. \textbf{124} (2000), 999-1005

\bibitem{Korepanov:2000aj}
I.~G.~Korepanov,
``Invariants of PL manifolds from metrized simplicial complexes: Three-dimensional case,''
J. Nonlin. Math. Phys. \textbf{8} (2001), 196-210
[arXiv:math/0009225 [math.GT]].

\bibitem{Baratin:2006yu}
A.~Baratin and L.~Freidel,
``Hidden Quantum Gravity in 3-D Feynman diagrams,''
Class. Quant. Grav. \textbf{24} (2007), 1993-2026

\bibitem{Baratin:2006gy}
A.~Baratin and L.~Freidel,
``Hidden Quantum Gravity in 4-D Feynman diagrams: Emergence of spin foams,''
Class. Quant. Grav. \textbf{24} (2007), 2027-2060
[arXiv:hep-th/0611042 [hep-th]].

\bibitem{Baratin:2014era}
A.~Baratin and L.~Freidel,
``A 2-categorical state sum model,''
J. Math. Phys. \textbf{56} (2015) no.1, 011705
[arXiv:1409.3526 [math.QA]].

\bibitem{Girelli} 
F.~Girelli, H.~Pfeiffer and E.~M.~Popescu,
``Topological Higher Gauge Theory - from BF to BFCG theory,''
J. Math. Phys. \textbf{49} (2008), 032503
[arXiv:0708.3051 [hep-th]].

\bibitem{Asante:2019lki}
S.~K.~Asante, B.~Dittrich, F.~Girelli, A.~Riello and P.~Tsimiklis,
``Quantum geometry from higher gauge theory,''
Class. Quant. Grav. \textbf{37} (2020) no.20, 205001
[arXiv:1908.05970 [gr-qc]].

\bibitem{Bahr:2009qc}
B.~Bahr and B.~Dittrich,
``Improved and Perfect Actions in Discrete Gravity,''
Phys. Rev. D \textbf{80} (2009), 124030
[arXiv:0907.4323 [gr-qc]].

\bibitem{NewRegge}
B.~Bahr and B.~Dittrich,
``Regge calculus from a new angle,''
New J. Phys. \textbf{12} (2010), 033010
[arXiv:0907.4325 [gr-qc]].

\bibitem{Sorkin1974} R.~Sorkin,
``Time Evolution Problem in Regge Calculus,''
Phys. Rev. D \textbf{12} (1975), 385-396
[erratum: Phys. Rev. D \textbf{23} (1981), 565-565]

\bibitem{Sorkin2019} R.~D.~Sorkin,
``Lorentzian angles and trigonometry including lightlike vectors,''
[arXiv:1908.10022 [gr-qc]].

\bibitem{EffSF1}
S.~K.~Asante, B.~Dittrich and H.~M.~Haggard,
``Effective Spin Foam Models for Four-Dimensional Quantum Gravity,''
Phys. Rev. Lett. \textbf{125} (2020) no.23, 231301
[arXiv:2004.07013 [gr-qc]].

\bibitem{Barrettetal}
J.~W.~Barrett, M.~Rocek and R.~M.~Williams,
``A Note on area variables in Regge calculus,''
Class. Quant. Grav. \textbf{16} (1999), 1373-1376
[arXiv:gr-qc/9710056 [gr-qc]].

\bibitem{ADHAreaR}
S.~K.~Asante, B.~Dittrich and H.~M.~Haggard,
``The Degrees of Freedom of Area Regge Calculus: Dynamics, Non-metricity, and Broken Diffeomorphisms,''
Class. Quant. Grav. \textbf{35} (2018) no.13, 135009
[arXiv:1802.09551 [gr-qc]].

\bibitem{HC1}
B.~Dittrich,
``Modified Graviton Dynamics From Spin Foams: The Area Regge Action,''
[arXiv:2105.10808 [gr-qc]].

\bibitem{HC2}
B.~Dittrich and A.~Kogios,
``From spin foams to area metric dynamics to gravitons,''
[arXiv:2203.02409 [gr-qc]].

\bibitem{DittrichSpeziale}
B.~Dittrich and S.~Speziale,
``Area-angle variables for general relativity,''
New J. Phys. \textbf{10} (2008), 083006
[arXiv:0802.0864 [gr-qc]].

\bibitem{BarrettFO}
J.~W.~Barrett,
``First order Regge calculus,''
Class. Quant. Grav. \textbf{11} (1994), 2723-2730
[arXiv:hep-th/9404124 [hep-th]].

\bibitem{DGS}
B.~Dittrich, S.~Gielen and S.~Schander,
``Lorentzian quantum cosmology goes simplicial,''
Class. Quant. Grav. \textbf{39} (2022) no.3, 035012
[arXiv:2109.00875 [gr-qc]].

\bibitem{Pachner:1991}
U.~Pachner, 
``P.L. Homeomorphic Manifolds are Equivalent by Elementary Shellings,"
Europ.\ J.\ Combinatorics {\bf 12} (1991), 129-145

\bibitem{DittrichHoehnCanSimp}
B.~Dittrich and P.~A.~Hohn,
``Canonical simplicial gravity,''
Class. Quant. Grav. \textbf{29} (2012), 115009
[arXiv:1108.1974 [gr-qc]].

\bibitem{TimeEvol}  B.~Dittrich and S.~Steinhaus,
``Time evolution as refining, coarse graining and entangling,''
New J. Phys. \textbf{16} (2014), 123041
[arXiv:1311.7565 [gr-qc]].

\bibitem{Kokkendorff:2007}
S.~L.~Kokkendorff,
``Polar duality and the generalized law of sines,''
Journal of Geometry {\bf 86} no. 1-2, (2007) 140--149

\bibitem{Baldazzi:2019kim}
A.~Baldazzi, R.~Percacci and V.~Skrinjar,
``Quantum fields without Wick rotation,''
Symmetry \textbf{11} (2019) no.3, 373
[arXiv:1901.01891 [gr-qc]].

\bibitem{Rocek:1981ama}
M.~Rocek and R.~M.~Williams,
``Quantum Regge Calculus,''
Phys. Lett. B \textbf{104} (1981), 31

\bibitem{Bahr:2011uj}
B.~Bahr, B.~Dittrich and S.~Steinhaus,
``Perfect discretization of reparametrization invariant path integrals,''
Phys. Rev. D \textbf{83} (2011), 105026
[arXiv:1101.4775 [gr-qc]].

\bibitem{Davids} S.~Davids,
``A State sum model for (2+1) Lorentzian quantum gravity,''
[arXiv:gr-qc/0110114 [gr-qc]].

\bibitem{Freidel} L.~Freidel,
``A Ponzano-Regge model of Lorentzian 3-dimensional gravity,''
Nucl. Phys. B Proc. Suppl. \textbf{88} (2000), 237-240
[arXiv:gr-qc/0102098 [gr-qc]].

\bibitem{Ponzano:1969}
G.~P.~Ponzano and T.~E.~Regge,
``Semiclassical limit of Racah Coefficients,''
in F Bloch (Ed.),  {\it Spectroscopic and
	Group Theoretical Methods in Physics}, pp. 1-58. North-Holland Publ. Co., Amsterdam, Netherlands, 1968.

\bibitem{Roberts:1998zka}
J.~Roberts,
``Classical 6j-symbols and the tetrahedron,''
Geom. Topol. \textbf{3} (1999) no.1, 21-66
[arXiv:math-ph/9812013 [math-ph]].

\bibitem{Davids:1998bp}
S.~Davids,
``Semiclassical limits of extended Racah coefficients,''
J. Math. Phys. \textbf{41} (2000), 924-943
[arXiv:gr-qc/9807061 [gr-qc]].

\bibitem{AmbjornSpikes}
J.~Ambjorn, J.~L.~Nielsen, J.~Rolf and G.~K.~Savvidy,
``Spikes in quantum Regge calculus,''
Class. Quant. Grav. \textbf{14} (1997), 3225-3241
[arXiv:gr-qc/9704079 [gr-qc]].

\bibitem{PL} E.~Witten,
``A New Look At The Path Integral Of Quantum Mechanics,''
[arXiv:1009.6032 [hep-th]].

\bibitem{Asante:2021blx}
S.~K.~Asante and B.~Dittrich,
``Perfect discretizations as a gateway to one-loop partition functions for 4D gravity,''
JHEP \textbf{05} (2022), 172
[arXiv:2112.03307 [gr-qc]].

\bibitem{Asante:2019ndj}
S.~K.~Asante, B.~Dittrich and F.~Hopfmueller,
``Holographic formulation of 3D metric gravity with finite boundaries,''
Universe \textbf{5} (2019) no.8, 181
[arXiv:1905.10931 [gr-qc]].

\bibitem{Bahr:2010cq}
B.~Bahr, B.~Dittrich and S.~He,
``Coarse graining free theories with gauge symmetries: the linearized case,''
New J. Phys. \textbf{13} (2011), 045009
[arXiv:1011.3667 [gr-qc]].

\bibitem{Asante:2018kfo}
S.~K.~Asante, B.~Dittrich and H.~M.~Haggard,
``Holographic description of boundary gravitons in (3+1) dimensions,''
JHEP \textbf{01} (2019), 144
[arXiv:1811.11744 [hep-th]].

\bibitem{Dittrich:2012jq}
B.~Dittrich,
``From the discrete to the continuous: Towards a cylindrically consistent dynamics,''
New J. Phys. \textbf{14} (2012), 123004
[arXiv:1205.6127 [gr-qc]].

\bibitem{Bonzom:2013ofa}
V.~Bonzom and B.~Dittrich,
``Bubble divergences and gauge symmetries in spin foams,''
Phys. Rev. D \textbf{88} (2013), 124021
[arXiv:1304.6632 [gr-qc]].

\bibitem{Riello}
A.~Riello,
``Self-energy of the Lorentzian Engle-Pereira-Rovelli-Livine and Freidel-Krasnov model of quantum gravity,''
Phys. Rev. D \textbf{88} (2013) no.2, 024011
doi:10.1103/PhysRevD.88.024011
[arXiv:1302.1781 [gr-qc]].

\bibitem{Banburski:2014cwa}
A.~Banburski, L.~Q.~Chen, L.~Freidel and J.~Hnybida,
``Pachner moves in a 4d Riemannian holomorphic Spin Foam model,''
Phys. Rev. D \textbf{92} (2015) no.12, 124014
[arXiv:1412.8247 [gr-qc]].

\bibitem{Dona:2023myv}
P.~Don\`a and P.~Frisoni,
``Summing bulk quantum numbers with Monte Carlo in spin foam theories,''
[arXiv:2302.00072 [gr-qc]].

\bibitem{EffSF3} S.~K.~Asante, B.~Dittrich and J.~Padua-Arguelles,
``Effective spin foam models for Lorentzian quantum gravity,''
Class. Quant. Grav. \textbf{38} (2021) no.19, 195002
[arXiv:2104.00485 [gr-qc]].

\bibitem{EffSF2} S.~K.~Asante, B.~Dittrich and H.~M.~Haggard,
``Discrete gravity dynamics from effective spin foams,''
Class. Quant. Grav. \textbf{38} (2021) no.14, 145023
[arXiv:2011.14468 [gr-qc]].7

\bibitem{EPRL}
J.~Engle, E.~Livine, R.~Pereira and C.~Rovelli,
``LQG vertex with finite Immirzi parameter,''
Nucl. Phys. B \textbf{799} (2008), 136-149
[arXiv:0711.0146 [gr-qc]].

\bibitem{Gozzini}
F.~Gozzini,
``A high-performance code for EPRL spin foam amplitudes,''
Class. Quant. Grav. \textbf{38} (2021) no.22, 225010
[arXiv:2107.13952 [gr-qc]].

\bibitem{Dona:2022yyn}
P.~Dona, M.~Han and H.~Liu,
``Spinfoams and high performance computing,''
[arXiv:2212.14396 [gr-qc]].

\bibitem{Qu1}
M.~Han, Z.~Huang, H.~Liu and D.~Qu,
``Complex critical points and curved geometries in four-dimensional Lorentzian spinfoam quantum gravity,''
Phys. Rev. D \textbf{106} (2022) no.4, 044005
[arXiv:2110.10670 [gr-qc]].

\bibitem{Qu2} 
M.~Han, H.~Liu and D.~Qu,
``Complex critical points in Lorentzian spinfoam quantum gravity: 4-simplex amplitude and effective dynamics on double-$\Delta_3$ complex,''
[arXiv:2301.02930 [gr-qc]].

\bibitem{Visser2} 
K.~Tate and M.~Visser,
``Realizability of the Lorentzian (n,1)-Simplex,''
JHEP \textbf{01} (2012), 028
[arXiv:1110.5694 [gr-qc]].

\bibitem{Hartle}
J.~B.~Hartle,
``Simplicial minisuperspace I. General discussion,''
J.\ Math.\ Phys. \textbf{26} (1985), 804-814.

\bibitem{Dittrich:2007wm}
B.~Dittrich, L.~Freidel and S.~Speziale,
``Linearized dynamics from the 4-simplex Regge action,''
Phys. Rev. D \textbf{76} (2007), 104020
[arXiv:0707.4513 [gr-qc]].


\end{thebibliography}

\endgroup
\end{document}